\documentclass[11pt]{article}
\usepackage{geometry}                
\usepackage{pifont}
\usepackage{xcolor}           
\usepackage{tikz}
\usepackage{soul}
\usepackage{enumitem}
\usepackage{marvosym}
\usepackage{tikzsymbols}
\usepackage{stfloats}
\usepackage{verbatim}
\usepackage{booktabs}
\usepackage{multirow,multicol}
\usepackage{adjustbox}
\usepackage{array}
\usetikzlibrary{patterns}
\usetikzlibrary{matrix}
\usetikzlibrary{tikzmark}
\usetikzlibrary{decorations.markings} 
\usetikzlibrary{calc,arrows} 

\tikzstyle{arrowmid}[0.75]=[decoration= 
{markings, 
	mark=at position #1 with {\arrow{>}}}, 
postaction={decorate}] 

\usetikzlibrary{fit}
\usetikzlibrary{shadows.blur}
\usetikzlibrary{shapes.symbols}
\usepackage{menukeys}
\usetikzlibrary{shapes.geometric, arrows}
\usepackage{cite}
\usepackage{caption}
\usepackage{mdframed}
\usepackage{subcaption}
\usetikzlibrary{positioning}
\usepackage{cancel}
\usepackage{amssymb, amsmath, amsthm}
\usepackage{dsfont}
\usepackage{epstopdf}
\usepackage{algorithmic}
\usepackage{textcomp}
\usepackage{xcolor}
\def\BibTeX{{\rm B\kern-.05em{\sc i\kern-.025em b}\kern-.08em
		T\kern-.1667em\lower.7ex\hbox{E}\kern-.125emX}}
\tikzstyle{arrow} = [thick,->,>=stealth]
\newtheoremstyle{mytheoremstyle}{0pt}{0pt}{\itshape}{}{\bfseries}{.}{.5em}{} 
\newtheorem{definition}{Definition}
\newtheorem{theorem}{Theorem}
\newtheorem{corollary}{Corollary}

\def\bc{{\mbox{\tiny \normalfont BC}}}
\def\ic{{\mbox{\tiny \normalfont IC}}}
\def\xlc{{\mbox{\tiny \normalfont XLC}}}
\def\x{{\mbox{\tiny \normalfont X}}}
\def\iclc{{\mbox{\tiny \normalfont ICLC}}}
\newcommand{\E}{\mathrm{E}}

\newcommand\blfootnote[1]{%
	\begingroup
	\renewcommand\thefootnote{}\footnote{#1}%
	\addtocounter{footnote}{-1}%
	\endgroup
}

\begin{document}
	\title{GDoF of Interference Channel with Limited Cooperation under Finite Precision CSIT}
	\author{Junge Wang, Bofeng Yuan, Lexiang Huang and Syed A. Jafar}
	\date{}
	\maketitle
	
	\blfootnote{Junge Wang (email: jungew@uci.edu) and Syed A. Jafar (email: syed@uci.edu) are with the Center of Pervasive Communications and Computing (CPCC) in the Department of Electrical Engineering and Computer Science (EECS) at the University of California Irvine. Bofeng Yuan (bofeng.yuan@samsung.com) and Lexiang Huang (lzh376@psu.edu) contributed to this work in 2017-18 while they were at UCI during their PhD and undergraduate studies, respectively. The results of this work will appear in part at IEEE GLOBECOM 2019. 
	}
	
	\begin{abstract}
		The Generalized Degrees of Freedom (GDoF) of the two user interference channel  are characterized for all parameter regimes under the assumption of finite precision channel state information at the transmitters (CSIT), when a limited amount of (half-duplex or full-duplex) cooperation is allowed between the transmitters in the form of   $\pi$ DoF of shared messages. In all cases, the number of over-the-air bits that each cooperation bit buys is shown to be equal to either $0, 1, 1/2$ or $1/3$. The most interesting aspect of the result is the  $1/3$ slope, which appears only under finite precision CSIT and strong interference, and as such has not been encountered in previous studies that invariably assumed perfect CSIT. Indeed, the achievability and converse for the parameter regimes with $1/3$ slope are the most challenging aspects of this work. In particular, the converse relies on non-trivial applications of Aligned Images bounds.
	\end{abstract}
	\newpage
	\section{Introduction}
	
	As  distributed computing applications become increasingly practical there is renewed interest in fundamental limits of cooperative communication in \emph{robust} settings. Partially overlapping message sets naturally arise as computing tasks are distributed with some redundancy, e.g.,  to account for straggling nodes and adverse channel conditions \cite{YU_StragglerMitagation}. Studies of cellular communication  with limited backhaul \cite{Sanderovich_Somekh_Poor_Shamai}, unreliable cooperating links \cite{Huleihel_Steinberg}, and variable delay constrained messages \cite{Nikbakht_Wigger_Shitz} lead to similar scenarios as well. An elementary model for information theoretic analysis of such settings is an interference network with a limited amount of shared messages between the  transmitters. While the body of literature on  information theoretic benefits of cooperative communication is too vast to survey here (e.g., see \cite{Simeone_coop}), it is notable that robust settings with finite precision CSIT remain underexplored, especially with limited cooperative capacities. Most closely related to this work are degrees of freedom (DoF) and generalized degrees of freedom (GDoF) studies in \cite{Lapidoth_Shamai_collapse, Arash_Jafar, Arash_Bofeng_Jafar_BC, Yoga_Junge_Jafar, Arash_Jafar_cooperation, Etkin_Tse_Wang, HsiangWang_DavidTse_TX}. Connections to these prior works are explained in the remainder of this section.
	
	Since exact capacity limits tend to be intractable, Generalized Degrees of Freedom (GDoF) studies have emerged as an alternative path to progress for understanding the fundamental limits of wireless networks. Robustness is enforced in GDoF studies by limiting the channel state information at the transmitters (CSIT) to finite precision. Until recently, a stumbling block for robust GDoF characterizations has been the difficulty of obtaining tight converse bounds under finite precision CSIT (cf. Lapidoth-Shamai-Wigger conjecture in \cite{Lapidoth_Shamai_collapse} and the PN conjecture in \cite{Tandon_Jafar_Shamai_Poor}). However, the introduction of aligned images bounds in \cite{Arash_Jafar} has made it possible to circumvent this challenge. Building upon this opportunity, in this work we pursue the  the GDoF of the interference channel under finite precision CSIT with limited cooperation between the transmitters.

	Perhaps the most powerful regime for cooperative communication is the strong interference regime, because the sharing of messages among transmitters allows essentially a re-routing of messages through stronger channels. However, this regime turns out to be also the most challenging regime for information theoretic GDoF characterizations under finite precision CSIT. For example, in \cite{Arash_Bofeng_Jafar_BC} the GDoF are characterized for the $K$ user broadcast channel obtained by full transmitter cooperation in a $K$ user symmetric interference channel with partial CSIT levels. Remarkably, while the GDoF are characterized for the weak interference regime, the strong interference regime remains open. More recently, the extremal GDoF benefits of transmitter cooperation under finite precision CSIT were characterized in \cite{Yoga_Junge_Jafar} for large interference networks. The benefits of cooperation are shown to be substantial, but the extremal analysis is again limited to weak interference settings. Evidently the strong interference regime poses some challenges. To gauge the difficulty of robust  GDoF characterizations in different parameter regimes with limited cooperation,  especially the strong interference regime, in this work we explore the $2$-user setting. 
	
	The main result of this work is the  exact GDoF characterization of the $2$ user interference channel under finite precision CSIT, when a limited amount of cooperation is allowed between the transmitters in the form of  $\pi$ DoF of shared messages. 
	To place this work in perspective, let us note that the GDoF region for the $2$-user broadcast channel (where all messages are shared) under finite-precision CSIT is found in \cite{Arash_Jafar_cooperation}, while the GDoF region of $2$-user interference channel  (where no  messages are shared) under finite-precision CSIT  is the same as that under perfect CSIT \cite{Etkin_Tse_Wang}.  This work  bridges the gap between these two extremes.  Finally, let us recall that under perfect CSIT, Wang and Tse found in \cite{HsiangWang_DavidTse_TX} that each bit of cooperation buys either $0, 1$ or $1/2$ bit over-the-air. In this work, with finite precision CSIT, for all parameter regimes we show that the number of over-the-air bits that each  bit of transmitter cooperation buys  is either $0, 1, 1/2$ or $1/3$. Remarkably, the $1/3$ factor shows up only in the strong interference regime and only under finite precision CSIT. Indeed, the central contribution of this work,  is the strong interference regime which requires the most sophisticated converse and achievability arguments. 
	
	\textit{Notation:} The notation $(x)^+$ represents $\max(x,0)$. Index set $\{1,2,\dots,n\}$ is represented as $[n]$. $f(x)=o(g(x))$ denotes that $\limsup_{x\rightarrow\infty}\frac{|f(x)|}{|g(x)|}=0$. Define $\lfloor x \rfloor$ as the largest integer that is smaller than or equal to $x$ when $x$ is nonnegative.  
	\section{System Model: Interference Channel with Limited Cooperation}
	\begin{figure}[ht]
		\centering
		\begin{tikzpicture}[scale=1.0]
		\node (Tx1) at (-3.4,1) [draw, thick, minimum width=1cm, minimum height=1cm] {Tx1};
		\node [left= 0.5cm of Tx1] (M11){$W_{11}, W_{01}$};
		\draw[thick, ->] (M11)--(Tx1);
		
		\node (Tx2) at (-3.4,-1.5) [draw, thick, minimum width=1cm, minimum height=1cm] {Tx2};
		\node [left= 0.5cm of Tx2] (M22){$W_{22}, W_{02}$};
		\draw[thick, ->] (M22)--(Tx2);
		\draw[thick, ->] ([xshift=-0.2cm]Tx1.south)--([xshift=-0.2cm]Tx2.north) node [left, midway]{$W_{01}$};
		\draw[thick, ->] ([xshift=0.2cm]Tx2.north)--([xshift=0.2cm]Tx1.south) node [right, midway]{$W_{02}$};
		
		\coordinate (T1) at (2cm,1cm);
		\coordinate (T2) at (2cm,-1.5cm);
		\coordinate (R1) at (5cm,1cm);
		\coordinate (R2) at (5cm,-1.5cm);
		\node [left=0.2cm of T1] (X1) {\small $X_1$};
		\node [left=0.2cm of T2] (X2) {\small $X_2$};
		\node [right=0.2cm of R1] (Y1) {\small $Y_1$};
		\node [right=0.2cm of R2] (Y2) {\small $Y_2$};
		\draw[thick] (2cm,1cm) circle(2mm);
		\draw[thick] (5,1) circle(2mm);
		\draw[thick] (2,-1.5) circle(2mm);
		\draw[thick] (5,-1.5) circle(2mm);
		\node [left=1.5cm of T1] (M1all) {$(W_{11}, W_{01},W_{02})$};
		\node [left=1.5cm of T2] (M2all) {$(W_{22}, W_{01},W_{02})$};
		\node [right=1.5cm of R1] (M1hat) {$(\widehat{W}_{11}, \widehat{W}_{01})$};
		\draw[thick,->](Y1)--(M1hat);
		
		\node [right=1.5cm of R2] (M2hat) {$(\widehat{W}_{22}, \widehat{W}_{02})$};
		\draw[thick,->](Y2)--(M2hat);
		\draw[thick, ->] (M1all)--(X1);
		\draw[thick, ->] (Tx1)--(M1all);
		\draw[thick, ->] (M2all)--(X2);
		\draw[thick, ->] (Tx2)--(M2all);

		\draw[thick,arrowmid] (2.2,1)--(4.8,1)node[above,pos=0.5]{$\alpha_{11}$};
		\draw[thick,arrowmid] (2.2,-1.5)--(4.8,-1.5)node[below,pos=0.5]{$\alpha_{22}$};
		\draw[thick,arrowmid] (2.2,1)--(4.8,-1.5)node[right,pos=0.8]{$\alpha_{21}$};
		\draw[thick,arrowmid] (2.2,-1.5)--(4.8,1)node[right,pos=0.8]{$\alpha_{12}$};

		\end{tikzpicture}
		\caption{\it\small Interference Channel with Limited Cooperation. The rates of cooperative messages $W_{01}, W_{02}$ are limited by the cooperation capability $\pi$.}
	\end{figure}
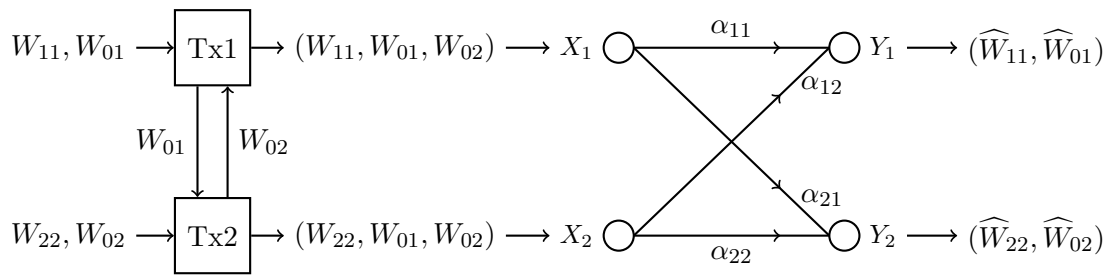
	The interference channel with limited cooperation is comprised of $4$ independent messages: $W_{11}, W_{22}, W_{01}, W_{02}$. Messages $W_{11}, W_{22}$ are the noncooperative messages that originate at Transmitters $1,2$, and are intended for Receivers $1,2$, respectively. Messages $W_{01}, W_{02}$ are the cooperative messages intended for Receivers $1,2$, respectively, with the distinction that these messages are assumed to be known to both transmitters because they are shared among the transmitters through the limited conference link. Specifically, message $W_{01}$ is sent through the cooperation link by Transmitter $1$ to Transmitter $2$, and message $W_{02}$ is sent through the cooperation link by Transmitter $2$ to Transmitter $1$. 
	
	For GDoF studies, the $2$-user interference channel with limited cooperation is described by the following input-output relationship.
	\begin{align}
	Y_1(t)&=\sqrt{P^{\alpha_{11}}}G_{11}(t)X_1(t)+\sqrt{P^{\alpha_{12}}}G_{12}(t)X_2(t)+Z_1(t)\label{icmodel_1}\\
	Y_2(t)&=\sqrt{P^{\alpha_{21}}}G_{21}(t)X_1(t)+\sqrt{P^{\alpha_{22}}}G_{22}(t)X_2(t)+Z_2(t)\label{icmodel_2}
	\end{align}
	During the $t^{th}$ use of the channel, $X_i(t)=f_{i,t}(W_{ii},W_{01},W_{02})\in\mathbb{C}$ is the symbol sent from Transmitter $i$, and  is  subject to unit transmit power constraint. The symbol observed by Receiver $i$ is denoted $Y_i(t)\in\mathbb{C}$, and $Z_i(t)\sim\mathcal{N_C}(0,1)$ is the zero mean unit variance additive white Gaussian noise (AWGN) at Receiver $i$. The variable $P$ is referred to as power and represents a nominal parameter that approaches infinity to define the GDoF limit. The parameters $\alpha_{ki}\in\mathbb{R}^+$ represent the coarse channel strength  between Transmitter $i$ and Receiver $k$, respectively. To understand the intuition behind the GDoF model, it is useful to think of  $\alpha_{ki}$ as the (approximate) capacity of the physical channel between Transmitter $i$ and Receiver $k$ in a given finite SNR setting that we wish to study. The GDoF model scales the capacity of every link by the same factor $\gamma=\log(P)$. Note that in the GDoF model the capacity of the physical channel between Transmitter $i$ and Receiver $k$ is approximately $\alpha_{ki}\log(P)$. Intuitively, the reason for this proportional scaling of capacities is the expectation of approximate scale invariance, i.e., when the the capacity of every link in a network is scaled by the same factor $\gamma$, then we expect that the capacity of the overall network should scale approximately by the same factor $\gamma$ as well. So normalizing the capacity of the network by  $\gamma=\log(P)$  yields an approximation to the capacity of the original finite SNR network; hence the normalization  by $\log(P)$ of the rates in the GDoF definition (see \eqref{def:GDoF}).
	
	The power $P$ and the channel strengths $\alpha_{ki}$ are known to all transmitters and receivers. $G_{ki}(t)\in\mathbb{C}$ are the channel coefficient values, known perfectly to the receivers. Robustness is enforced by the assumption that the channel coefficients are only available to transmitters with \emph{finite precision}. Recall that under the finite precision CSIT assumption,  as defined in \cite{Arash_Jafar}, the transmitters are only aware of the probability density functions of the channel coefficients, and it is assumed that all joint and conditional probability density functions of channel coefficients exist and are bounded. As in  \cite{Arash_Jafar}, to avoid degenerate conditions, the channel coefficients are also assumed to be bounded away from $0$ and infinity, i.e., all $G_{ki}(t)\in [1/\Delta, \Delta]$ for some positive finite constant $\Delta$. The set of all channel coefficient random variables is denoted $\mathcal{G}=\{G_{ki}(t)~|~ i,k\in\{1,2\}, t\in\mathbb{Z}_+\}$.

	The  rates associated with messages $W_{11}, W_{22}, W_{01}, W_{02}$ are denoted as $R_{11}, R_{22}, R_{01}, R_{02}$, respectively.  The definitions of probability of error, achievable rate tuples $(R_{11},R_{22},R_{01},R_{02})$, codebooks and capacity region $\mathcal{C}$ are all in the standard Shannon-theoretic sense (see for example \cite{NIT}).
	The GDoF region is defined as,
	\begin{align} \label{def:GDoF}
	\mathcal{D}=\left\{(d_{11},d_{22},d_{01},d_{02}): 	
	\begin{array}{lll}\\
	\exists((R_{11}(P),R_{22}(P),R_{01}(P),R_{02}(P))\in\mathcal{C}(P)\vspace{0.25cm}\\ 
	\begin{array}{lll}
	s.t.\quad d_{11}=\lim\limits_{P\rightarrow\infty}\frac{R_{11}(P)}{\log(P)},&d_{22}=\lim\limits_{P\rightarrow\infty}\frac{R_{22}(P)}{\log(P)},\\
	\quad\quad d_{01}=\lim\limits_{P\rightarrow\infty}\frac{R_{01}(P)}{\log(P)},&
	d_{02}=\lim\limits_{P\rightarrow\infty}\frac{R_{02}(P)}{\log(P)}
	\end{array}
	\end{array}\right\}
	\end{align}

	The total cooperation capability of the system is fixed by a given parameter $\pi$. We focus in particular on two models for cooperation, half-duplex and full-duplex, represented by the following assumptions.
	\begin{align}
	\mbox{Half-duplex Assumption: }&&d_{01}+d_{02}\leq \pi,\label{eq:pi}\\
	\mbox{Full-duplex Assumption: }&&d_{01}\leq\frac{\pi}{2},~~~d_{02}\leq \frac{\pi}{2}.\label{eq:pi'}
	\end{align}
	Thus, the half-duplex assumption implies that the capacity of the cooperation link is limited to $\pi$ GDoF, which can be divided arbitrarily between the two one-way modes, while the full-duplex assumption implies that the capacity of the cooperation link is limited to $\frac{\pi}{2}$, which can be simultaneously utilized in both directions without mutual interference. Correspondingly, the sum-GDoF value  of the interference channel with limited cooperation is denoted as $\mathcal{D}_{\Sigma,\iclc}$ for the half-duplex model, and as $\mathcal{D}_{\Sigma,\iclc}'$ for the full-duplex model. In each case, the sum-GDoF value is the maximum value of $d_{11}+d_{22}+d_{01}+d_{02}$ across all $(d_{11},d_{22},d_{01},d_{02})$ tuples the GDoF region.

	\subsection{Interference Channel}
	The interference channel corresponds to the setting with no cooperation, i.e., $\pi=0$, so there are no cooperative messages $W_{01},W_{02}$. In \cite{Etkin_Tse_Wang}, the GDoF region of the interference channel is characterized under perfect CSIT. As noted in \cite{Arash_Jafar_IC}, for the $2$-user interference channel, GDoF  under finite precision CSIT are the same as that under perfect CSIT. The sum-GDoF value, denoted $\mathcal{D}_{\Sigma,\ic}$ is found to be,
	\begin{align}\label{ic}
	\mathcal{D}_{\Sigma,\ic}=\min\bigg(&\max(\alpha_{11}-\alpha_{21} ,\alpha_{12})+\max(\alpha_{22}-\alpha_{12},\alpha_{21}),\notag\\ &\max(\alpha_{11},\alpha_{12})+(\alpha_{22}-\alpha_{12})^+,\notag\\ &\max(\alpha_{21},\alpha_{22})+(\alpha_{11}-\alpha_{21})^+,\notag\\ &\alpha_{11}+\alpha_{22}\bigg)
	\end{align}
	
	\subsection{Broadcast Channel}
	The broadcast channel corresponds to unlimited cooperation, i.e., $\pi\rightarrow\infty$, so that only cooperative messages $W_{01}, W_{02}$ are needed for the sum-GDoF characterization. The sum-GDoF value, denoted $\mathcal{D}_{\Sigma,\bc}$ under finite-precision CSIT is found in \cite{Arash_Jafar_cooperation} as,
	\begin{align}\label{bc}
	\mathcal{D}_{\Sigma,\bc}=\min\big(&\max(\alpha_{11},\alpha_{12})+\max(\alpha_{21}-\alpha_{11},\alpha_{22}-\alpha_{12})^+,\notag\\ &\max(\alpha_{21},\alpha_{22})+\max(\alpha_{11}-\alpha_{21},\alpha_{12}-\alpha_{22})^+\big)
	\end{align}
	Note that unlike the interference channel, the broadcast channel suffers a loss in GDoF due to finite precision CSIT as compared to perfect CSIT. 
	
	\subsection{Weak, Mixed and Strong Interference Regimes}
	The range of values of $\alpha_{ki}$ parameters is partitioned into three regimes, labeled weak, mixed and strong interference. These regimes are defined as follows.
	\begin{align}
	\text{Weak interference: }& \max(\alpha_{12},\alpha_{21})\leq\min(\alpha_{11},\alpha_{22}) \\
	\text{Mixed interference: }&\min(\alpha_{12},\alpha_{21})\leq\max(\alpha_{11},\alpha_{22}),\max(\alpha_{12},\alpha_{21})\geq\min(\alpha_{11},\alpha_{22})\\
	\text{Strong interference: }& \max(\alpha_{11},\alpha_{22})\leq\min(\alpha_{12},\alpha_{21})
	\end{align}
	The boundaries between  regimes may be considered to belong to either regime.
	
	\subsection{Sub-Messages} In the description of the achievable scheme, we partition messages into sub-messages, and in labeling these sub-messages we use subscripts to indicate transmitter cooperation, while the superscripts are associated with the decodability of the message. Specifically, if the subscript contains a $0$ then that part of the message is shared between the two transmitters, otherwise it is not. Similarly, if the superscript is a $p$ then that part of the message is private, i.e., only decodable at its desired receiver, otherwise it is common, i.e., decodable by both receivers. Specifically,  the noncooperative message $W_{ii}$ and cooperative message $W_{0i}$ are split into  common and private parts, so that $W_{ii}=(W_{ii}^c,W_{ii}^p),W_{0i}=(W_{0i}^c,W_{0i}^p)$, and we have the following sub-messages:\\
	{\small $W_{11}^p$: Noncooperative private message, encoded by Transmitter $1$ and decoded by Receiver $1$.\\
		$W_{22}^p$: Noncooperative private message, encoded by Transmitter $2$ and decoded by Receiver $2$.\\
		$W_{11}^c$: Noncooperative common message, encoded by Transmitter $1$, decoded by both receivers.\\
		$W_{22}^c$: Noncooperative common message, encoded by Transmitter $2$, decoded by both receivers.\\
		$W_{01}^p$: Cooperative private message,  private part of $W_{01}$, encoded\footnote{Note that even though $W_{01}^p$ is a cooperative message, i.e., it is known to both transmitters and as such could be jointly encoded by both transmitters, our achievable schemes only require it to be encoded by Transmitter $2$. Similar observation holds for $W_{02}^p$ as well.} by Transmitter $2$, decoded by Receiver $1$.\\
		$W_{02}^p$: Cooperative private message,  private part of $W_{02}$, encoded by Transmitter $1$, decoded by Receiver $2$.\\
		$W_{01}^c$: Cooperative common message,  common part of $W_{01}$,  encoded by both transmitters, decoded by both receivers.\\
		$W_{02}^c$: Cooperative common message,  common part of $W_{02}$, encoded by both transmitters, decoded by both receivers.\\
		$W_{0}^c$: Combination of common parts of $W_{01},W_{02}$, i.e., $W_0^c=(W_{01}^c,W_{02}^c)$. 	
	}

	\section{Results}\label{results}
	Under the half-duplex model, the sum-GDoF value for the interference channel with limited transmitter cooperation under finite precision CSIT is characterized in the following theorem.
	\begin{theorem}\label{theorem:GDoF} Under the half-duplex model, in the weak and mixed interference regime, we have
		\begin{align}
		\mathcal{D}_{\Sigma,\iclc}&=\min\Big(\mathcal{D}_{\Sigma,\ic}+\pi,\mathcal{D}_{\Sigma,\bc}\Big), \label{eq:other}
		\intertext{and in the strong interference regime}
		\mathcal{D}_{\Sigma,\iclc}&=\min\Big(\mathcal{D}_{\Sigma,\ic}+\pi,\frac{\mathcal{D}_{2e}+\pi}{2},\frac{\mathcal{D}_{3e}+\pi}{3},\mathcal{D}_{\Sigma,\bc}\Big) \label{eq:strong}
		\end{align}
		where \begin{align}
		\mathcal{D}_{2e}&=\alpha_{12}+\alpha_{21}\\
		\mathcal{D}_{3e}&=\min(\alpha_{21}-\alpha_{22},\alpha_{11})+2\max(\alpha_{21}-\alpha_{11},\alpha_{22})+\alpha_{12}+\max(\alpha_{12}-\alpha_{22},\alpha_{11}). \label{D_3e}
		\end{align}
	\end{theorem}
	As an immediate corollary, we obtain the minimum value of $\pi$ needed for the interference channel to achieve the same sum-GDoF  value as the broadcast channel.
	\begin{corollary}
		Let $\pi^*$ denote the minimum half-duplex cooperation GDoF needed to achieve the broadcast channel bound.  In the strong interference regime with an assumption $\alpha_{12}\geq\alpha_{21}$, $\pi^*>\mathcal{D}_{\Sigma,\bc}-\mathcal{D}_{\Sigma,\ic}$, and its value is given below
		\begin{align}
		\pi^*=\left\{
		\begin{array}{lcl}
		N-2\max(\alpha_{11},\alpha_{22}) && \alpha_{12},\alpha_{21}\leq M, N\leq M+\alpha_{11}\\
		2N-M-3\max(\alpha_{11},\alpha_{22}) && \alpha_{12},\alpha_{21}\leq M, N\geq M+\alpha_{11}\\
		N+\alpha_{21}-3\max(\alpha_{11},\alpha_{22}) &&\alpha_{12}\geq M, \alpha_{21}\leq M\\
		N+M-3\max(\alpha_{11},\alpha_{22}) &&\alpha_{12}\geq M, \alpha_{21}\geq M\\
		\end{array}
		\right.
		\end{align}
		where $M=\alpha_{11}+\alpha_{22},N=\alpha_{12}+\alpha_{21}$. In all other parameter regimes, $\pi^*=\mathcal{D}_{\Sigma,\bc}-\mathcal{D}_{\Sigma,\ic}$.
	\end{corollary} 
	Our next result is the sum-GDoF characterization of the interference channel with limited \emph{full-duplex} transmitter cooperation, under finite precision CSIT, as presented in the following theorem.
	\begin{theorem}\label{theorem:FDGDoF}Under the full-duplex model, in the weak interference regime, we have	\begin{align}
		\mathcal{D}_{\Sigma,\iclc}'&=\min\Big(\mathcal{D}_{\Sigma,\ic}+\pi,\mathcal{D}_{\Sigma,\bc}\Big), \label{eq:weak2}
		\end{align}
		in the mixed interference regime we have,
		\begin{align}
		\mathcal{D}_{\Sigma,\iclc}'&=\min\Big(\mathcal{D}_{\Sigma,\ic}+\frac{\pi}{2},\mathcal{D}_{\Sigma,\bc}\Big) \label{eq:mixed2}
		\end{align}
		and in the strong interference regime we have,
		\begin{align}
		\mathcal{D}_{\Sigma,\iclc}'&=\min\Big(\mathcal{D}_{\Sigma,\ic}+\pi,\min(\alpha_{12},\alpha_{21})+\frac{\pi}{2},\frac{\mathcal{D}_{3e}+\pi}{3},\mathcal{D}_{\Sigma,\bc}\Big) \label{eq:strong2}
		\end{align}
		where $\mathcal{D}_{3e}$ is the same as in \eqref{D_3e}.
	\end{theorem}
	Similarly, as a  corollary we obtain the minimum value of $\pi$ needed for the interference channel with full-duplex cooperation to achieve the same sum-GDoF  value as the broadcast channel.	
	\begin{corollary}
		Let $\pi^+$ denote the minimum full-duplex cooperation GDoF needed to achieve the broadcast channel bound. In the weak interference regime, $\pi^+=(\mathcal{D}_{\Sigma,\bc}-\mathcal{D}_{\Sigma,\ic})$. In the mixed interference regime, $\pi^+=2(\mathcal{D}_{\Sigma,\bc}-\mathcal{D}_{\Sigma,\ic})$. In the strong interference regime, where we assume $\alpha_{21}\leq\alpha_{12}$ without loss of generality, the value of $\pi^+$ is given below
		\begin{align}
		\pi^+=\left\{
		\begin{array}{lcl}
		2N-M-3\max(\alpha_{11},\alpha_{22}) && \alpha_{12},\alpha_{21}\leq M,2\alpha_{21}\geq M+\max(\alpha_{11},\alpha_{22})\\
		N+\alpha_{21}-3\max(\alpha_{11},\alpha_{22}) &&\alpha_{12}\geq M, \alpha_{21}\leq M,\alpha_{12}\leq2\alpha_{21}-\max(\alpha_{11},\alpha_{22})\\
		N+M-3\max(\alpha_{11},\alpha_{22}) &&\alpha_{12}\geq M, \alpha_{21}\geq M,\alpha_{12}\leq\alpha_{21}+\min(\alpha_{11},\alpha_{22})\\
		2\alpha_{12}-2\max(\alpha_{11},\alpha_{22})&&otherwise,\\
		\end{array}
		\right.
		\end{align}
		where $M=\alpha_{11}+\alpha_{22},N=\alpha_{12}+\alpha_{21}$. 
	\end{corollary}

	To place the results in perspective, let us present some observations and examples.
	\begin{enumerate}
		\item A comparison of Theorem \ref{theorem:FDGDoF} with Theorem \ref{theorem:GDoF} reveals that the sum-GDoF of the full-duplex setting are identical to the half-duplex setting, i.e., for the same amount of total cooperation capability, with only two exceptions -- the mixed interference regime where the full-duplex bound $\mathcal{D}_{\Sigma,\ic}+\frac{\pi}{2}$ is different from the half-duplex bound $\mathcal{D}_{\Sigma,\ic}+\pi$, and the strong interference regime where the full-duplex bound $\min(\alpha_{12},\alpha_{21})+\frac{\pi}{2}$ is different from the half-duplex bound $0.5(\alpha_{12}+\alpha_{21}+\pi)$. A notable insight here is that when either of these bounds is active in the full-duplex setting, then only one-way cooperation is needed, i.e., half of the cooperation capability is wasted in the full-duplex setting.
		\begin{figure}[h]
			\includegraphics[width=10cm]{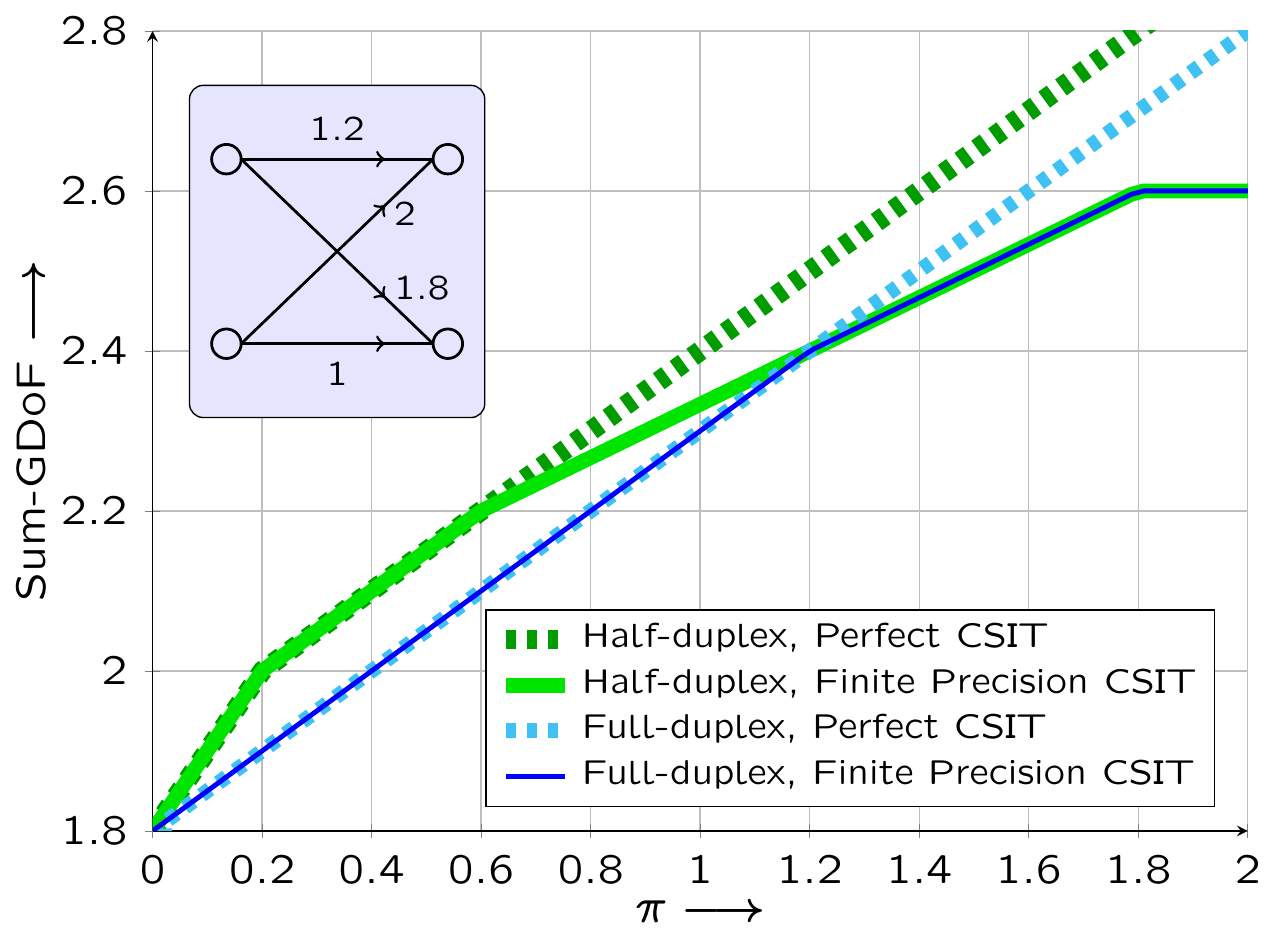}
			\centering
			\caption{\small \it Sum-GDoF of the interference channel ($\alpha_{11}=1.2,\alpha_{22}=1,\alpha_{12}=2,\alpha_{21}=1.8)$ with limited cooperation for half-duplex and full-duplex settings, under perfect \cite{HsiangWang_DavidTse_TX} and finite precision CSIT (this work).}
			\label{fig2}
		\end{figure}
		\item The slope of sum-GDoF with respect to $\pi$ for full-duplex and half-duplex, respectively, represents how many over-the-air bits are bought with each  bit of total cooperation capability. Based on Theorem \ref{theorem:GDoF} and Theorem \ref{theorem:FDGDoF} the slope only takes values $0,1,1/2$, or $1/3$. Figure \ref{fig2} shows an example where the slopes $0, 1, 1/2, 1/3$ can all be seen. Note that in Figure \ref{fig2}, half-duplex cooperation has greater slope than full-duplex cooperation for $0\leq\pi\leq 0.2$, and  smaller slope than full-duplex cooperation for $0.6\leq\pi\leq 1.2$. Thus, the incremental benefit from each additional bit of cooperation capability may be greater for either half-duplex or full-duplex cooperation in different regimes. Also note that the benefits of cooperation saturate much more quickly under finite precision CSIT.
		\item In general, for both full-duplex and half-duplex settings, each incremental bit of cooperation capability buys either $0,1,1/2$ or $1/3$ additional over-the-air bit. Compare this to the findings in \cite{HsiangWang_DavidTse_TX} for perfect CSIT, where each  incremental bit of cooperation capability buys either $0,1$, or $1/2$ additional bit over-the-air. The $1/3$ slope  appears only under finite precision CSIT and only under strong interference. In fact,  the GDoF bounds with slope $1/3$ are the only\footnote{Of course, the bound corresponding to the sum-GDoF of the broadcast channel takes different values under perfect CSIT and finite precision CSIT. Under perfect CSIT, we have $\mathcal{D}_{\Sigma,\bc}=\max(\alpha_{11}+\alpha_{22}, \alpha_{12}+\alpha_{21})$, while under finite precision CSIT, the value is given by \eqref{bc}.} bounds in Theorem 1 and Theorem 2 that do not appear in the perfect CSIT setting studied in \cite{HsiangWang_DavidTse_TX}. Indeed, the converse and achievability for the parameter regimes where the $1/3$ slope appears are the central contributions of this work.
		\item A notable insight here is that when the $1/3$ slope appears, it is because each incremental $\epsilon$ increase in GDoF corresponds to an $\epsilon$ increase in the GDoF of each of the three cooperative messages $W_{01}^p,W_{02}^p,W_0^c$, and a simultaneous $\epsilon$ decrease in the GDoF of each of the two noncooperative messages $W_{11},W_{22}$. Therefore, the total increase in GDoF is $\epsilon$, while the total increase in the required cooperation capability is $3\epsilon$, which gives us the $1/3$ slope.  
		
		\begin{figure}[h]
			\includegraphics[width=15cm]{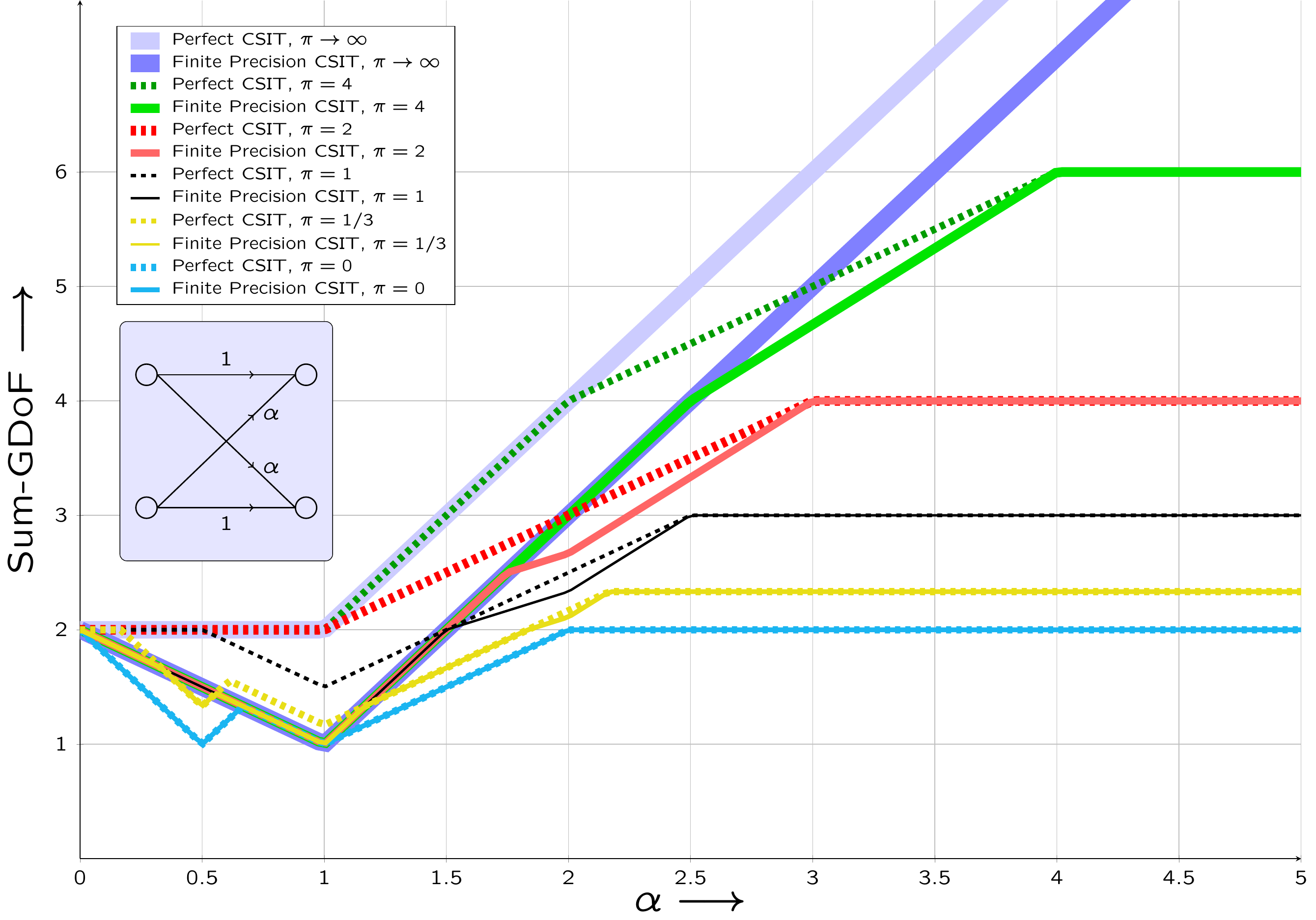}
			\centering
			\caption{\it\small Sum-GDoF of the symmetric interference channel ($\alpha_{11}=\alpha_{22}=1,\alpha_{12}=\alpha_{21}=\alpha)$ with limited cooperation for various half-duplex and full-duplex settings, under perfect \cite{HsiangWang_DavidTse_TX} and finite precision CSIT (this work).}
			\label{fig1}
		\end{figure}
		\item Figure \ref{fig1} plots the sum-GDoF value of the $2$ user interference channel with limited cooperation under the symmetric setting $(\alpha_{11}=\alpha_{22}=1, \alpha_{12}=\alpha_{21}=\alpha)$  for both half-duplex and full-duplex cooperation models, under both perfect CSIT \cite{HsiangWang_DavidTse_TX} and finite-precision CSIT (this work). Note that in this symmetric setting, full-duplex cooperation and half-duplex cooperation have identical sum-GDoF as a function of $\pi$. This is because the mixed interference regime does not appear in the symmetric setting, and  in the strong interference regime the full-duplex bound $\min(\alpha_{12},\alpha_{21})+\frac{\pi}{2}$ matches the half-duplex bound $0.5(\alpha_{12}+\alpha_{21}+\pi)$. Note that there is no cooperation gain for  $2/3\leq\alpha\leq1$, which recovers the results in \cite{Arash_Jafar_cooperation}.
		Furthermore, for any fixed  cooperation capability $\pi$, as $\alpha$ increases, eventually the sum-GDoF with perfect CSIT match the sum-GDoF of finite precision CSIT, as they both converge to $2+\pi$.  In fact, this is true more generally (even with asymmetric settings) in the following sense. For any fixed values of $(\pi, \alpha_{11},\alpha_{22})$, as the cross-channels $\alpha_{12}, \alpha_{21}$ become stronger, the sum-GDoF for both finite precision CSIT and perfect CSIT must converge to $\alpha_{11}+\alpha_{22}+\pi$. This is because as the cross-channels become stronger, each receiver is able to decode all interference and desired signals without interference, so the sum-GDoF for each user are only limited by the min-cut between its transmitter and receiver. Thus, the total GDoF of User $1$, $d_1$ is only limited by $\alpha_{11}+d_{01}$, and similarly $d_2$ is only limited by $\alpha_{22}+d_{02}$, so that the sum-GDoF are only limited by $\alpha_{11}+\alpha_{22}+\pi$. 
		
		\item From Theorems $1$ and $2$ we can find the minimum amount of cooperation capability needed to achieve any given sum-GDoF value. In particular, the minimum amount of cooperation needed to achieve the same sum-GDoF as with unlimited cooperation, i.e., $\mathcal{D}_{\Sigma,\bc}$, is specified in Corollaries $1$ and $2$. Since the $\mathcal{D}_{\Sigma,\bc}$ was characterized previously in \cite{Arash_Jafar_cooperation}, a natural question is to gauge the efficiency of the achievable schemes used in \cite{Arash_Jafar_cooperation}. Since cooperation efficiency is not a concern in \cite{Arash_Jafar_cooperation}, understandably the schemes from  \cite{Arash_Jafar_cooperation} that achieve $\mathcal{D}_{\Sigma,\bc}$  are in general not the most efficient in terms of the amount of cooperation needed. This is shown explicitly through the examples in Figure \ref{fig:ex1} and Figure \ref{fig:ex2}. Evidently, even for settings where the sum-GDoF are already known, the most efficient solution in terms of the minimum required level of cooperation is a non-trivial question that is answered by Theorems $1$ and $2$.
	\end{enumerate}
	
	\section{Proof of General Converse}
	Let us recall some definitions that are needed for aligned images bounds.
	\begin{definition}[Power Levels]\label{def:plevel} Consider the integer valued random variables $X_i$ over alphabet $\mathcal{X}_{\lambda_i}$
		\begin{align}
		\mathcal{X}_{\lambda_i}\triangleq\{0,1,2,\cdots,\bar{P}^{\lambda_i}-1\}
		\end{align}
		where $\bar{P}^{\lambda_i}\triangleq\lfloor \sqrt{P^{\lambda_i}}\rfloor $. We are primarily interested in limits as $P\rightarrow\infty$, where $P\in \mathbb{R}_+$ is denoted as power. The constant $\lambda_i$ refers to the power level of $X_i$.
	\end{definition}

	\begin{definition} Consider integer valued random variables  $X\in\mathcal{X}_{\lambda}$, and any nonnegative real numbers $\lambda_1,\lambda_2$ such that $0\leq\lambda_1\leq\lambda_2\leq\lambda$, define
		\begin{align}
		(X)^{\lambda_2}&\triangleq \Bigl\lfloor\frac{X}{\bar{P}^{\lambda-\lambda_2}} \Bigr\rfloor\label{def_top}\\
		(X)_{\lambda_1}&\triangleq X-\bar{P}^{\lambda_1}\Bigl\lfloor\frac{X}{\bar{P}^{\lambda_1}}\Bigr\rfloor\\
		(X)^{\lambda_2}_{\lambda_1}&\triangleq \Bigl\lfloor \frac{(X)_{\lambda_2}}{\bar{P}^{\lambda_1}} \Bigr\rfloor
		\end{align}
		In other words, $(X)^{\lambda_2}$ retrieves the top $\lambda_2$ power levels of X, $(X)_{\lambda_1}$ retrieves the bottom $\lambda_1$ power levels of X, $(X)^{\lambda_2}_{\lambda_1}$ retrieves the partition of X between levels $\lambda_1$ and $\lambda_2$. 
	\end{definition}

	Let us prove the outer bounds on the GDoF region of the interference channel with limited cooperation under finite precision CSIT, for arbitrary levels of cooperation,  $d_{01}\leq\pi_{01},d_{02}\leq\pi_{02}$. These bounds can then be specialized to obtain the tight converse for both half-duplex  and full-duplex models. As noted previously, with the exception of the bounds that have slope $1/3$ (as a function of $\pi$), all other bounds that we need for Theorem 1 and Theorem 2 also hold under perfect CSIT, so they can be obtained from \cite{HsiangWang_DavidTse_TX}. However, for the sake of completeness we will prove all the bounds here. 
	
	The bound $\mathcal{D}_{\Sigma,\iclc}\leq \mathcal{D}_{\Sigma,\bc}$ is  trivial because full cooperation cannot reduce GDoF. The bound $\mathcal{D}_{\Sigma,\iclc}\leq\mathcal{D}_{\Sigma,\ic}+\pi_{01}+\pi_{02}$ is also trivial because $d_{11}+d_{22}\leq\mathcal{D}_{\Sigma,\ic}$ and $d_{01}\leq\pi_{01},d_{02}\leq\pi_{02}$ by assumption. These bounds suffice for the weak interference regime in both half-duplex and full-duplex settings. 
	
	Next, let us consider the bounds that are needed for the mixed and strong interference regimes. Here we will use the Aligned Images bounds approach, starting with  the deterministic model of \cite{Arash_Jafar} whose GDoF region contains the GDoF region of the original channel model from above.
	\begin{align}
	\bar{Y}_1(t)&=\lfloor{\sqrt{P^{\alpha_{11}-\max(\alpha_{11},\alpha_{21})}}}G_{11}(t){\bar X}_1(t)\rfloor+\lfloor {\sqrt{P^{\alpha_{12}-\max(\alpha_{12},\alpha_{22})}}G_{12}(t)\bar{X}_2(t)}\rfloor\label{deticmodel_1}\\
	\bar{Y}_2(t)&=\lfloor{ \sqrt{P^{\alpha_{21}-\max(\alpha_{11},\alpha_{21})}}G_{21}(t)\bar{X}_1(t)}\rfloor+\lfloor{\sqrt{P^{\alpha_{12}-\max(\alpha_{12},\alpha_{22})}}}G_{22}(t)\bar{X}_2(t)\rfloor\label{deticmodel_2}
	\end{align}
	where $\bar{X}_{i}(t)=\bar{X}_{iR}(t)+j\bar{X}_{iI}(t)$, $i\in\{1,2\}$, and $\bar{X}_{1R},\bar{X}_{1I}\in\{0,1,2,\cdots, \lceil\sqrt{P^{\max(\alpha_{11},\alpha_{21})}}\rceil\}$, while $\bar{X}_{2R},\bar{X}_{2I}\in\{0,1,2,\cdots, \lceil\sqrt{P^{\max(\alpha_{12},\alpha_{22})}}\rceil\}$.
	
	Applying  Fano's inequality and ignoring the $o(\log(P))$ terms that are inconsequential for GDoF, we have
	\begin{align}
	nR_{22}+nR_{02}&\leq I(W_{22},W_{02};\bar{Y}^{[n]}_2\mid \mathcal{G})\\
	&\leq H(W_{22},W_{02};\bar{Y}^{[n]}_2\mid W_{01},\mathcal{G})\\
	&=H(\bar{Y}^{[n]}_2\mid W_{01},\mathcal{G})-H(\bar{Y}^{[n]}_2\mid W_{22},W_{01},W_{02},\mathcal{G})\\
	&=H(\bar{Y}^{[n]}_2\mid W_{01},\mathcal{G})-H((\bar{X}^{[n]}_1)^{\alpha_{21}}\mid W_{22},W_{01},W_{02},\mathcal{G}) \label{mx_1}
	\end{align}
	where (\ref{mx_1}) holds because $\bar{X}^{[n]}_2$ is a function of $(W_{22},W_{01},W_{02})$, and $\bar{Y}^{[n]}_2$ is a function of $(\bar{X}^{[n]}_1)^{\alpha_{21}}$ and $(\bar{X}^{[n]}_2)^{\alpha_{22}}$ because Receiver $2$ only hears the top $\alpha_{21}$ dimensional space of $\bar{X}^{[n]}_1$ and top $\alpha_{22}$ dimensional space of $\bar{X}^{[n]}_2$ above the noise floor. In addition, from Fano's inequality, 
	\begin{align}
	nR_{11}&\leq I(W_{11};\bar{Y}^{[n]}_1\mid \mathcal{G})\\
	&\leq I(W_{11};\bar{Y}^{[n]}_1\mid W_{01},W_{02},W_{22},\mathcal{G})\\
	&\leq H(\bar{Y}^{[n]}_1\mid W_{01},W_{02},W_{22},\mathcal{G})\label{32} \\
	&=H(\lfloor{\sqrt{P^{\alpha_{11}-\max(\alpha_{11},\alpha_{21})}}}G_{11}^{[n]}{\bar X}^{[n]}_1\rfloor\mid W_{01},W_{02},W_{22},\mathcal{G})\label{mx_2}\\
		&\leq H(\lfloor{\sqrt{P^{\alpha_{11}-\max(\alpha_{11},\alpha_{21})}}}G_{11}^{[n]}{\bar X}^{[n]}_1\rfloor, (\bar{X}^{[n]}_1)^{\alpha_{11}}\mid W_{22},W_{01},W_{02},\mathcal{G})  \label{mx_2a}\\
				&\leq H( (\bar{X}^{[n]}_1)^{\alpha_{11}}\mid W_{22},W_{01},W_{02},\mathcal{G})  +no(\log(P))\label{mx_2b}		
	\end{align}
	where (\ref{mx_2}) holds because given $(W_{22},W_{01},W_{02})$ and $\mathcal{G}$, Receiver $1$ is able to reconstruct $\bar{X}^{[n]}_2$ and then subtract $\lfloor {\sqrt{P^{\alpha_{12}-\max(\alpha_{12},\alpha_{22})}}G_{12}^{[n]}\bar{X}^{[n]}_2}\rfloor$ from received signal, and the remaining received signal at Receiver $1$ is $\lfloor{\sqrt{P^{\alpha_{11}-\max(\alpha_{11},\alpha_{21})}}}G_{11}^{[n]}{\bar X}^{[n]}_1\rfloor$. Then \eqref{mx_2b} holds because conditioned on any value of  $(\bar{X}^{[n]}_1)^{\alpha_{11}}=\lfloor{\sqrt{P^{\alpha_{11}-\max(\alpha_{11},\alpha_{21})}}}{\bar X}^{[n]}_1\rfloor$, there can be\footnote{Since $\lfloor b \bar{X}\rfloor\leq b\bar{X}=(b/a)a\bar{X}\leq (b/a)(\lfloor a\bar{X}\rfloor+1)$ and $\lfloor b\bar{X}\rfloor \geq b\bar{X}-1= (b/a)(a\bar{X})-1\geq (b/a)\lfloor a\bar{X}\rfloor -1$, it follows that given $a,b,\lfloor a \bar{X}\rfloor$ there are no more than $|b/a|+2$ possible values for $\lfloor b\bar{X}\rfloor$. Also recall that $|G_{11}(t)|\leq \Delta$.} no more than $(\Delta+2)^n$ possible values of $\lfloor{\sqrt{P^{\alpha_{11}-\max(\alpha_{11},\alpha_{21})}}}G_{11}^{[n]}{\bar X}^{[n]}_1\rfloor$, so $H(\lfloor{\sqrt{P^{\alpha_{11}-\max(\alpha_{11},\alpha_{21})}}}G_{11}^{[n]}{\bar X}^{[n]}_1\rfloor\mid (\bar{X}^{[n]}_1)^{\alpha_{11}}, W_{22},W_{01},W_{02},\mathcal{G})\leq n\log(\Delta+2)=no(\log(P))$.
		
	Adding \eqref{mx_1} and \eqref{mx_2}, then applying the Aligned image inequalities (Lemma 1 in \cite{Arash_Jafar_IC}), we have
	\begin{align}
	nR_{11}+nR_{22}+nR_{02}&\leq H(\bar{Y}^{[n]}_2|W_{01}, \mathcal{G})+
	\notag\\
	&\big[H((\bar{X}^{[n]}_1)^{\alpha_{11}}\mid W_{22},W_{01},W_{02},\mathcal{G})-H((\bar{X}^{[n]}_1)^{\alpha_{21}}\mid W_{22},W_{01},W_{02},\mathcal{G})\big]\\
	&\leq n\max(\alpha_{21},\alpha_{22})\log(P)+n(\alpha_{11}-\alpha_{21})^+\log(P).\label{eq:mixed_1}
	\end{align}
	Similarly, 
	\begin{align}
	nR_{11}+nR_{22}+nR_{01}\leq n\max(\alpha_{12},\alpha_{11})\log(P)+n(\alpha_{22}-\alpha_{12})^+\log(P).\label{eq:mixed_2}
	\end{align}
	Dividing  both sides in \eqref{eq:mixed_1} and \eqref{eq:mixed_2} by $n\log(P)$, and applying the GDoF limit, we obtain the following GDoF bounds:
	\begin{align}
	d_{11}+d_{22}+d_{02}\leq \max(\alpha_{21},\alpha_{22})+(\alpha_{11}-\alpha_{21})^+\label{GDoF_mixed_1}\\ 
	d_{11}+d_{22}+d_{01}\leq \max(\alpha_{12},\alpha_{11})+(\alpha_{22}-\alpha_{12})^+ \label{GDoF_mixed_2}
	\end{align}
	Thus,  the following bound is obtained. 
	\begin{align}
	\mathcal{D}_{\Sigma,\iclc}\leq\min(\max(\alpha_{21},\alpha_{22})+(\alpha_{11}-\alpha_{21})^++\pi_{01},\max(\alpha_{12},\alpha_{11})+(\alpha_{22}-\alpha_{12})^++\pi_{02})
	\end{align}
	This bound is useful in the mixed and strong regimes. Note that in the strong interference regime, the bound can be simplified as 
	\begin{align}
	\mathcal{D}_{\Sigma,\iclc}\leq\min(\alpha_{21}+\pi_{01}, \alpha_{12}+\pi_{02}).
	\end{align}
	Finally, consider the strong interference regime, and in particular, the case $\alpha_{12}\geq\alpha_{21}$. The alternative setting of $\alpha_{12}\leq\alpha_{21}$ will follow similarly.
	For ease of notation, define 
	\begin{align}
	A&=\left\{
	\begin{array}{lcl}
	(\bar{X}_1^{[n]})^{\alpha_{21}}_{\alpha_{22}} &&\alpha_{21}\leq\alpha_{11}+\alpha_{22}\\
	(\bar{X}_1^{[n]})^{\alpha_{21}}_{\alpha_{21}-\alpha_{11}}  && \alpha_{21}\geq\alpha_{11}+\alpha_{22}
	\end{array}
	\right.\\
	B&=\left\{
	\begin{array}{lcl}
	(\bar{X}_1^{[n]})^{\alpha_{22}}_{\alpha_{21}-\alpha_{11}} &&\alpha_{21}\leq\alpha_{11}+\alpha_{22}\\
	0  && \alpha_{21}\geq\alpha_{11}+\alpha_{22}
	\end{array}
	\right.\\
	C&=(\bar{X}_2^{[n]})^{\alpha_{12}}_{\alpha_{12}-\alpha_{22}}
	\end{align}
	\begin{figure}[t]
		\centering
		\begin{tikzpicture}[scale=1.0]
		\begin{scope}[shift={(0,0.5)}]
		\draw[ thick] (0,0) rectangle (1,1.6);
		\draw (0.5, 0) node[below]{$\bar{X}_1$};
		\draw[ thick, fill=cyan!30!white, pattern color=blue] (0,0.6) rectangle (1,0.8)node [pos=.5] {\tiny $B$};
		\draw[arrows=<->] (-0.15,0.6)--(-0.15,0.8)node[left,pos=0.5]{\small $\theta$};
		\draw[ thick, fill=blue!30!white, pattern color=blue] (0,0.8) rectangle (1,1.6)node [pos=.5] {\tiny$A$};
		\draw[arrows=<->] (-0.15,0.8)--(-0.15,1.6)node[left,pos=0.5]{\small $\eta$};
		\draw[ thick] (0,-0.8) rectangle (1,-2.8);
		\draw (0.5, -2.8) node[below]{$\bar{X}_2$};
		\draw[ thick, fill=red!20!white, pattern color=blue] (0,-0.8) rectangle (1,-1.6) node [pos=.5] {\tiny$C$};
		\draw[arrows=<->] (-0.15,-0.8)--(-0.15,-1.6)node[left,pos=0.5]{\small $\alpha_{22}$};
		\end{scope}		
		\draw[thick] (2,1) circle(2mm);
		\draw[thick] (5,1) circle(2mm);
		\draw[thick] (2,-1.5) circle(2mm);
		\draw[thick] (5,-1.5) circle(2mm);
		
		\draw[thick,arrowmid] (2.2,1)--(4.8,1)node[above,pos=0.5]{$\alpha_{11}$};
		\draw[thick,arrowmid] (2.2,-1.5)--(4.8,-1.5)node[below,pos=0.5]{$\alpha_{22}$};
		\draw[thick,arrowmid] (2.2,1)--(4.8,-1.5)node[right,pos=0.8]{$\alpha_{21}$};
		\draw[thick,arrowmid] (2.2,-1.5)--(4.8,1)node[right,pos=0.8]{$\alpha_{12}$};
		
		\begin{scope}[shift={(6.5,0.5)}]
		\draw[thick, fill=cyan!30!white, pattern color=blue] (0,0) rectangle (1,0.2)node [pos=.5] {\tiny $B$};
		\draw[arrows=<->] (-0.15,0)--(-0.15,0.2)node[left,pos=0.5]{\small $\theta$};
		\draw[ thick, fill=blue!30!white, pattern color=blue] (0,0.2) rectangle (1,1.0) node [pos=.5] {\tiny $A$};
		\draw[arrows=<->] (-0.15,0.2)--(-0.15,1.0)node[left,pos=0.5]{\small $\eta$};
		\draw (0.4, 1.0) node[above]{\tiny $(X_1)^{\alpha_{21}}_{\delta}$};
		
		\draw[thick] (1,0) rectangle (2,1.2);
		\draw[thick, fill=red!20!white, pattern color=blue] (1,1.2) rectangle (2,2) node [pos=.5] {\tiny$C$};
		\draw[arrows=<->] (2.15,1.2)--(2.15,2)node[right,pos=0.5]{\small $\alpha_{22}$};
		\draw (1.5, 2) node[above]{\tiny $X_2$};
		
		\draw[thick] (1,0) node[below]{$\bar{Y}_1$};
		\end{scope}
		
		\begin{scope}[shift={(6.5,-2.3)}]
		\draw[thick] (0,0) rectangle (1,0.6);
		\draw[thick, fill=cyan!30!white, pattern color=blue] (0,0.6) rectangle (1,0.8)node [pos=.5] {\tiny $B$};
		\draw[arrows=<->] (-0.15,0.6)--(-0.15,0.8)node[left,pos=0.5]{\small $\theta$};
		\draw[ thick, fill=blue!30!white, pattern color=blue] (0,0.8) rectangle (1,1.6) node [pos=.5] {\tiny $A$};
		\draw[arrows=<->] (-0.15,0.8)--(-0.15,1.6)node[left,pos=0.5]{\small $\eta$};
		\draw (0.5, 1.6) node[above]{\tiny $X_1$};
		
		\draw[thick, fill=red!20!white, pattern color=blue] (1,0) rectangle (2,0.8) node [pos=.5] {\tiny$C$};
		\draw[arrows=<->] (2.15,0)--(2.15,0.8)node[right,pos=0.5]{\small $\alpha_{22}$};
		\draw (1.8, 0.8) node[above]{\tiny  $(X_2)^{\alpha_{12}}_{\gamma}$};
		
		\draw[thick] (1,0) node[below]{$\bar{Y}_2$};
		\end{scope}
		\end{tikzpicture}
		\caption{\it\small Power level partitions $A,B,C$ where $\eta=\alpha_{21}-\alpha_{22},\theta=\alpha_{11}+\alpha_{22}-\alpha_{21},\delta=\alpha_{21}-\alpha_{11},\gamma=\alpha_{12}-\alpha_{22}  ,\bar{X}_1\in\mathcal{X}_{\alpha_{21}},\bar{X}_2\in\mathcal{X}_{\alpha_{12}}$ and $\alpha_{21}\leq\alpha_{11}+\alpha_{22}$.}\label{def_AB}
	\end{figure}
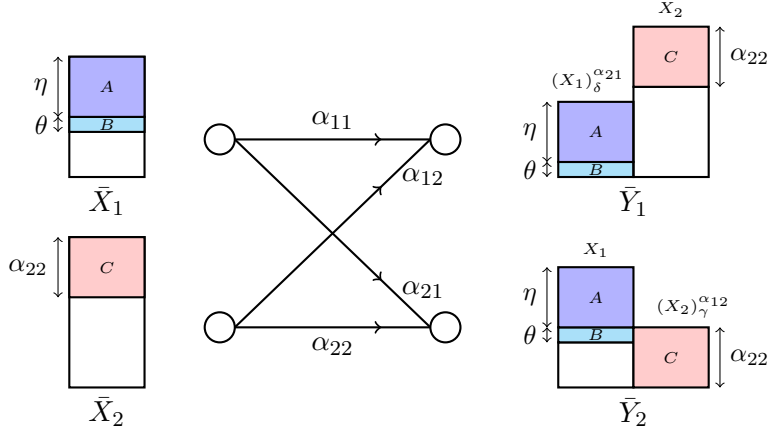

	Figure \ref{def_AB} illustrates the definitions for the case $\alpha_{21}\leq\alpha_{11}+\alpha_{22}$, where the notation ${[n]}$ is omitted for simplicity. The case $\alpha_{21}\geq\alpha_{11}+\alpha_{22}$ can be shown similarly. 
	Note that if $\alpha_{21}\leq \alpha_{11}+\alpha_{22}$, then $A$ represents the top $\alpha_{21}-\alpha_{22}$ power levels of $\bar{X}_1^{[n]}$, and $B$ represents the remaining power level partition of $\bar{X}_1^{[n]}$ that appears above the noise floor at Receiver $1$. Otherwise, if  $\alpha_{21}\geq \alpha_{11}+\alpha_{22}$, then $A$ represents the top $\alpha_{11}$ levels of $\bar{X}_1^{[n]}$ and $B$ is zero. The combination of $A,B$ is the partition of $\bar{X}_1^{[n]}$ that is heard by Receiver $1$ above the noise floor. Note that in both cases, $A$ represents the power level partition of $\bar{X}_1^{[n]}$ that is heard  above the signal due to $\bar{X}_2^{[n]}$ at Receiver $2$, i.e., 
	\begin{align}
	H(A\mid \bar{Y}_2^{[n]},\mathcal{G})&=no(\log(P))
	\end{align}
	$C$ represents the top $\alpha_{22}$ power levels of $\bar{X}_2^{[n]}$, which is all that Receiver $2$ is able to hear from Transmitter 2. Note that the sum of power levels of $A$ and $C$ is always less than $\alpha_{12}$, which will be important when applying the sum-set inequality. 
	
	Because $C$ is a function of $W_{22}, W_{01}, W_{02}$,
	\begin{align}
	&H(C\mid W_{22},W_{02},\mathcal{G})\notag\\
	&=I(C;W_{01}\mid W_{22},W_{02},\mathcal{G})\\
	&\leq I(A,C;W_{01}\mid W_{22},W_{02},\mathcal{G})\notag\\
	&=I(A;W_{01}\mid W_{22},W_{02},\mathcal{G})+I(C;W_{01}\mid W_{22},W_{02},A,\mathcal{G})\\
	&\leq I(A;W_{01}\mid W_{22},W_{02},\mathcal{G})+H(C\mid W_{22},W_{02},A,\mathcal{G})
	\end{align}
	Thus,
	\begin{align}
	I(A;W_{01}\mid W_{22},W_{02},\mathcal{G})\geq H(C\mid W_{22},W_{02},\mathcal{G})-H(C\mid W_{22},W_{02},A,\mathcal{G}) \label{inequ_2}
	\end{align}
	At the same time, we also have the following bound,
	\begin{align}
	H(C\mid W_{22},W_{02},\mathcal{G})&\geq H(C\mid W_{02},\mathcal{G})-H(W_{22}|W_{02},\mathcal{G})\\
	&= H(C\mid W_{02},\mathcal{G})-H(W_{22}\mid \mathcal{G})\\
	&\geq H(C\mid W_{02},\mathcal{G})-I(C; W_{22}\mid W_{11},W_{01},W_{02},\mathcal{G})\\
	&\geq H(C\mid W_{02},\mathcal{G})-H(C\mid W_{11},W_{01},W_{02},\mathcal{G})\\
	&=I(C; W_{11}, W_{01}\mid W_{02},\mathcal{G})\\
	&=I(C,\bar{Y_1}^{[n]}; W_{11}, W_{01}\mid W_{02},\mathcal{G})-I(\bar{Y_1}^{[n]}; W_{11}, W_{01}\mid C,W_{02},\mathcal{G})\\
	&\geq I(\bar{Y_1}^{[n]};W_{11},W_{01}\mid W_{02},\mathcal{G})+I(C;W_{11},W_{01}\mid \bar{Y_1}^{[n]},\mathcal{G})\notag\\
	&\hspace{1cm}-H(\bar{Y_1}^{[n]}\mid C,W_{02},\mathcal{G})+H(\bar{Y_1}^{[n]}\mid C,W_{11},W_{01},W_{02},\mathcal{G})\label{inequ1}\\
	&\geq  I(\bar{Y_1}^{[n]};W_{11},W_{01}\mid W_{02},\mathcal{G})-H(\bar{Y_1}^{[n]}\mid C,W_{02},\mathcal{G}) \label{inequ2}\\
	&\geq I(\bar{Y_1}^{[n]};W_{11},W_{01}\mid \mathcal{G})-H(\bar{Y_1}^{[n]}\mid C,W_{02},\mathcal{G}) \\
	&\geq I(\bar{Y_1}^{[n]};W_{11},W_{01}\mid \mathcal{G})-H(\bar{Y_1}^{[n]}\mid C,\mathcal{G}) \label{inequ3}\\
	&\geq nR_{11}+nR_{01}-H(\bar{Y_1}^{[n]}\mid C,
	\mathcal{G}) \label{inequ_3}
	\end{align}
	Where \eqref{inequ1} is because mutual information and entropy are no less than zero. \eqref{inequ2} is because $(W_{11},W_{01})$ is independent from $W_{02}$ and for any three random variables $U,V,T$, if $V$ and $T$ are independent, then
	\begin{align}
	I(U;V)\leq I(U;V\mid T)
	\end{align}
	\eqref{inequ3} is because conditioning cannot increase entropy.
	
	Next, from Fano's inequality, we have 
	\begin{align}
	nR_{11}+nR_{01}
	&\leq I(W_{11},W_{01};\bar{Y_1}^{[n]}\mid \mathcal{G})\\
	&\leq H(\bar{Y_1}^{[n]}\mid \mathcal{G})-H(A,C|W_{11},W_{01},\mathcal{G}) \label{sumset}\\
	&\leq H(\bar{Y_1}^{[n]}\mid \mathcal{G})-H(A,C|W_{11},W_{01},\mathcal{G}) \\
	&= H(\bar{Y_1}^{[n]}\mid \mathcal{G})-H(A\mid W_{11},W_{01},\mathcal{G})-H(C|W_{11},W_{01},A,\mathcal{G})\\
	&\leq H(\bar{Y_1}^{[n]}\mid \mathcal{G})-H(A\mid W_{11},W_{01},\mathcal{G})-H(C|W_{11},W_{01},A, W_{02},\mathcal{G})\\
	&= H(\bar{Y_1}^{[n]}\mid \mathcal{G})-H(A\mid W_{11},W_{01},\mathcal{G})-H(C\mid W_{11},W_{01},W_{02},\mathcal{G})\\
	&= H(\bar{Y_1}^{[n]}\mid \mathcal{G})-H(A\mid W_{11},W_{01},\mathcal{G})-nR_{22} \\
	&= H(\bar{Y_1}^{[n]}\mid \mathcal{G})-I(A;W_{22},W_{02}\mid W_{11},W_{01},\mathcal{G})-nR_{22}\\
	&\leq H(\bar{Y_1}^{[n]}\mid \mathcal{G})-I(A;W_{22},W_{02}\mid \mathcal{G})-nR_{22}
	\end{align}
	where (\ref{sumset}) is due to the sumset inequality (Theorem 1 in \cite{Arash_Jafar_sumset}). Rearranging the above inequality we get 
	\begin{align}
	I(A;W_{22},W_{02}\mid \mathcal{G})\leq H(\bar{Y_1}^{[n]}\mid \mathcal{G})-n(R_{11}+R_{22}+R_{01}) \label{fa_1}
	\end{align}
	Next, applying Fano's inequality at Receiver $2$, we have
	\begin{align}
	nR&_{22}+nR_{02}\notag\\
	&\leq I(W_{22},W_{02};\bar{Y_2}^{[n]}\mid \mathcal{G})\\
	&\leq I(W_{22},W_{02};\bar{Y_2}^{[n]},A\mid \mathcal{G})\\
	&=I(W_{22},W_{02};A\mid \mathcal{G})+I(W_{22},W_{02};\bar{Y_2}^{[n]}\mid A, \mathcal{G})\\
	&=I(W_{22},W_{02};A\mid \mathcal{G})+H(\bar{Y_2}^{[n]}\mid A, \mathcal{G})-H(\bar{Y_2}^{[n]}\mid A,W_{22},W_{02}, \mathcal{G})\\
	&=I(W_{22},W_{02};A\mid \mathcal{G})+H(\bar{Y_2}^{[n]}\mid A, \mathcal{G})-H(\bar{Y_2}^{[n]}\mid W_{22},W_{02}, \mathcal{G})+\notag\\&\hspace{8cm}I(\bar{Y}_2^{[n]}; A\mid W_{22},W_{02},  \mathcal{G})\label{fa_3}\\
	&\leq I(W_{22},W_{02};A\mid \mathcal{G})+H(\bar{Y}_2^{[n]}\mid A,  \mathcal{G})-H(A,C\mid W_{22},W_{02}, \mathcal{G})+\notag\\&\hspace{8cm}I(\bar{Y}_2^{[n]}; A\mid W_{22},W_{02}, \mathcal{G})\label{eq:sumsetac}\\
	&\leq I(W_{22},W_{02};A\mid \mathcal{G})+H(\bar{Y_2}^{[n]}\mid A, \mathcal{G})-H(C\mid A,W_{22},W_{02},\mathcal{G})\label{fa_2}
	\end{align}
	where in \eqref{eq:sumsetac}, we used sum-set inequality from Theorem 1 in \cite{Arash_Jafar_sumset}.
	Combining (\ref{fa_1}) and (\ref{fa_2}),
	\begin{align}
	H(C\mid A,W_{22},W_{02},\mathcal{G})&\leq H(\bar{Y}^{[n]}_1\mid \mathcal{G})+H(\bar{Y}_2^{[n]}\mid A, \mathcal{G})-n(R_{11}+2R_{22}+R_{01}+R_{02})\label{inequ_4}
	\end{align}
	Combining (\ref{inequ_2}), (\ref{inequ_3}), (\ref{inequ_4}), we have,
	\begin{align}
	I(A;W_{01}\mid W_{22},W_{02},\mathcal{G})&\geq n(2R_{11}+2R_{22}+2R_{01}+R_{02})\notag\\
	&\hspace{1cm}-H(\bar{Y}_1^{[n]}\mid C, \mathcal{G})-H(\bar{Y}_1^{[n]}\mid \mathcal{G})-H(\bar{Y}_2^{[n]}\mid A,  \mathcal{G})\label{eq:lowerbound}
	\end{align}
	Using again the sum-set inequality from Theorem 1 in \cite{Arash_Jafar_sumset} we have,
	\begin{align}
	H(\bar{Y}_2^{[n]}\mid W_{22},W_{02},\mathcal{G})\geq H(A,B\mid W_{22},W_{02},\mathcal{G}) \label{eq:sumsetab}
	\end{align}
	Combining \eqref{eq:sumsetab} with (\ref{fa_3}), and rearranging the terms we get
	\begin{align}
	H(B\mid W_{22},W_{02},A)&\leq H(A)+H(\bar{Y}_2^{[n]}\mid A, \mathcal{G})-n(R_{22}+R_{02})\notag\\
	&\hspace{1cm}-H(A\mid W_{22},W_{02},\mathcal{G}) \label{ine_1}
	\end{align}
	Message $W_{11}$ can only be transmitted through $A,B$ because it needs to be successfully decoded by User 1. Therefore,
	\begin{align}
	nR_{11}&\leq H(A,B\mid W_{22},W_{02},W_{01},\mathcal{G})\\
	&=
	H(A\mid W_{22}, W_{02}, W_{01},\mathcal{G})+H(B\mid W_{22}, W_{02}, W_{01},A,\mathcal{G})\\
	&\leq H(A\mid W_{22}, W_{02}, W_{01},\mathcal{G})+H(B\mid W_{22}, W_{02},A,,\mathcal{G}) \label{ine_2}
	\end{align}
	Combining (\ref{ine_1}) and (\ref{ine_2}), we get
	\begin{align}
	I(A;W_{01}\mid &W_{02},W_{22},\mathcal{G})\leq H(A\mid \mathcal{G})+H(\bar{Y}_2^{[n]}\mid A,\mathcal{G})-n({R_{11}+R_{22}+R_{02}}).\label{eq:upperbound}
	\end{align}
	Because \eqref{eq:upperbound} and \eqref{eq:lowerbound} are upper and lower bound on the same mutual information, combining them we have
	\begin{align}
	3n(R_{11}&+R_{22})+2n(R_{01}+R_{02})\leq\notag \\
	&H(A\mid \mathcal{G})+2H(\bar{Y}_2^{[n]}\mid A,\mathcal{G})+H(\bar{Y}_1^{[n]}\mid \mathcal{G})+H(\bar{Y}_1^{[n]}\mid C,\mathcal{G}) \label{bd3}
	\end{align}
	Note that the following bounds hold, with $o(\log(P))$ terms omitted.
	\begin{align}
	H(A\mid \mathcal{G})&\leq n\min(\alpha_{21}-\alpha_{22},\alpha_{11})\log(P)\\
	H(\bar{Y}_2^{[n]}|A,\mathcal{G})&\leq n\max(\alpha_{21}-\alpha_{11},\alpha_{22})\log(P)\\
	H(\bar{Y}_1^{[n]}\mid \mathcal{G})&\leq n\alpha_{12}\log(P)\\
	H(\bar{Y}_1^{[n]}|C,\mathcal{G})&\leq n\max(\alpha_{12}-\alpha_{22},\alpha_{11})\log(P)
	\end{align}
	Thus, (\ref{bd3}) yields the GDoF bound, 
	\begin{align}
	3d_{11}+3d_{22}+2d_{01}+2d_{02}\leq D_{3e}
	\end{align}
	Combining it with the assumption $d_{01}\leq\pi_{01},d_{02}\leq\pi_{02}$, we get the bound
	\begin{align}
	\mathcal{D}_{\Sigma,\iclc}\leq\frac{D_{3e}+\pi_{01}+\pi_{02}}{3}.
	\end{align}
	Proceeding similarly, the same bound is obtained for $\alpha_{21}\geq \alpha_{12}$.
	
	At this point, let us list the bounds that we have shown along with the regimes where they are useful.
	\begin{align}
	\intertext{\sc Weak Interference Regime:}	
	\mathcal{D}_{\Sigma,\iclc}\leq\min\Big(\mathcal{D}_{\Sigma,\ic}+\pi_{01}+\pi_{02},\mathcal{D}_{\Sigma,\bc}\Big)\label{weak}\\ 
	\intertext{\sc Mixed Interference Regime:}
	\mathcal{D}_{\Sigma,\iclc}\leq\min\Big(\mathcal{D}_{\Sigma,\ic}+\pi_{01}+\pi_{02},\max(\alpha_{21},\alpha_{22})+(\alpha_{11}-\alpha_{21})^++\pi_{01},\notag\\
	\hspace{2cm}\max(\alpha_{12},\alpha_{11})+(\alpha_{22}-\alpha_{12})^++\pi_{02},\mathcal{D}_{\Sigma,\bc}\label{mixed}\Big) 
	\intertext{\sc Strong Interference Regime:}
	\mathcal{D}_{\Sigma,\iclc}\leq\min\Big(\mathcal{D}_{\Sigma,\ic}+\pi_{01}+\pi_{02},\alpha_{12}+\pi_{02},\alpha_{21}+\pi_{01},\frac{\mathcal{D}_{3e}+\pi_{01}+\pi_{02}}{3},\mathcal{D}_{\Sigma,\bc}\label{strong}\Big)
	\end{align}
	
	Next, we show how these bounds provide a tight converse for Theorem 1 as well as Theorem 2. 
	\subsection{Converse for Theorem 1 and Theorem 2}
	\subsubsection{Weak Interference}
	First consider the weak interference regime where we apply the bound \eqref{weak}. Setting $\pi_{01}+\pi_{02}\leq \pi$ for the half-duplex setting we recover the tight converse bound,  $\mathcal{D}_{\Sigma,\iclc}\leq \min(\mathcal{D}_{\Sigma,\bc}, \mathcal{D}_{\Sigma,\ic}+\pi)$. Similarly, setting $\pi_{01}\leq\frac{\pi}{2}, \pi_{02}\leq\frac{\pi}{2}$ for the full-duplex setting, we obtain the tight converse bound, $\mathcal{D}_{\Sigma,\iclc}'\leq \min(\mathcal{D}_{\Sigma,\bc}, \mathcal{D}_{\Sigma,\ic}+\pi)$.
	
	\subsubsection{Mixed Interference}\label{converse:mixed}
	Next consider the mixed interference regime. The converse for the half-duplex case with mixed interference is trivial because the bounds are identical to the weak interference regime. So let us focus on the full-duplex case. It follows from the sum-GDoF  under finite precision CSIT of  the broadcast channel (reference \cite{Arash_Jafar_cooperation}, summarized in \eqref{bc}), and the interference channel without cooperation (reference \cite{HsiangWang_DavidTse_TX,Arash_Jafar_cooperation}, summarized in \eqref{ic}),  that in the mixed interference regime, there is no cooperation gain, i.e., $\mathcal{D}_{\Sigma,\bc}=\mathcal{D}_{\Sigma,\ic}$, when either of the following conditions holds. 
	\begin{enumerate}
	\item $\alpha_{11}+\alpha_{22}\geq\alpha_{12}+\alpha_{21}$
	\item $\min(\alpha_{11},\alpha_{22})\leq\min(\alpha_{12},\alpha_{21})\leq\max(\alpha_{12},\alpha_{21})\leq\max(\alpha_{11},\alpha_{22})$
	\end{enumerate}
	In both cases  the trivial bound $\mathcal{D}_{\Sigma,\iclc}\leq \mathcal{D}_{\Sigma,\bc}$ is tight. Henceforth in this section we will only consider the remainder of the mixed interference regime, which excludes $\alpha_{11}+\alpha_{22}\geq\alpha_{12}+\alpha_{21}$ and $\min(\alpha_{11},\alpha_{22})\leq\min(\alpha_{12},\alpha_{21})\leq\max(\alpha_{12},\alpha_{21})\leq\max(\alpha_{11},\alpha_{22})$.
	
	Let us assume without loss of generality that $\alpha_{22}\leq\alpha_{11}$. Next, let us define $\max(\alpha_{21},\alpha_{22})+(\alpha_{11}-\alpha_{21})^+$ as $\Lambda_1$, and similarly $\max(\alpha_{12},\alpha_{11})+(\alpha_{22}-\alpha_{12})^+$ as $\Lambda_2$, so the bound \eqref{mixed} can be written as:
	\begin{align*}
	\mathcal{D}_{\Sigma,\iclc}\leq\min\left(\mathcal{D}_{\Sigma,\ic}+\pi_{01}+\pi_{02},\Lambda_1+\pi_{01}, \Lambda_2+\pi_{02}, \mathcal{D}_{\Sigma,\bc}\right)
	\end{align*}
	Now let us show that  one of $\Lambda_1, \Lambda_2$ is equal to $\mathcal{D}_{\Sigma,\ic}$ and the other is equal to $\mathcal{D}_{\Sigma,\bc}$. This will be useful to simplify the bound later. We have the following four cases.
		\begin{itemize}
		\item $\alpha_{21}\leq\alpha_{22}\leq\alpha_{11}\leq\alpha_{12}$ \\
		$\Lambda_1$ is $\alpha_{11}+\alpha_{22}-\alpha_{21}=\mathcal{D}_{\Sigma,\ic}$, and $\Lambda_2$ is $\alpha_{12}=\mathcal{D}_{\Sigma,\bc}$.
		\item $\alpha_{12}\leq\alpha_{22}\leq\alpha_{11}\leq\alpha_{21}$ \\
		$\Lambda_1$ is $\alpha_{21}=\mathcal{D}_{\Sigma,\bc}$, and $\Lambda_2$ is $\alpha_{11}+\alpha_{22}-\alpha_{12}=\mathcal{D}_{\Sigma,\ic}$.
		\item $\alpha_{22}\leq\alpha_{12}\leq\alpha_{11}\leq\alpha_{21}$\\
		$\Lambda_1$ is $\alpha_{21}=\mathcal{D}_{\Sigma,\bc}$, and $\Lambda_2$ is $\alpha_{11}=\mathcal{D}_{\Sigma,\ic}$.
		\item $\alpha_{22}\leq\alpha_{21}\leq\alpha_{11}\leq\alpha_{12}$\\
		$\Lambda_1$ is $\alpha_{11}=\mathcal{D}_{\Sigma,\ic}$, and $\Lambda_2$ is $\alpha_{12}=\mathcal{D}_{\Sigma,\bc}$.
	\end{itemize}
	
	Next, let us apply the bound \eqref{mixed} to the full-duplex setting \eqref{eq:pi'} which corresponds to $\pi_{01}\leq \frac{\pi}{2}, \pi_{02}\leq\frac{\pi}{2}$. In the mixed interference regime with $\alpha_{11}+\alpha_{22}\leq\alpha_{12}+\alpha_{21}$, we have,
	\begin{align*}
	\mathcal{D}_{\Sigma,\iclc}'&\leq\min\left(\mathcal{D}_{\Sigma,\ic}+\pi_{01}+\pi_{02},\Lambda_1+\pi_{01}, \Lambda_2+\pi_{02}, \mathcal{D}_{\Sigma,\bc}\right)\\
	&\leq \min\left(\mathcal{D}_{\Sigma,\ic}+\pi,\mathcal{D}_{\Sigma,\ic}+\frac{\pi}{2}, \mathcal{D}_{\Sigma,\bc}+\frac{\pi}{2}, \mathcal{D}_{\Sigma,\bc}\right)\\
	&\leq \min\left(\mathcal{D}_{\Sigma,\ic}+\frac{\pi}{2},  \mathcal{D}_{\Sigma,\bc}\right)
	\end{align*}
	Thus, a tight converse for the full-duplex setting is obtained in the mixed interference regime.
	\subsubsection{Strong Interference}
	Let us apply the bound \eqref{strong} to the half-duplex setting \eqref{eq:pi} which corresponds to $\pi_{01}+\pi_{02}\leq\pi$. Here we have,
	\begin{align}
	\mathcal{D}_{\Sigma,\iclc}&\leq\min\Big(\mathcal{D}_{\Sigma,\ic}+\pi_{01}+\pi_{02},\alpha_{12}+\pi_{02},\alpha_{21}+\pi_{01},\frac{\mathcal{D}_{3e}+\pi_{01}+\pi_{02}}{3},\mathcal{D}_{\Sigma,\bc}\Big)\\
	&\leq \min\Big(\mathcal{D}_{\Sigma,\ic}+\pi,\frac{\alpha_{12}+\alpha_{21}+\pi_{01}+\pi_{02}}{2},\frac{\mathcal{D}_{3e}+\pi}{3},\mathcal{D}_{\Sigma,\bc}\Big)\\
	&\leq \min\Big(\mathcal{D}_{\Sigma,\ic}+\pi,\frac{\alpha_{12}+\alpha_{21}+\pi}{2},\frac{\mathcal{D}_{3e}+\pi}{3},\mathcal{D}_{\Sigma,\bc}\Big)\\
	&\leq \min\Big(\mathcal{D}_{\Sigma,\ic}+\pi,\frac{D_{2e}+\pi}{2},\frac{\mathcal{D}_{3e}+\pi}{3},\mathcal{D}_{\Sigma,\bc}\Big)
	\end{align}
	which is the tight converse bound for the half-duplex setting in the strong interference regime.
	
	Next, let us apply the bound \eqref{strong} to the full-duplex setting \eqref{eq:pi'} which corresponds to $\pi_{01}\leq\frac{\pi}{2}, \pi_{02}\leq\frac{\pi}{2}$. Here we have,
	\begin{align}
	\mathcal{D}_{\Sigma,\iclc}'&\leq\min\Big(\mathcal{D}_{\Sigma,\ic}+\pi_{01}+\pi_{02},\alpha_{12}+\pi_{02},\alpha_{21}+\pi_{01},\frac{\mathcal{D}_{3e}+\pi_{01}+\pi_{02}}{3},\mathcal{D}_{\Sigma,\bc}\Big)\\
	&\leq\min\Big(\mathcal{D}_{\Sigma,\ic}+\pi,\alpha_{12}+\frac{\pi}{2},\alpha_{21}+\frac{\pi}{2},\frac{\mathcal{D}_{3e}+\pi}{3},\mathcal{D}_{\Sigma,\bc}\Big)\\
	&\leq\min\Big(\mathcal{D}_{\Sigma,\ic}+\pi,\min(\alpha_{12},\alpha_{21})+\frac{\pi}{2},\frac{\mathcal{D}_{3e}+\pi}{3},\mathcal{D}_{\Sigma,\bc}\Big)
	\end{align}
	which is the tight converse bound for the full-duplex setting in the strong interference regime. This completes the proof of converse for both Theorem 1 and Theorem 2.
	
	\section{Achievability for Weak and Mixed Interference} \label{lowerbound}
	In this section, we specify the achievable schemes for the weak and mixed interference regimes, for both the half-duplex setting and the full-duplex setting. Without loss of generality, we will assume throughout this section that
	\begin{align}
	\alpha_{11}&\geq\alpha_{22}.
	\end{align}
	\subsection{Weak Interference Regime: $\max(\alpha_{12},\alpha_{21})\leq\min(\alpha_{11},\alpha_{22})$}\label{ach:weak}
	\begin{table}
		\centering
		\begin{adjustbox}{angle=0,width=\columnwidth,center}
			\begin{tabular}{|p{2.5cm}<{\centering}|p{1.5cm}<{\centering}|p{2cm}<{\centering}|p{2cm}<{\centering}|p{4cm}<{\centering}|p{1cm}<{\centering}|p{1cm}<{\centering}|}
				\hline
				\multicolumn{2}{|c|}{Subcases}                                              & \multicolumn{3}{c|}{Codewords' GDoF and Transmitted Power}                                                                                                                                                                                                                                                                                                                                                                                                                                                                                                                                                                                                           & \multicolumn{2}{c|}{Received Power}                                                                                    \\ \hline
				\multicolumn{2}{|c|}{}                                                  &$X_{11}$
				&$X_{22}$                                                               &$X_0^c$                                                                &User~1                                                                &User~2                                                                \\ \hline
				\multicolumn{2}{|c|}{\begin{tabular}[c]{@{}c@{}}$\alpha_{11},\alpha_{22}\geq N$\\ $\pi\leq\mathcal{D}_{\Sigma,\bc}-\mathcal{D}_{\Sigma,\ic}$				\\ $=\min(\alpha_{12},\alpha_{21})$\end{tabular}}                         & \multicolumn{1}{c|}{\begin{tabular}[c]{@{}c@{}}$d_{11}=\alpha_{11}-\alpha_{21}$\\ $E|X_{11}|^2=P^{-\alpha_{21}}$\end{tabular}}                                                                                                              & \multicolumn{1}{c|}{\begin{tabular}[c]{@{}c@{}}$d_{22}=\alpha_{22}-\alpha_{12}$\\ $E|X_{22}|^2=P^{-\alpha_{12}}$\end{tabular}}                                                                                            & \begin{tabular}[c]{@{}c@{}}$d_0^c=\pi$\\ $E|X_0^c|^2=$\\ $\mbox{\tt Diag}(1-P^{-\alpha_{21}},$\\ $1-P^{-\alpha_{12}})$\end{tabular}                                 & \multicolumn{1}{l|}{\begin{tabular}[c]{@{}l@{}}$X_0^c:\sim P^{\alpha_{11}}$\\$X_{11}:\sim P^{\alpha_{11}-\alpha_{21}}$\\ $X_{22}:\sim P^0$\\ \end{tabular}} & \multicolumn{1}{l|}{\begin{tabular}[c]{@{}l@{}}$X_0^c:\sim P^{\alpha_{22}}$\\ $X_{22}:\sim P^{\alpha_{22}-\alpha_{12}}$\\ $X_{11}:\sim P^0$\\  \end{tabular}}                             \\ \hline
				\multirow{2}{*}{\begin{tabular}[c]{@{}c@{}}$\alpha_{11}\geq N$\\ $\alpha_{22}\leq N$\\ $\pi\leq\mathcal{D}_{\Sigma,\bc}$ \\ $-\mathcal{D}_{\Sigma,\ic}$				\\ $=\alpha_{22}-$\\ $\max(\alpha_{12},\alpha_{21})$ \end{tabular}} & $\alpha_{12}\geq\alpha_{21}$ & \multicolumn{1}{c|}{\begin{tabular}[c]{@{}c@{}}$d_{11}^p=\alpha_{11}-\alpha_{21}$\\ $E|X_{11}^p|^2=P^{-\alpha_{21}}$     \\ $d_{11}^c=$\\ $\alpha_{12}+\alpha_{21}-\alpha_{22}$\\ $E|X_{11}^c|^2=$\\ $1-P^{-d_{11}^c}$\end{tabular}}               & \multicolumn{1}{c|}{\begin{tabular}[c]{@{}c@{}}$d_{22}=\alpha_{22}-\alpha_{12}$\\ $E|X_{22}|^2=P^{-\alpha_{12}}$\end{tabular}}                                                                                            & \begin{tabular}[c]{@{}c@{}}$d_0^c=\pi$\\ $E|X_0^c|^2=$\\ $\mbox{\tt Diag}(P^{-d_{11}^c}-P^{-\alpha_{21}},$\\ $1-P^{-\alpha_{12}})$\end{tabular}                     & \multicolumn{1}{l|}{\begin{tabular}[c]{@{}l@{}}$X_{11}^c:\sim P^{\alpha_{11}}$\\ $X_0^c:\sim P^{-d_{11}^c+\alpha_{11}}$ \\$X_{11}^p:\sim P^{\alpha_{11}-\alpha_{21}}$\\ $X_{22}:\sim P^0$ \end{tabular}} & \multicolumn{1}{l|}{\begin{tabular}[c]{@{}l@{}}$X_0^c:\sim P^{\alpha_{22}}$ \\ $X_{11}^c:\sim P^{\alpha_{21}}$\\$X_{22}:\sim P^{\alpha_{22}-\alpha_{12}}$\\ $X_{11}^p:\sim P^{0}$ \end{tabular}} \\ \cline{2-7} 
				& $\alpha_{12}\leq\alpha_{21}$ & \multicolumn{1}{c|}{\begin{tabular}[c]{@{}c@{}}$d_{11}^p=\alpha_{11}-\alpha_{21}$\\ $E|X_{11}^p|^2=P^{-\alpha_{21}}$      \\ $d_{11}^c=2\alpha_{21}-\alpha_{22}$\\ $E|X_{11}^c|^2=$\\ $1-P^{-d_{11}^c}$\end{tabular}}           & \multicolumn{1}{c|}{\begin{tabular}[c]{@{}c@{}}$d_{22}=\alpha_{22}-\alpha_{21}$\\ $E|X_{22}|^2=P^{-\alpha_{21}}$\end{tabular}}                                                                                            & \begin{tabular}[c]{@{}c@{}}$d_0^c=\pi$\\ $E|X_0^c|^2=$\\ $\mbox{\tt Diag}(P^{-d_{11}^c}-P^{-\alpha_{21}},$\\ $1-P^{-\alpha_{21}})$\end{tabular}                                                                                                               & \multicolumn{1}{l|}{\begin{tabular}[c]{@{}l@{}}$X_{11}^c:\sim P^{\alpha_{11}}$\\ $X_0^c:\sim P^{-d_{11}^c+\alpha_{11}}$\\ $X_{11}^p:\sim P^{\alpha_{11}-\alpha_{21}}$\\  $X_{22}:\sim P^{0}$\\  \end{tabular}} & \multicolumn{1}{l|}{\begin{tabular}[c]{@{}l@{}}$X_0^c:\sim P^{\alpha_{22}}$\\ $X_{11}^c:\sim P^{\alpha_{21}}$\\  $X_{22}:\sim P^{\alpha_{22}-\alpha_{21}}$ \\  $X_{11}^p:\sim P^{0}$\end{tabular}} \\ \hline
				\multicolumn{2}{|c|}{\begin{tabular}[c]{@{}c@{}}$\alpha_{11},\alpha_{22}\leq N$\\    $N+\max(\alpha_{12},\alpha_{21})\leq M$\\                $\pi\leq\mathcal{D}_{\Sigma,\bc}-\mathcal{D}_{\Sigma,\ic}$				\\ $=M-N-\max(\alpha_{12},\alpha_{21})$\end{tabular}}                                                              & \multicolumn{1}{c|}{\begin{tabular}[c]{@{}c@{}}$d_{11}^p=\alpha_{11}-\alpha_{21}$\\ $E|X_{11}^p|^2=P^{-\alpha_{21}}$\\ $d_{11}^c=$\\ $\alpha_{12}+\alpha_{21}-\alpha_{22}$\\ $E|X_{11}^c|^2=$\\ $1-P^{-d_{11}^c}$\end{tabular}}          & \multicolumn{1}{c|}{\begin{tabular}[c]{@{}c@{}}$d_{22}^p=\alpha_{22}-\alpha_{12}$\\ $E|X_{22}^p|^2=P^{-\alpha_{12}}$\\ $d_{22}^c=$\\ $\alpha_{12}+\alpha_{21}-\alpha_{11}$\\ $E|X_{22}^c|^2=$\\ $1-P^{-d_{22}^c}$\end{tabular}} & \begin{tabular}[c]{@{}c@{}}$d_0^c=\pi$\\ $E|X_0^c|^2=$\\ $\mbox{\tt Diag}(P^{-d_{11}^c}-P^{-\alpha_{21}},$\\ $P^{-d_{22}^c}-P^{-\alpha_{12}})$\end{tabular}       & \begin{tabular}[c]{@{}l@{}}$X_{11}^c:\sim P^{\alpha_{11}}$\\$X_0^c:\sim P^{-d_{11}^c+\alpha_{11}}$\\ $X_{22}^c:\sim P^{\alpha_{12}}$\\ $X_{11}^p:\sim P^{\alpha_{11}-\alpha_{21}}$ \\ $X_{22}^p:\sim P^0$ \end{tabular} & \begin{tabular}[c]{@{}l@{}}$X_{22}^c:\sim P^{\alpha_{22}}$\\ $X_0^c:\sim P^{-d_{22}^c+\alpha_{22}}$ \\$X_{11}^c:\sim P^{\alpha_{21}}$\\   $X_{22}^p:\sim P^{\alpha_{22}-\alpha_{12}}$\\ $X_{11}^p:\sim P^{0}$ \end{tabular} \\ \hline
			\end{tabular}
		\end{adjustbox}
		\caption{\small \it \label{tab:weak}The achievability for weak interference regime under both half-duplex and full-duplex settings, where $M\triangleq\alpha_{11}+\alpha_{22},N\triangleq\alpha_{12}+\alpha_{21}$, and $\pi\leq\mathcal{D}_{\Sigma,\bc}-\mathcal{D}_{\Sigma,\ic}$. The received powers of different codewords at each receiver are specified in decreasing order, which also corresponds to the successive decoding order at that receiver.}
	\end{table}
	We will assume $\pi\leq\mathcal{D}_{\Sigma,\bc}-\mathcal{D}_{\Sigma,\ic}$. There is no loss of generality in this assumption because the achievability for $\pi>\mathcal{D}_{\Sigma,\bc}-\mathcal{D}_{\Sigma,\ic}$ is the same as that for $\pi=\mathcal{D}_{\Sigma,\bc}-\mathcal{D}_{\Sigma,\ic}$, i.e., the upperbound of $\mathcal{D}_{\Sigma,\bc}$ is achieved without need for further cooperation. The achievable schemes for both half-duplex and full-duplex settings are shown in the Table \ref{tab:weak}. This is because in the weak interference regime the cooperative messages $W_{01},W_{02}$ are combined into one common message $W_{0}^c=(W_{01},W_{02})$, which carries $d_0^c$ DoF and can be decoded by both users. Therefore, without loss of generality we can assume $d_{01}=d_{02}=d_0^c/2$. Since the total cooperation capability is shared equally in the two directions, there is no distinction between the half-duplex and full-duplex settings in the weak interference regime.

	As shown in Table \ref{tab:weak} the achievable schemes are partitioned into three sub-cases. To complement Table \ref{tab:weak}, let us explicitly note the sum-GDoF of the  interference channel  \cite{Etkin_Tse_Wang} and the broadcast channel \cite{Arash_Jafar_cooperation}  for each sub-case as follows.
	\begin{itemize}
		\item $\alpha_{11},\alpha_{22}\geq N$ \\
		$\mathcal{D}_{\Sigma,\ic}=M-N$, $\mathcal{D}_{\Sigma,\bc}=M-\max(\alpha_{12},\alpha_{21})$.
		\item $\alpha_{11}\geq N, \alpha_{22}\leq N$ \\
		$\mathcal{D}_{\Sigma,\ic}=\alpha_{11}$, $\mathcal{D}_{\Sigma,\bc}=M-\max(\alpha_{12},\alpha_{21})$.
		\item $\alpha_{11},\alpha_{22}\leq N$ \\
		$\mathcal{D}_{\Sigma,\ic}=\min(N,M-\max(\alpha_{12},\alpha_{21}))$, $\mathcal{D}_{\Sigma,\bc}=M-\max(\alpha_{12},\alpha_{21})$. Note that if $N+\max(\alpha_{12},\alpha_{21})> M$ then there is no cooperation gain as $\mathcal{D}_{\Sigma,\ic}=\mathcal{D}_{\Sigma,\bc}=M-\max(\alpha_{12},\alpha_{21})$. This is why we have the constraint $N+\max(\alpha_{12},\alpha_{21})\leq M$ in the last row of the table.
	\end{itemize}
	In order to illustrate how the entries in the table describe the achievable scheme for each case, let us explain the last row of the table. The achievability for all other cases follows from the description in Table \ref{tab:weak}  in a similar fashion.
	
	In the subcase corresponding to the last row of Table \ref{tab:weak}, the noncooperative messages $W_1$ and $W_2$ are both split into private and common components, $W_{11}=(W_{11}^c,W_{11}^p)$, $W_{22}=(W_{22}^c,W_{22}^p)$. The submessages $W_{11}^c$, $W_{11}^p$, $W_{22}^c$, $W_{22}^p$ carry $\alpha_{12}+\alpha_{21}-\alpha_{22}$, $\alpha_{11}-\alpha_{21}$, $\alpha_{12}+\alpha_{21}-\alpha_{11}$, and $\alpha_{22}-\alpha_{12}$ GDoF respectively. $W_{11}^p,W_{11}^c,W_{22}^p,W_{22}^c$ are encoded into independent Gaussian codebooks producing codewords $X_{11}^p,X_{11}^c,X_{22}^p,X_{22}^c$ with power levels $P^{-\alpha_{21}},~1-P^{d_{11}^c}, ~P^{-\alpha_{12}}$, and $1-P^{d_{22}^c}$,  respectively. Message $W_0^c$ carries $\pi$ GDoF and is encoded into the vector Gaussian codeword $X_0^c=(X_{01}^{c},X_{02}^{c})$ with  covariance matrix $\mbox{\tt Diag}(P^{-d_{11}^c}-P^{-\alpha_{21}},P^{-d_{22}^c}-P^{-\alpha_{12}})$. The transmitted symbols are $X_1=X_{11}^c+X_{01}^{c}+X_{11}^p,X_2=X_{22}^c+X_{02}^{c}+X_{22}^p$. Next let us describe the decoding. User $1$ (resp. User $2$) decodes $X_{11}^c,X_0^c,X_{22}^c,X_{11}^p$ (resp. $X_{22}^c,X_0^c,X_{11}^c,X_{22}^p$) successively. Specifically, for User $1$, the received power of $X_{11}^c$ is $\sim P^{\alpha_{11}}$ while the interference power is $\sim P^{-d_{11}^c+\alpha_{11}}$, so that the SINR is $\sim P^{d_{11}^c}$. Therefore $X_{11}^c$ for message $W_{11}^c$ can be successfully decoded. Then User $1$ reconstructs and subtracts the contribution of $X_{11}^c$ and starts to decode $X_0^c$. The desired power for $X_0^c$ is $\sim P^{-d_{11}^c+\alpha_{11}}$ while the interference power is $P^{\alpha_{12}}$, so that SINR is $\sim P^{-d_{11}^c+\alpha_{11}-\alpha_{12}}=P^{\alpha_{11}+\alpha_{22}-2\alpha_{12}-\alpha_{21}}=P^{M-N-\alpha_{12}}$. Since $d_0^c=\pi\leq M-N-\max(\alpha_{12},\alpha_{21})\leq M-N-\alpha_{12}$, it follows that  message $W_0^c$ can be successfully decoded. Proceeding similarly, by using successive interference cancellation, User $1$ can decode $X_{22}^c,X_{11}^p,X_{22}^p$ in that order for messages $W_{22}^c,W_{11}^p,W_{22}^p$. Note that the decoding order corresponds to the decreasing order of power levels at the receiver, which is also the order in which the codewords are listed in the last two columns of Table \ref{tab:weak}.
	\subsection{Mixed Interference Regime: $\min(\alpha_{12},\alpha_{21})\leq\max(\alpha_{11},\alpha_{22}),\max(\alpha_{12},\alpha_{21})\geq\min(\alpha_{11},\alpha_{22})$} \label{ach:mixed}
	\begin{table}
		\centering
		\begin{adjustbox}{angle=0,width=\columnwidth,center}
			\begin{tabular}{|p{3.0cm}<{\centering}|p{1.5cm}<{\centering}|p{1.5cm}<{\centering}|p{1.5cm}<{\centering}|p{1.5cm}<{\centering}|p{2.5cm}<{\centering}|p{2.5cm}<{\centering}|}
				\hline
				Subcases                                                                                                                                                                                     & \multicolumn{4}{c|}{Codeswords' GDoF and Transmitted Power}                                                                                                                                                                                                                                                                                                                                         & \multicolumn{2}{c|}{Received Power}                                                                                                                                                                         \\ \hline
				& $X_{11}$                                                                                                  & $X_{22}$                                                                                                  & $X_{01}$                                                                            & $X_{02}$                                                                            & User $1$                                                                                             & User $2$                                                                                             \\ \hline
				\begin{tabular}[c]{@{}c@{}}$\alpha_{21}\leq\alpha_{22}\leq$\\ \hspace{0.5cm}$\alpha_{11}\leq\alpha_{12}$\\ \hspace{-0.2cm}$\pi\leq\mathcal{D}_{\Sigma,\bc}-\mathcal{D}_{\Sigma,\ic}$\\ $=N-M$\end{tabular} & \begin{tabular}[c]{@{}c@{}}\hspace{-0.2cm}$d_{11}=\alpha_{11}$
					\\$-\alpha_{21}$\\ \hspace{-0.2cm}$E|X_{11}|^2=$\\
					$P^{-\alpha_{21}}$\end{tabular} & \begin{tabular}[c]{@{}c@{}}\hspace{-0.2cm}$d_{22}=\alpha_{22}$\\ \hspace{-0.2cm}$E|X_{22}|^2=$\\\hspace{-0.1cm}$1-P^{-\alpha_{22}}$\end{tabular}           & \begin{tabular}[c]{@{}c@{}}\hspace{-0.5cm}$d_{01}=\pi$\\ \hspace{-0.2cm}$E|X_{01}|^2=$\\$P^{-\alpha_{22}}$\end{tabular} & $\empty$                                                                             & \begin{tabular}[c]{@{}c@{}}$X_{22}:\sim P^{\alpha_{12}}$\\ \hspace{-0.2cm}$X_{01}:\sim P^{\alpha_{12}-\alpha_{22}}$\\ \hspace{-0.2cm}$X_{11}:\sim P^{\alpha_{11}-\alpha_{21}}$\end{tabular} & \begin{tabular}[c]{@{}c@{}}$X_{22}:\sim P^{\alpha_{22}}$\\ $X_{01}:\sim P^{0}$\\ $X_{11}:\sim P^{0}$\end{tabular} \\ \hline
				\begin{tabular}[c]{@{}c@{}}$\alpha_{12}\leq\alpha_{22}\leq$\\ \hspace{0.5cm}$\alpha_{11}\leq\alpha_{21}$\\ \hspace{-0.2cm}$\pi\leq\mathcal{D}_{\Sigma,\bc}-\mathcal{D}_{\Sigma,\ic}$\\ $=N-M$\end{tabular}                     & \begin{tabular}[c]{@{}c@{}}\hspace{-0.2cm}$d_{11}=\alpha_{11}$\\ \hspace{-0.2cm}$E|X_{11}|^2=$\\\hspace{-0.1cm}$1-P^{-\alpha_{11}}$\end{tabular}           & \begin{tabular}[c]{@{}c@{}}\hspace{-0.2cm}$d_{22}=\alpha_{22}$\\
					$-\alpha_{12}$\\ \hspace{-0.2cm}$E|X_{22}|^2=$\\$P^{-\alpha_{12}}$\end{tabular} & $\empty$                                                                             & \begin{tabular}[c]{@{}c@{}}$d_{02}=\pi$\\ $E|X_{02}|^2=$\\$P^{-\alpha_{11}}$\end{tabular} & \begin{tabular}[c]{@{}c@{}}$X_{11}:\sim P^{\alpha_{11}}$\\ $X_{02}:\sim P^{0}$\\ $X_{22}:\sim P^{0}$\end{tabular} & \begin{tabular}[c]{@{}c@{}}$X_{11}:\sim P^{\alpha_{21}}$\\ \hspace{-0.2cm}$X_{02}:\sim P^{\alpha_{21}-\alpha_{11}}$\\ \hspace{-0.2cm}$X_{22}:\sim P^{\alpha_{22}-\alpha_{12}}$\end{tabular} \\ \hline
				\begin{tabular}[c]{@{}c@{}}$\alpha_{22}\leq\alpha_{12}\leq$\\ \hspace{0.5cm}$\alpha_{11}\leq\alpha_{21}$\\ \hspace{-0.2cm}$\pi\leq\mathcal{D}_{\Sigma,\bc}-\mathcal{D}_{\Sigma,\ic}$\\ $=\alpha_{21}-\alpha_{11}$\end{tabular} & \begin{tabular}[c]{@{}c@{}}\hspace{-0.2cm}$d_{11}=\alpha_{11}$\\ \hspace{-0.2cm}$E|X_{11}|^2=$\hspace{-0.1cm}\\$1-P^{-\alpha_{11}}$\end{tabular}           & $\empty$                                                                                                  & $\empty$                                                                             & \begin{tabular}[c]{@{}c@{}}$d_{02}=\pi$\\ $E|X_{02}|^2=$\\\hspace{-0.2cm}$P^{-\alpha_{11}}$\end{tabular} & \begin{tabular}[c]{@{}c@{}}$X_{11}:\sim P^{\alpha_{11}}$\\ $X_{02}:\sim P^{0}$\end{tabular}                      & \begin{tabular}[c]{@{}c@{}}$X_{11}:\sim P^{\alpha_{21}}$\\ \hspace{-0.2cm}$X_{02}:\sim P^{\alpha_{21}-\alpha_{11}}$\end{tabular}                      \\ \hline
				\begin{tabular}[c]{@{}c@{}}$\alpha_{22}\leq\alpha_{21}\leq$\\ \hspace{0.5cm}$\alpha_{11}\leq\alpha_{12}$\\ \hspace{-0.2cm}$\pi\leq\mathcal{D}_{\Sigma,\bc}-\mathcal{D}_{\Sigma,\ic}$\\ $=\alpha_{12}-\alpha_{11}$\end{tabular} & \begin{tabular}[c]{@{}c@{}}\hspace{-0.2cm}$d_{11}=\alpha_{11}$\\ \hspace{-0.2cm}$E|X_{11}|^2$\\$=1$\end{tabular}                            & $\empty$                                                                                                  & \begin{tabular}[c]{@{}c@{}}\hspace{-0.5cm}$d_{01}=\pi$\\ \hspace{-0.2cm}$E|X_{01}|^2$\\$=1$\end{tabular}               & $\empty$                                                                             & \begin{tabular}[c]{@{}c@{}}$X_{01}:\sim P^{\alpha_{12}}$\\$X_{11}:\sim P^{\alpha_{11}}$ \end{tabular}                      & \begin{tabular}[c]{@{}c@{}}$X_{11}:\sim P^{\alpha_{21}}$\\ $X_{01}:\sim P^{\alpha_{22}}$\end{tabular}                      \\ \hline
			\end{tabular}
		\end{adjustbox}
		\caption{\small \it \label{tab:mixed}The achievability for mixed interference regime under half-duplex setting and full-duplex setting. It is assumed that $\pi\leq\mathcal{D}_{\Sigma,\bc}-\mathcal{D}_{\Sigma,\ic}$, because any further cooperation is redundant. The table also applies to the full-duplex setting, provided that $\pi$ is replaced by $\frac{\pi}{2}$. This is because one of $W_{01},W_{02}$ is wasted. }
	\end{table}
	In the mixed interference regime, as explained in Section \ref{converse:mixed}, there are only four cases where a cooperation gain exists. Therefore, the description of  achievable schemes in Table \ref{tab:mixed} shows only these four sub-cases. 
	
	The achievability for the full-duplex setting follows by replacing $\pi$ with $\pi/2$. This is because only one-sided cooperation is needed in the mixed interference regime, i.e., either $W_{01}$ or $W_{02}$ is not used, thereby wasting one-half of the cooperation capability.
	To illustrate how the table describes the achievable scheme, let us consider the first row. In this regime, User $1$ is strictly stronger than User $2$. Messages $W_{11},W_{22},W_{01}$  carry $\alpha_{11}-\alpha_{21},\alpha_{22}$, $\pi$ GDoF, and they are encoded into independent Gaussian codebooks $X_{11},X_{22},X_{01}$ with powers $P^{-\alpha_{21}},1-P^{-\alpha_{22}}$, and $P^{-\alpha_{22}}$, respectively. The transmitted signals are $X_1=X_{11},X_2=X_{22}+X_{01}$. User $1$ decodes $X_{22}$ for $W_{22}$ first, while treating everything else as noise. For this decoding stage, the desired signal power is $\sim P^{\alpha_{12}}$ while the interference power is $\sim P^{\alpha_{12}-\alpha_{22}}$ so that SINR is $\sim P^{\alpha_{12}}$, which gives us the GDoF value $d_{22}=\alpha_{22}$. After successfully decoding $W_{22}$, Receiver $1$ is able to reconstruct the codeword $X_{22}$ and subtract its contribution from the received signal.  Then it decodes the codeword $X_{01}$ for its message $W_{01}$ while treating the remaining signal as noise. The desired power is $\sim P^{\alpha_{12}-\alpha_{22}}$ while interference power is $\sim P^{\alpha_{11}-\alpha_{21}}$, so that the SINR  for this decoding is $\sim P^{N-M}$. Since $\pi\leq N-M$, $W_{01}$ can be successfully decoded. After reconstructing and subtracting the contribution of codeword $X_{01}$, User $1$ decodes $X_{11}$ for its desired message $W_{11}$, while treating the remaining signal as noise. The desired signal power is $\sim P^{\alpha_{11}-\alpha_{21}}$ while interference power is $\sim P^0$. Since $d_{11}=\alpha_{11}-\alpha_{21}$, message $W_{11}$ can be successfully decoded. Receiver $2$ is able to decode $X_{22}$ by treating everything else as noise.
	
	\section{Achievability for Strong interference: $\min(\alpha_{12},\alpha_{21})\geq\max(\alpha_{11},\alpha_{22})$}
	In this section, we describe the achievable schemes for the strong interference regime, which are separated into half-duplex and full-duplex settings. The broadcast channel bound for the strong interference, which is found in \cite{Arash_Jafar_cooperation}, is $\mathcal{D}_{\Sigma,\bc}=\alpha_{12}+\alpha_{21}-\max(\alpha_{11},\alpha_{22})$. In this section, we  no longer assume that $\alpha_{11}\geq\alpha_{22}$. Instead, in the strong interference regime, it is more convenient to  assume $\alpha_{12}\geq\alpha_{21}$ without loss of generality.
	\subsection{Half-duplex Setting}\label{ach:hdstrong}
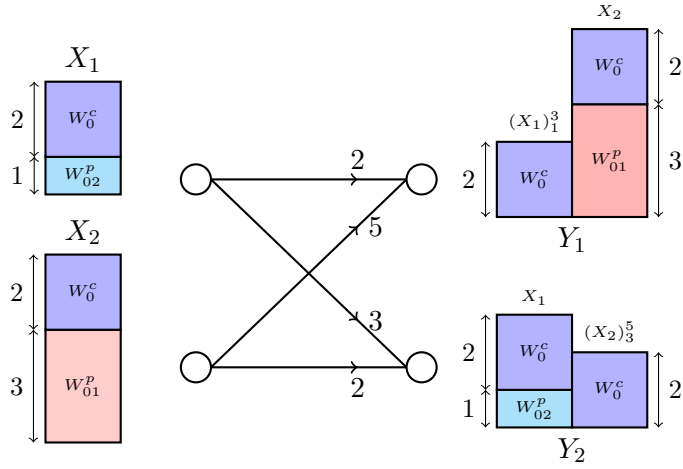
\begin{figure}[ht]
		\centering
		\begin{tikzpicture}[scale=1.0]
		\begin{scope}[shift={(0,0.8)}]
		\draw (0.5, 1.5) node[above]{${X}_1$};
		\draw[ thick, fill=cyan!30!white, pattern color=blue] (0,0) rectangle (1,0.5) node [pos=.5] {\tiny $W_{02}^p$};
		\draw[arrows=<->] (-0.15,0)--(-0.15,0.5)node[left,pos=0.5]{\small $1$};
		\draw[ thick, fill=blue!30!white, pattern color=blue] (0,0.5) rectangle (1,1.5) node [pos=.5] {\tiny $W_0^c$};
		\draw[arrows=<->] (-0.15,0.5)--(-0.15,1.5)node[left,pos=0.5]{\small $2$};
		\end{scope}

		\begin{scope}[shift={(-1,-2.5)}]
		\draw (1.5, 2.5) node[above]{${X}_2$};
		\draw[ thick, fill=red!20!white, pattern color=blue] (1,0) rectangle (2,1.5)node [pos=.5] {\tiny $W_{01}^p$};
		\draw[arrows=<->] (0.84,0)--(0.84,1.5)node[left,pos=0.5]{\small $3$};
		\draw[ thick, fill=blue!30!white, pattern color=blue] (1,1.5) rectangle (2,2.5) node [pos=.5] {\tiny $W_0^c$};
		\draw[arrows=<->] (0.84,1.5)--(0.84,2.5)node[left,pos=0.5]{\small $2$};
		\end{scope}		
		\draw[thick] (2,1) circle(2mm);
		\draw[thick] (5,1) circle(2mm);
		\draw[thick] (2,-1.5) circle(2mm);
		\draw[thick] (5,-1.5) circle(2mm);
		
		\draw[thick, arrowmid] (2.2,1)--(4.8,1)node[above,pos=0.75]{$2$};
		\draw[thick,arrowmid] (2.2,-1.5)--(4.8,-1.5)node[below,pos=0.75]{$2$};
		\draw[thick, arrowmid] (2.2,1)--(4.8,-1.5)node[right,pos=0.75]{$3$};
		\draw[thick,arrowmid] (2.2,-1.5)--(4.8,1)node[right,pos=0.75]{$5$};
		\begin{scope}[shift={(6,0.5)}]
		\draw[ thick, fill=blue!30!white, pattern color=blue] (0,0) rectangle (1,1) node [pos=.5] {\tiny $W_0^c$};
		\draw[arrows=<->] (-0.15,0)--(-0.15,1)node[left,pos=0.5]{\small $2$};
		\draw[ thick, fill=red!30!white, pattern color=blue] (1,0) rectangle (2,1.5)node [pos=.5] {\tiny $W_{01}^p$};
		\draw[arrows=<->] (2.15,0)--(2.15,1.5)node[right,pos=0.5]{\small $3$};
		\draw[ thick, fill=blue!30!white, pattern color=blue] (1,1.5) rectangle (2,2.5) node [pos=.5] {\tiny $W_0^c$};
		\draw[arrows=<->] (2.15,1.5)--(2.15,2.5)node[right,pos=0.5]{\small $2$};
		
		\draw[thick] (1,0) node[below]{$Y_1$};
		\draw (0.5, 1) node[above]{\tiny $(X_1)^3_1$};
		\draw (1.5, 2.5) node[above]{\tiny $X_2$};
		\end{scope}
		
		\begin{scope}[shift={(6,-2.3)}]
		\draw[ thick, fill=cyan!30!white, pattern color=blue] (0,0) rectangle (1,0.5) node [pos=.5] {\tiny $W_{02}^p$};
		\draw[arrows=<->] (-0.15,0)--(-0.15,0.5)node[left,pos=0.5]{\small $1$};
		\draw[ thick, fill=blue!30!white, pattern color=blue] (0,0.5) rectangle (1,1.5) node [pos=.5] {\tiny $W_0^c$};
		\draw[arrows=<->] (-0.15,0.5)--(-0.15,1.5)node[left,pos=0.5]{\small $2$};
		
		\draw[ thick, fill=blue!30!white, pattern color=blue] (1,0) rectangle (2,1.0) node [pos=.5] {\tiny $W_0^c$};
		\draw[arrows=<->] (2.15,0)--(2.15,1.0)node[right,pos=0.5]{\small $2$};
		
		\draw[thick] (1,0) node[below]{$Y_2$};
		\draw (0.5, 1.5) node[above]{\tiny $X_1$};
		\draw (1.5,1) node[above]{\tiny $(X_2)^{5}_{3}$};
		\end{scope}
		\end{tikzpicture}
		\caption{\it \small The  scheme from \cite{Arash_Jafar_cooperation} requires $\pi=6$ GDoF of cooperation to achieve the broadcast channel bound.}\label{fig:ex1}
	\end{figure}

\begin{figure}[!h]
		\centering
		\begin{tikzpicture}[scale=1.0]
		\begin{scope}[shift={(0,0.8)}]
		\draw (0.5, 1.5) node[above]{${X}_1$};
		\draw[ thick, fill=cyan!30!white, pattern color=blue] (0,0) rectangle (1,0.5) node [pos=.5] {\tiny$W_{02}^p$};
		\draw[<->] (-0.15,0)--(-0.15,0.5)node[left,pos=0.5]{\small $1$};
		\draw[ thick, fill=white, pattern color=blue] (0,0.5) rectangle (1,1.0);
		\draw[ thick, fill=blue!30!white, pattern color=blue] (0,1.0) rectangle (1,1.5) node [pos=.5] {\tiny $W_0^c$};
		\draw[<->] (-0.15,1.0)--(-0.15,1.5)node[left,pos=0.5]{\small $1$};
		\end{scope}
		\begin{scope}[shift={(0,-2.5)}]
		\draw (0.5, 2.5) node[above]{${X}_2$};
		\draw[ thick, fill=red!30!white, pattern color=blue] (0,0) rectangle (1,1.5)node [pos=.5] {\tiny$W_{01}^p$};
		\draw[<->] (-0.15,0)--(-0.15,1.5)node[left,pos=0.5]{\small $3$};
		\draw[ thick, fill=blue!30!white, pattern color=blue] (0,1.5) rectangle (1,2.0) node [pos=.5] {\tiny$W_0^c$};
		\draw[<->] (-0.15,1.5)--(-0.15,2)node[left,pos=0.5]{\small $1$};
		\draw[ thick, fill=orange!30!white, pattern color=blue] (0,2.0) rectangle (1,2.5) node [pos=.5] {\tiny$W_{22}$};
		\draw[<->] (-0.15,2)--(-0.15,2.5)node[left,pos=0.5]{\small $1$};
		\end{scope}	
		
		\draw[thick] (2,1) circle(2mm);
		\draw[thick] (5,1) circle(2mm);
		\draw[thick] (2,-1.5) circle(2mm);
		\draw[thick] (5,-1.5) circle(2mm);
		
		\draw[thick,arrowmid] (2.2,1)--(4.8,1)node[above,pos=0.5]{$2$};
		\draw[thick,arrowmid] (2.2,-1.5)--(4.8,-1.5)node[below,pos=0.5]{$2$};
		\draw[thick,arrowmid] (2.2,1)--(4.8,-1.5)node[right,pos=0.8]{$3$};
		\draw[thick,arrowmid] (2.2,-1.5)--(4.8,1)node[right,pos=0.8]{$5$};
		
		\begin{scope}[shift={(6,0.5)}]
		\draw[ thick, fill=white, pattern color=blue] (0,0) rectangle (1,0.5);
		\draw[ thick, fill=blue!30!white, pattern color=blue] (0,0.5) rectangle (1,1) node [pos=.5] {\tiny$W_0^c$};
		\draw[arrows=<->] (-0.15,0.5)--(-0.15,1)node[left,pos=0.5]{\small $1$};		
		\draw[ thick, fill=red!30!white, pattern color=blue] (1,0) rectangle (2,1.5)node [pos=.5] {\tiny$W_{01}^p$};
		\draw[arrows=<->] (2.15,0)--(2.15,1.5)node[right,pos=0.5]{\small $3$};
		\draw[ thick, fill=blue!30!white, pattern color=blue] (1,1.5) rectangle (2,2.0) node [pos=.5] {\tiny$W_0^c$};
		\draw[arrows=<->] (2.15,1.5)--(2.15,2)node[right,pos=0.5]{\small $1$};
		\draw[ thick, fill=orange!30!white, pattern color=blue] (1,2.0) rectangle (2,2.5) node [pos=.5] {\tiny$W_{22}$};
		\draw[arrows=<->] (2.15,2)--(2.15,2.5)node[right,pos=0.5]{\small $1$};
		
		\draw[thick] (1,0) node[below]{$Y_1$};
		\draw (0.5, 1) node[above]{\tiny $(X_1)^3_1$};
		\draw (1.5, 2.5) node[above]{\tiny $X_2$};
		\end{scope}	
		\begin{scope}[shift={(6,-2.3)}]
		\draw[  thick, fill=cyan!30!white, pattern color=blue] (0,0) rectangle (1,0.5)node [pos=.5] {\tiny$W_{02}^p$};
		\draw[arrows=<->] (-0.15,0)--(-0.15,0.5)node[left,pos=0.5]{\small $1$};
		\draw[ thick, fill=white, pattern color=blue] (0,0.5) rectangle (1,1.0);
		\draw[ thick, fill=blue!30!white, pattern color=blue] (0,1.0) rectangle (1,1.5) node [pos=.5] {\tiny$W_0^c$};	
		\draw[arrows=<->] (-0.15,1)--(-0.15,1.5)node[left,pos=0.5]{\small $1$};

		\draw[ thick, fill=blue!30!white, pattern color=blue] (1,0) rectangle (2,0.5) node [pos=.5] {\tiny$W_0^c$};
		\draw[arrows=<->] (2.15,0)--(2.15,0.5)node[right,pos=0.5]{\small $1$};
		\draw[ thick, fill=orange!30!white, pattern color=blue] (1,0.5) rectangle (2,1.0) node [pos=.5] {\tiny$W_{22}$};
		\draw[arrows=<->] (2.15,0.5)--(2.15,1.0)node[right,pos=0.5]{\small $1$};
		
		\draw[thick] (1,0) node[below]{$Y_2$};
		\draw (0.5, 1.5) node[above]{\tiny $X_1$};
		\draw (1.5, 1) node[above]{\tiny $(X_2)^5_3$};
		\end{scope}	
		\end{tikzpicture}
		\caption{\small \it The optimally efficient achievable scheme  achieves the broadcast channel bound with only $\pi=5$ GDoF of cooperation.}\label{fig:ex2}
	\end{figure}
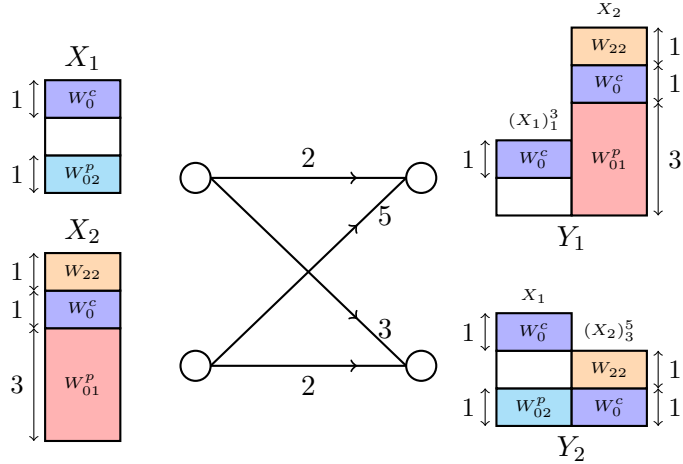

	Let us begin with an illustrative example where $\alpha_{11}=\alpha_{22}=2,\alpha_{12}=5,\alpha_{21}=3$. For this setting,  $\mathcal{D}_{\Sigma,\bc}=6$ according to \cite{Arash_Jafar_cooperation} and $D_{\Sigma,\ic}=3$ according to \cite{Etkin_Tse_Wang}. Let us consider how much cooperation is needed in this case to achieve $\mathcal{D}_{\Sigma,\bc}$. The achievable scheme of \cite{Arash_Jafar_cooperation} summarized in Figure \ref{fig:ex1}, requires $\pi=6$ GDoF of cooperation, i.e., all messages must be shared between the two transmitters. This is because in order to take advantage of the strong interference links, the private messages of Users $1$ and $2$,  are sent from opposing transmitters, i.e., Transmitters $2$ and $1$, respectively. These are  messages $W_{01}^p, W_{02}^p$ in Figure \ref{fig:ex1}. The common message $W_o^c$ that is decoded by both users  is sent from both transmitters, so it is shared as well. 
	However, as shown in Theorem \ref{theorem:GDoF} in this paper, the sum-GDoF of limited cooperation interference channel for this example is $D_{\Sigma,\iclc}=\min(3+\pi,\frac{8+\pi}{2},\frac{13+\pi}{3},6)$. Therefore, $\pi^*=5$  is the minimum value of cooperative GDoF needed to achieve the BC bound. The optimally efficient scheme is shown in Figure \ref{fig:ex2}. The improvement in efficiency comes from the observation that part of the common message (in this case, $W_{22}$) can be transmitted from only one transmitter (in this case, Transmitter $2$), and therefore requires no cooperation.
	
	The achievable scheme is described as follows: The cooperative messages $W_{01},W_{02}$ are split into a cooperative common\footnote{The cooperative common message may be arbitrarily divided among the two users, e.g., without loss of generality, we can assume that half of $W_0^c$ is the desired message for User $1$ and the other half of $W_0^c$ is the desired message for User $2$. } message $W_0^c=(W_{01}^c,W_{02}^c)$ and the cooperative private messages $W_{01}^p,W_{02}^p$. Messages $W_{22},W_{0}^c,W_{01}^p,W_{02}^p$  carry $1, 1, 3, 1$ GDoF respectively such that $\pi=5$. $W_{22},W_{01}^p,W_{02}^p$ are encoded into independent Gaussian codebooks $X_{22}, X_{01}^p,X_{02}^p$ respectively with powers $\E|X_{22}|^2=1-P^{-1}$, $\E|X_{01}^p|^2=P^{-2}$, $\E|X_{02}^p|^2=P^{-2}$.
	Message $W_0^c$ carries $1$ GDoF and is encoded to a vector Gaussian codebook $X_0^c=(X_{01}^c,X_{02}^c)$ with power covariance matrix $\mbox{\tt Diag}(1-P^{-2},P^{-1}-P^{-2})$. The transmitted symbols are $X_1=X_{01}^c+X_{02}^p,X_2=X_{22}+X_{02}^c+X_{01}^p$.
	Suppressing the time index for clarity, the received signals are:
	{\small
		\begin{align}	Y_1&=\sqrt{P^2}G_{11}(X_{01}^c+X_{02}^p)+\sqrt{P^5}G_{12}(X_{22}+X_{02}^c+X_{01}^p)+Z_1\notag\\	
		Y_2&=\sqrt{P^3}G_{21}(X_{01}^c+X_{02}^p)+\sqrt{P^2}G_{22}(X_{22}+X_{02}^c+X_{01}^p)+Z_2\notag
		\end{align}}
	
	When decoding, User $1$ first decodes $X_{22}$ for $W_{22}$ while treating everything else as Gaussian noise. Since $X_{22}$ is received at power level $\sim P^5$ while all other signals are received with power levels $\sim P^4$ or lower, the SINR for decoding $W_{22}$ is $\sim P^1$, which gives us the GDoF value $d_{22}=1$. After decoding $W_{22}$, Receiver $1$ is able to reconstruct codeword $X_{22}$ and subtract its contribution from the received signal. After this, Receiver $1$ decodes the codeword $X_{0}^c$ for message $W_{0}^c$, while treating the remaining signals as Gaussian noise.  Since the desired signal for this decoding is received with power level $\sim P^4$ while all other signals are received with power levels $\sim P^3$ or less, the SINR for decoding $W_{0}^c$ is $\sim P^1$ which gives GDoF value $d_0^c=1$. Then Receiver $1$ subtracts the contribution of $X_{0}^c$ and decodes message $W_{01}^p$ while treating all other remaining signals as Gaussian noise. As evident from Figure \ref{fig:ex2}, the SINR for this decoding is $\sim P^3$ which gives us GDoF value $d_{01}^p=3$. Receiver $2$ proceeds similarly by successively decoding $W_{0}^c,W_{22},W_{02}^p$.
	
	In general, to prove the achievability for the strong interference regime completely, there are 4 subcases, which cover all possibilities. Note that we assume $\pi\leq\pi^*$ because the achievable scheme for $\pi>\pi^*$ is the same as $\pi=\pi^*$, since $\pi^*$ already achieves the broadcast channel bound. 
	
	\subsubsection*{\bf Case 1: $\alpha_{12}\leq\alpha_{11}+\alpha_{22},\alpha_{21}\leq\alpha_{11}+\alpha_{22},\alpha_{12}+\alpha_{21}\leq\alpha_{11}+\alpha_{22}+\max(\alpha_{11},\alpha_{22})$}
	\begin{figure}[htb]
		\centering
		\begin{tikzpicture}[scale=1.0, baseline=(current bounding box.center)]
		\begin{scope}[shift={(0,1.0)}]
		\draw (0.5, 2.1) node[above]{${X}_1$};
		\draw[ thick, fill=cyan!30!white, pattern color=blue] (0,0) rectangle (1,0.3) node [pos=.5] {\tiny $W_{02}^p$};
		\draw[arrows=<->] (-0.15,0)--(-0.15,0.3)node[left,pos=0.5]{\small $d_{02}^p$};
		\draw[ thick, fill=white, pattern color=blue] (0,0.3) rectangle (1,1.2);
		\draw[ thick, fill=green!30!white, pattern color=blue] (0,1.2) rectangle (1,2.1) node [pos=.5] {\tiny $W_{11}$};
		\draw[arrows=<->] (-0.15,1.2)--(-0.15,2.1)node[left,pos=0.5]{\small $d_{11}$};
		\end{scope}
		\begin{scope}[shift={(-1,-2.5)}]
		\draw (1.5, 2.7) node[above]{${X}_2$};
		\draw[ thick, fill=red!30!white, pattern color=blue] (1,0) rectangle (2,0.9)node [pos=.5] {\tiny $W_{01}^p$};
		\draw[arrows=<->] (0.85,0)--(0.85,0.9)node[left,pos=0.5]{\small $d_{01}^p$};
		\draw[ thick, fill=white, pattern color=blue] (1,0.9) rectangle (2,1.8);
		\draw[ thick, fill=orange!30!white, pattern color=blue] (1,1.8) rectangle (2,2.7) node [pos=.5] {\tiny $W_{22}$};
		\draw[arrows=<->] (0.85,1.8)--(0.85,2.7)node[left,pos=0.5]{\small $d_{22}$};
		\end{scope}		
		\draw[thick] (2,1) circle(2mm);
		\draw[thick] (5,1) circle(2mm);
		\draw[thick] (2,-1.5) circle(2mm);
		\draw[thick] (5,-1.5) circle(2mm);
		
		\draw[thick,arrowmid] (2.2,1)--(4.8,1)node[above,pos=0.5]{$\alpha_{11}$};
		\draw[thick,arrowmid] (2.2,-1.5)--(4.8,-1.5)node[below,pos=0.5]{$\alpha_{22}$};
		\draw[thick,arrowmid] (2.2,1)--(4.8,-1.5)node[right,pos=0.8]{$\alpha_{21}$};
		\draw[thick,arrowmid] (2.2,-1.5)--(4.8,1)node[right,pos=0.8]{$\alpha_{12}$};
		
		\begin{scope}[shift={(6.5,0.5)}]
		\draw[ thick, fill=white, pattern color=blue] (0,0) rectangle (1,0.9);
		\draw[ thick, fill=green!30!white, pattern color=red] (0,0.9) rectangle (1,1.8)node [pos=.5] {\tiny $W_{11}$};
		\draw[arrows=<->] (-0.15,0.9)--(-0.15,1.8)node[left,pos=0.5]{\small $d_{11}$};
		
		\draw[ thick, fill=red!30!white, pattern color=blue] (1,0) rectangle (2,0.9)node [pos=.5] {\tiny $W_{01}^p$};
		\draw[arrows=<->] (2.15,0)--(2.15,0.9)node[right,pos=0.5]{\small $d_{01}^p$};
		\draw[ thick, fill=white, pattern color=blue] (1,0.9) rectangle (2,1.8);
		\draw[ thick, fill=orange!30!white, pattern color=blue] (1,1.8) rectangle (2,2.7) node [pos=.5] {\tiny $W_{22}$};
		\draw[arrows=<->] (2.15,1.8)--(2.15,2.7)node[right,pos=0.5]{\small $d_{22}$};
		
		\draw[thick] (1,0) node[below]{$Y_1$};
		\draw (0.5, 1.8) node[above]{\tiny $(X_1)^{\alpha_{21}}_{\delta}$};
		\draw (1.5, 2.7) node[above]{\tiny $X_2$};
		\end{scope}
		
		\begin{scope}[shift={(6.5,-2.3)}]
		\draw[ thick, fill=cyan!30!white, pattern color=blue] (0,0) rectangle (1,0.3) node [pos=.5] {\tiny $W_{02}^p$};
		\draw[arrows=<->] (-0.15,0)--(-0.15,0.3)node[left,pos=0.5]{\small $d_{02}^p$};
		\draw[ thick, fill=white, pattern color=blue] (0,0.3) rectangle (1,1.2);
		\draw[ thick, fill=green!30!white, pattern color=blue] (0,1.2) rectangle (1,2.1) node [pos=.5] {\tiny $W_{11}$};
		\draw[arrows=<->] (-0.15,1.2)--(-0.15,2.1)node[left,pos=0.5]{\small $d_{11}$};
		
		\draw[ thick, fill=white, pattern color=blue] (1,0) rectangle (2,0.6);
		\draw[ thick, fill=orange!30!white, pattern color=blue] (1,0.6) rectangle (2,1.5) node [pos=.5] {\tiny $W_{22}$};
		\draw[arrows=<->] (2.15,0.6)--(2.15,1.5)node[right,pos=0.5]{\small $d_{22}$};
		
		\draw[thick] (1,0) node[below]{$Y_2$};
		\draw (0.5, 2.1) node[above]{\tiny$X_1$};
		\draw (1.5, 1.5) node[above]{\tiny$(X_2)^{\alpha_{12}}_{\gamma}$};
		\end{scope}
		\end{tikzpicture}
		\caption{\small \it Signal partition in the regime $\alpha_{12}\leq \alpha_{11}+\alpha_{22},\alpha_{21}\leq \alpha_{11}+\alpha_{22},\alpha_{12}+\alpha_{21}\leq\alpha_{11}+\alpha_{22}+\max(\alpha_{11},\alpha_{22}),\delta=\alpha_{21}-\alpha_{11},\gamma=\alpha_{12}-\alpha_{22}$.}\label{fig:case_1}
	\end{figure}
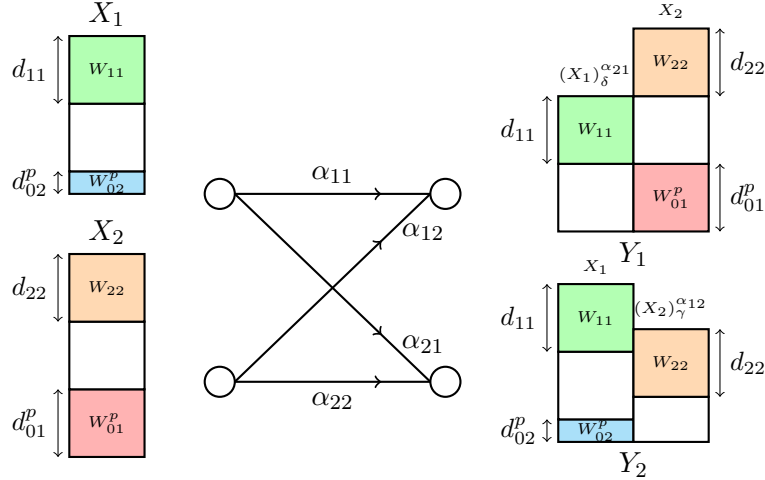
	The sum-GDoF value in this case is characterized as:
	\begin{align}
	\mathcal{D}_{\Sigma,\iclc}=\min\Big(\alpha_{21}+\pi,\frac{\alpha_{12}+\alpha_{21}+\pi}{2},
	\mathcal{D}_{\Sigma,\bc}\Big)
	\end{align}
	
	\begin{itemize}
		\item When $\pi\leq \alpha_{12}-\alpha_{21}$, the first bound is tight, which is achieved by having $W_{11},W_{22},W_{01}^p$ carry $\alpha_{21}-\alpha_{22},\alpha_{22},\pi$ GDoF respectively. They are encoded into independent Gaussian codebooks producing codewords $X_{11},X_{22},X_{01}^p$ with powers $\E|X_{11}|^2=1,\E|X_{22}|^2=1-P^{-\alpha_{21}},\E|X_{01}^p|^2=P^{-\alpha_{21}}$. The transmitted signals are $X_1=X_{11},X_2=X_{22}+X_{01}^p$. When decoding, User $1$ first \emph{jointly} (acting as the receiver in a multiple access channel) decodes $X_{11}$ and $X_{22}$ while treating everything else as noise, while the noise floor due to $X_{01}^p$ is $\sim P^{\alpha_{12}-\alpha_{21}}$. The GDoF region for this multiple access channel is the following.
		\begin{align}
		\{(d_{11},d_{22}):d_{11}\leq\alpha_{11}+\alpha_{21}-\alpha_{12}, d_{11}+d_{22}\leq \alpha_{21} \}.
		\end{align}
		Since $d_{11}=\alpha_{21}-\alpha_{22}\leq \alpha_{11}+\alpha_{21}-\alpha_{12}, d_{11}+d_{22}= \alpha_{21}$ belongs to the GDoF region of the multiple access channel, User $1$ is able to decode $X_{11},X_{22}$ for messages $W_{11},W_{22}$.  After this, User $1$ subtracts the reconstructed codewords $X_{11},X_{22}$ and then decodes $X_{01}^p$. The SINR for this decoding is $\sim P^{\alpha_{12}-\alpha_{21}}$, such that $d_{01}^p=\pi\leq\alpha_{12}-\alpha_{21}$ and the decoding is successful. User $2$ decodes $X_{11},X_{22}$ successively. The SINR values for $X_{11},X_{22}$ are $\sim P^{\alpha_{21}-\alpha_{22}},\sim P^{\alpha_{22}}$ respectively, which give us $d_{11}=\alpha_{21}-\alpha_{22},d_{22}=\alpha_{22}$. Therefore $X_{11},X_{22}$ are successfully decoded at User $2$.
		\item When $\alpha_{12}-\alpha_{21}\leq\pi\leq \pi^*$, where $\pi^*=\alpha_{12}+\alpha_{21}-2\max(\alpha_{11},\alpha_{22})$  according to Corollary 1, the second bound is tight and is achieved as follows: $W_{11},W_{22},W_{01}^p,W_{02}^p$ carry $d_{11}=(\alpha_{21}+2\alpha_{11}-\alpha_{12}-\pi)/2,~d_{22}=\alpha_{12}-\alpha_{11},~d_{01}^p=(\alpha_{12}-\alpha_{21}+\pi)/2,~d_{02}^p=(\alpha_{21}-\alpha_{12}+\pi)/2$ GDoF respectively. They are encoded into independent Gaussian codewords $X_{11},X_{22},X_{01}^p,X_{02}^p$ with powers $\E|X_{11}|^2=1-P^{-d_{11}-d_{22}},\E|X_{22}|^2=1-P^{-d_{11}-d_{22}},\E|X_{01}^p|^2=P^{-d_{11}-d_{22}},\E|X_{02}^p|^2=P^{-d_{11}-d_{22}}$. The transmitted symbols are $X_1=X_{11}+X_{02}^p,X_2=X_{22}+X_{01}^p$. When decoding, User $1$ decodes $X_{22},X_{11},X_{01}^p$ successively, The SINRs for these codewords are $\sim P^{\alpha_{12}-\alpha_{11}},\sim P^{\alpha_{11}-\alpha_{12}+d_{11}+d_{22}}= P^{d_{11}},\sim P^{\alpha_{12}-d_{11}-d_{22}}=P^{d_{01}^p}$ respectively. User $2$ acts as a multiple access receiver, it jointly decodes $X_{11}$ and $X_{22}$ while treating everything else as noise, where the noise floor due to $X_{02}^p$ is $\alpha_{21}-d_{11}-d_{22}=\frac{\alpha_{21}-\alpha_{12}+\pi}{2}$. Hence the GDoF region for this multiple access channel is the following.
		\begin{align}
		\bigg\{(d_{11},d_{22}):d_{22}\leq \alpha_{22}-\frac{\alpha_{21}-\alpha_{12}+\pi}{2},~d_{11}+d_{22}\leq \frac{\alpha_{12}+\alpha_{21}-\pi}{2}\bigg\}.
		\end{align}
		 Since $d_{22}=\alpha_{12}-\alpha_{11}\leq\alpha_{22}+
			\max(\alpha_{11},\alpha_{22})-\alpha_{21}=\alpha_{22}-\frac{\alpha_{21}-\alpha_{12}+\pi^*}{2}\leq\alpha_{22}-\frac{\alpha_{21}-\alpha_{12}+\pi}{2}, d_{11}+d_{22}=\frac{\alpha_{12}+\alpha_{21}-\pi}{2}$  belongs to the GDoF region of the multiple access channel, the messages $W_{11},W_{22}$ can be jointly decoded successfully by User $2$. After this, User $2$ subtracts the contribution of $X_{11},X_{22}$ and decodes $X_{02}^p$, whose SINR is $\sim P^{\alpha_{21}-d_{11}-d_{22}}=P^{d_{02}^p}$, such that $X_{02}^p$ for $W_{02}^p$ can be successfully decoded. The signal partitioning  is shown in Figure \ref{fig:case_1}. The cooperation capability beyond $\pi^*$ is redundant because with $\pi^*$ cooperation the broadcast GDoF are already achieved.
	\end{itemize}
	
	\subsubsection*{\bf Case 2: $\alpha_{12}\leq\alpha_{11}+\alpha_{22},\alpha_{21}\leq\alpha_{11}+\alpha_{22},\alpha_{12}+\alpha_{21}\geq\alpha_{11}+\alpha_{22}+\max(\alpha_{11},\alpha_{22})$} 
	\begin{figure}[htb]
		\centering
		\begin{tikzpicture}[scale=1.0]
		\begin{scope}[shift={(0,0.5)}]
		\draw[ thick] (0,0) rectangle (1,1.6);
		\draw (0.5, 1.8) node[above]{${X}_1$};
		\draw[ thick, fill=cyan!30!white, pattern color=blue] (0,0) rectangle (1,0.6) node [pos=.5] {\tiny$W_{02}^p$};
		\draw[arrows=<->] (-0.15,0)--(-0.15,0.6)node[left,pos=0.5]{\small $d_{02}^p$};
		\draw[ thick, fill=white, pattern color=blue] (0,0.6) rectangle (1,1);
		\draw[ thick, fill=blue!30!white, pattern color=blue] (0,1) rectangle (1,1.4) node [pos=.5] {\tiny$W_0^c$};
		\draw[arrows=<->] (-0.15,1)--(-0.15,1.4)node[left,pos=0.5]{\small $d_0^c$};
		\draw[ thick, fill=green!30!white, pattern color=blue] (0,1.4) rectangle (1,1.8) node [pos=.5] {\tiny$W_{11}$};
		\draw[arrows=<->] (-0.15,1.4)--(-0.15,1.8)node[left,pos=0.5]{\small $d_{11}$};
		\end{scope}
		
		\begin{scope}[shift={(-1,-2.5)}]
		\draw (1.5, 2) node[above]{${X}_2$};
		\draw[ thick, fill=red!30!white, pattern color=blue] (1,0) rectangle (2,0.8)node [pos=.5] {\tiny$W_{01}^p$};
		\draw[arrows=<->] (0.85,0)--(0.85,0.8)node[left,pos=0.5]{\small $d_{01}^p$};
		\draw[ thick, fill=white, pattern color=blue] (1,0.8) rectangle (2,1.2);
		\draw[ thick, fill=blue!30!white, pattern color=blue] (1,1.2) rectangle (2,1.6) node [pos=.5] {\tiny$W_0^c$};
		\draw[arrows=<->] (0.85,1.2)--(0.85,1.6)node[left,pos=0.5]{\small $d_0^c$};
		\draw[ thick, fill=orange!30!white, pattern color=blue] (1,1.6) rectangle (2,2) node [pos=.5] {\tiny$W_{22}$};
		\draw[arrows=<->] (0.85,1.6)--(0.85,2)node[left,pos=0.5]{\small $d_{22}$};
		\end{scope}		
		\draw[thick] (2,1) circle(2mm);
		\draw[thick] (5,1) circle(2mm);
		\draw[thick] (2,-1.5) circle(2mm);
		\draw[thick] (5,-1.5) circle(2mm);
		
		\draw[thick,arrowmid] (2.2,1)--(4.8,1)node[above,pos=0.5]{$\alpha_{11}$};
		\draw[thick,arrowmid] (2.2,-1.5)--(4.8,-1.5)node[below,pos=0.5]{$\alpha_{22}$};
		\draw[thick,arrowmid] (2.2,1)--(4.8,-1.5)node[right,pos=0.8]{$\alpha_{21}$};
		\draw[thick,arrowmid] (2.2,-1.5)--(4.8,1)node[right,pos=0.8]{$\alpha_{12}$};
		
		\begin{scope}[shift={(6.5,0.5)}]
		\draw[ thick, fill=white, pattern color=blue] (0,0) rectangle (1,0.4);
		\draw[ thick, fill=blue!30!white, pattern color=blue] (0,0.4) rectangle (1,0.8) node [pos=.5] {\tiny$W_0^c$};
		\draw[arrows=<->] (-0.15,0.4)--(-0.15,0.8)node[left,pos=0.5]{\small $d_0^c$};
		\draw[ thick, fill=green!30!white, pattern color=red] (0,0.8) rectangle (1,1.2)node [pos=.5] {\tiny$W_{11}$};
		\draw[arrows=<->] (-0.15,0.8)--(-0.15,1.2)node[left,pos=0.5]{\small $d_{11}$};
		
		\draw[ thick, fill=red!30!white, pattern color=blue] (1,0) rectangle (2,0.8)node [pos=.5] {\tiny$W_{01}^p$};
		\draw[arrows=<->] (2.15,0)--(2.15,0.8)node[right,pos=0.5]{\small $d_{01}^p$};
		\draw[ thick, fill=white, pattern color=blue] (1,0.8) rectangle (2,1.2);
		\draw[ thick, fill=blue!30!white, pattern color=blue] (1,1.2) rectangle (2,1.6) node [pos=.5] {\tiny$W_0^c$};
		\draw[arrows=<->] (2.15,1.2)--(2.15,1.6)node[right,pos=0.5]{\small $d_0^c$};
		\draw[ thick, fill=orange!30!white, pattern color=blue] (1,1.6) rectangle (2,2) node [pos=.5] {\tiny$W_{22}$};
		\draw[arrows=<->] (2.15,1.6)--(2.15,2)node[right,pos=0.5]{\small $d_{22}$};
		
		\draw[thick] (1,0) node[below]{$Y_1$};
		\draw (0.4, 1.2) node[above]{\tiny $(X_1)^{\alpha_{21}}_{\delta}$};
		\draw (1.5, 2) node[above]{\tiny $X_2$};
		\end{scope}
		
		\begin{scope}[shift={(6.5,-2.3)}]
		\draw[ thick, fill=cyan!30!white, pattern color=blue] (0,0) rectangle (1,0.6) node [pos=.5] {\tiny$W_{02}^p$};
		\draw[arrows=<->] (-0.15,0)--(-0.15,0.6)node[left,pos=0.5]{\small $d_{02}^p$};
		\draw[ thick, fill=white, pattern color=blue] (0,0.6) rectangle (1,1);
		\draw[ thick, fill=blue!30!white, pattern color=blue] (0,1) rectangle (1,1.4) node [pos=.5] {\tiny$W_0^c$};
		\draw[arrows=<->] (-0.15,1)--(-0.15,1.4)node[left,pos=0.5]{\small $d_0^c$};
		\draw[ thick, fill=green!30!white, pattern color=blue] (0,1.4) rectangle (1,1.8) node [pos=.5] {\tiny$W_{11}$};
		\draw[arrows=<->] (-0.15,1.4)--(-0.15,1.8)node[left,pos=0.5]{\small $d_{11}$};
		
		\draw[ thick, fill=white, pattern color=blue] (1,0) rectangle (2,0.2);
		\draw[ thick, fill=blue!30!white, pattern color=blue] (1,0.2) rectangle (2,0.6) node [pos=.5] {\tiny$W_0^c$};
		\draw[arrows=<->] (2.15,0.2)--(2.15,0.6)node[right,pos=0.5]{\small $d_0^c$};
		\draw[ thick, fill=orange!30!white, pattern color=blue] (1,0.6) rectangle (2,1) node [pos=.5] {\tiny$W_{22}$};
		\draw[arrows=<->] (2.15,0.6)--(2.15,1)node[right,pos=0.5]{\small $d_{22}$};
		
		\draw[thick] (1,0) node[below]{$Y_2$};
		\draw (0.5, 1.8) node[above]{\tiny $X_1$};
		\draw (1.8, 1) node[above]{\tiny  $(X_2)^{\alpha_{12}}_{\gamma}$};
		\end{scope}
		\end{tikzpicture}
		\caption{\it\small Signal partition  in the regime $\alpha_{12},\alpha_{21}\leq \alpha_{11}+\alpha_{22},\alpha_{12}+\alpha_{21}\geq\alpha_{11}+\alpha_{22}+\max(\alpha_{11},\alpha_{22})$, where $\delta=\alpha_{21}-\alpha_{11},\gamma=\alpha_{12}-\alpha_{22}$.}\label{fig:case_2}
	\end{figure}
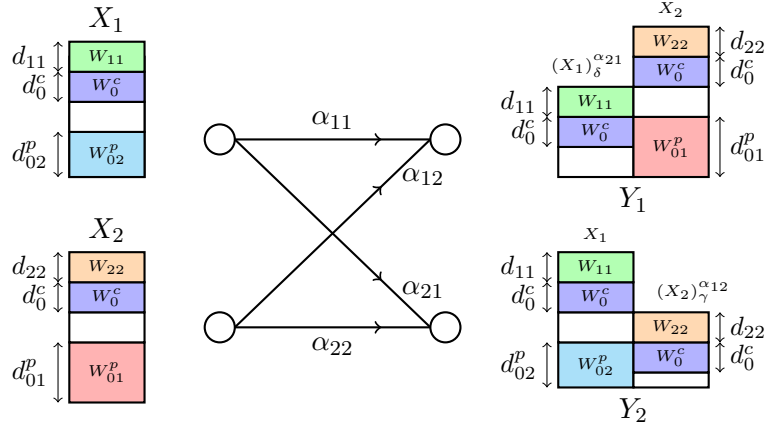
	In this regime, the sum-GDoF value, as characterized in \eqref{eq:strong}, is:
	\begin{align}
	\mathcal{D}_{\Sigma,\iclc}=\min\Big(\alpha_{21}+\pi,\frac{\alpha_{12}+\alpha_{21}+\pi}{2},
	\frac{\alpha_{11}+\alpha_{12}+\alpha_{21}+\alpha_{22}+\pi}{3},\mathcal{D}_{\Sigma,\bc}\Big)
	\end{align}
	\begin{itemize}
		\item  When $\pi\leq \alpha_{12}-\alpha_{21}$, the first bound is active. The achievable scheme is the same as the achievable scheme in Case $1$ which achieves the first bound for the corresponding $\pi$ value. 
		\item When  $\alpha_{21}-\alpha_{12}\leq\pi\leq2\alpha_{11}+2\alpha_{22}-\alpha_{12}-\alpha_{21}$, the second bound is active and also achieved with the same scheme as in Case 1 for corresponding $\pi$ value. 
		\item When $2\alpha_{11}+2\alpha_{22}-\alpha_{12}-\alpha_{21}\leq\pi\leq\pi^*$, where according to Corollary 1 we have $\pi^*=2\alpha_{12}+2\alpha_{21}-\alpha_{11}-\alpha_{22}-3\max(\alpha_{11},\alpha_{22})$, the third bound is tight. It is achieved by the following: $W_{11},W_{22},W_{01}^p,W_{02}^p$  carry $(2\alpha_{21}-\alpha_{12}+2\alpha_{11}-\alpha_{22}-\pi)/3,(2\alpha_{12}-\alpha_{21}+2\alpha_{22}-\alpha_{11}-\pi)/3,(\alpha_{11}+\alpha_{22}+\alpha_{12}-2\alpha_{21}+\pi)/3,(\alpha_{11}+\alpha_{22}+\alpha_{21}-2\alpha_{12}+\pi)/3$ GDoF respectively. They are encoded into independent Gaussian codebooks $X_{11},X_{22},X_{01}^p,X_{02}^p$ with powers $\E|X_{11}|^2=1-P^{-d_{11}},\E|X_{22}|^2=1-P^{-d_{22}},\E|X_{01}^p|^2=P^{-d_{11}-d_{22}-d_0^c}=P^{(\alpha_{11}+\alpha_{22}-2\alpha_{12}-2\alpha_{21}+\pi)/3},\E|X_{02}^p|^2=P^{-d_{11}-d_{22}-d_0^c}=P^{(\alpha_{11}+\alpha_{22}-2\alpha_{12}-2\alpha_{21}+\pi)/3}$. $W_0^c$ carries $(\alpha_{12}+\alpha_{21}-2\alpha_{11}-2\alpha_{22}+\pi)/3$ GDoF and it is encoded to a vector Gaussian codebook $X_0^c=(X_{01}^c,X_{02}^c)$ with power covariance matrix  $\mbox{\tt Diag}(P^{-d_{11}}-P^{(\alpha_{11}+\alpha_{22}-2\alpha_{12}-2\alpha_{21}+\pi)/3},$ $P^{-d_{22}}-P^{(\alpha_{11}+\alpha_{22}-2\alpha_{12}-2\alpha_{21}+\pi)/3})$. The transmitted symbols are $X_1=X_{11}+X_{01}^c+X_{02}^p$, $X_2=X_{22}+X_{02}^c+X_{01}^p$. When decoding, User $1$ decodes $X_{22}.X_0^c,X_{11},X_{01}^p$ for messages $W_{22},W_0^c,W_{11},W_{01}^p$ successively, whose SINR values are $\sim P^{d_{22}},\sim P^{-d_{22}+\alpha_{12}-\alpha_{11}}=P^{d_0^c},\sim P^{\alpha_{11}-(\alpha_{11}+\alpha_{22}+\alpha_{12}-2\alpha_{21}+\pi)/3}=P^{d_{11}},\sim P^{\alpha_{12}+(\alpha_{11}+\alpha_{22}-2\alpha_{12}-2\alpha_{21}+\pi)/3}=P^{d_{01}^p}$ respectively. User $2$ proceeds similarly by successively decoding $W_{11},W_0^c,W_{22},W_{02}^p$. See Figure \ref{fig:case_2} for an illustration. 
	\end{itemize}
	
	\subsubsection*{\bf Case 3: $\alpha_{12}\geq\alpha_{11}+\alpha_{22},\alpha_{21}\leq\alpha_{11}+\alpha_{22}$}
	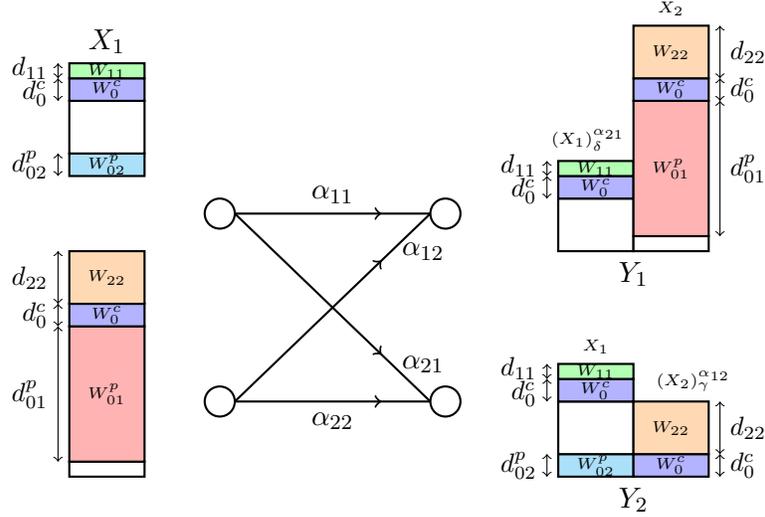
\begin{figure}[htb]
		\centering
		\begin{tikzpicture}[scale=1.0, baseline=(current bounding box.center)]
		\begin{scope}[shift={(0,1.5)}]
		\draw (0.5, 1.5) node[above]{${X}_1$};
		\draw[ thick, fill=cyan!30!white, pattern color=blue] (0,0) rectangle (1,0.3) node [pos=.5] {\tiny $W_{02}^p$};
		\draw[arrows=<->] (-0.15,0)--(-0.15,0.3)node[left,pos=0.5]{\small $d_{02}^p$};
		\draw[ thick, fill=white, pattern color=blue] (0,0.3) rectangle (1,1);
		\draw[ thick, fill=blue!30!white, pattern color=blue] (0,1) rectangle (1,1.3) node [pos=.5] {\tiny $W_0^c$};
		\draw[arrows=<->] (-0.15,1)--(-0.15,1.3)node[left,pos=0.35]{\small $d_0^c$};
		\draw[ thick, fill=green!30!white, pattern color=blue] (0,1.3) rectangle (1,1.5) node [pos=.5] {\tiny $W_{11}$};
		\draw[arrows=<->] (-0.15,1.3)--(-0.15,1.5)node[left,pos=0.7]{\small $d_{11}$};
		\end{scope}
		
		\begin{scope}[shift={(-1,-2.5)}]
		\draw (1.5, 2) node[above]{${X}_2$};
		
		\draw[ thick, fill=white, pattern color=blue] (1,0) rectangle (2,0.2);
		\draw[ thick, fill=red!30!white, pattern color=blue] (1,0.2) rectangle (2,2)node [pos=.5] {\tiny $W_{01}^p$};
		\draw[arrows=<->] (0.85,0.2)--(0.85,2)node[left,pos=0.5]{\small $d_{01}^p$};
		\draw[ thick, fill=blue!30!white, pattern color=blue] (1,2) rectangle (2,2.3) node [pos=.5] {\tiny $W_0^c$};
		\draw[arrows=<->] (0.85,2)--(0.85,2.3)node[left,pos=0.5]{\small $d_0^c$};
		\draw[ thick, fill=orange!30!white, pattern color=blue] (1,2.3) rectangle (2,3) node [pos=.5] {\tiny $W_{22}$};
		\draw[arrows=<->] (0.85,2.3)--(0.85,3)node[left,pos=0.5]{\small $d_{22}$};
		\end{scope}		
		
		\draw[thick] (2,1) circle(2mm);
		\draw[thick] (5,1) circle(2mm);
		\draw[thick] (2,-1.5) circle(2mm);
		\draw[thick] (5,-1.5) circle(2mm);
		
		\draw[thick,arrowmid] (2.2,1)--(4.8,1)node[above,pos=0.5]{$\alpha_{11}$};
		\draw[thick,arrowmid] (2.2,-1.5)--(4.8,-1.5)node[below,pos=0.5]{$\alpha_{22}$};
		\draw[thick,arrowmid] (2.2,1)--(4.8,-1.5)node[right,pos=0.8]{$\alpha_{21}$};
		\draw[thick,arrowmid] (2.2,-1.5)--(4.8,1)node[right,pos=0.8]{$\alpha_{12}$};
		
		\begin{scope}[shift={(6.5,0.5)}]
		\draw[ thick, fill=white, pattern color=blue] (0,0) rectangle (1,0.7);
		\draw[ thick, fill=blue!30!white, pattern color=blue] (0,0.7) rectangle (1,1) node [pos=.5] {\tiny $W_0^c$};
		\draw[arrows=<->] (-0.15,0.7)--(-0.15,1.0)node[left,pos=0.35]{\small $d_0^c$};
		\draw[ thick, fill=green!30!white, pattern color=red] (0,1) rectangle (1,1.2)node [pos=.5] {\tiny $W_{11}$};
		\draw[arrows=<->] (-0.15,1)--(-0.15,1.2)node[left,pos=0.7]{\small $d_{11}$};
		
		\draw[ thick, fill=white, pattern color=blue] (1,0) rectangle (2,0.2);
		\draw[ thick, fill=red!30!white, pattern color=blue] (1,0.2) rectangle (2,2)node [pos=.5] {\tiny $W_{01}^p$};
		\draw[arrows=<->] (2.15,0.2)--(2.15,2)node[right,pos=0.5]{\small $d_{01}^p$};
		\draw[ thick, fill=blue!30!white, pattern color=blue] (1,2) rectangle (2,2.3) node [pos=.5] {\tiny $W_0^c$};
		\draw[arrows=<->] (2.15,2)--(2.15,2.3)node[right,pos=0.5]{\small $d_0^c$};
		\draw[ thick, fill=orange!30!white, pattern color=blue] (1,2.3) rectangle (2,3) node [pos=.5] {\tiny $W_{22}$};
		\draw[arrows=<->] (2.15,2.3)--(2.15,3)node[right,pos=0.5]{\small $d_{22}$};
		
		\draw[thick] (1,0) node[below]{$Y_1$};
		\draw (0.4, 1.2) node[above]{\tiny $(X_1)^{\alpha_{21}}_{\delta}$};
		\draw (1.5, 3) node[above]{\tiny $X_2$};
		\end{scope}
		
		\begin{scope}[shift={(6.5,-2.5)}]
		\draw[ thick, fill=cyan!30!white, pattern color=blue] (0,0) rectangle (1,0.3) node [pos=.5] {\tiny $W_{02}^p$};
		\draw[arrows=<->] (-0.15,0)--(-0.15,0.3)node[left,pos=0.5]{\small $d_{02}^p$};
		\draw[ thick, fill=white, pattern color=blue] (0,0.3) rectangle (1,1);
		\draw[ thick, fill=blue!30!white, pattern color=blue] (0,1) rectangle (1,1.3) node [pos=.5] {\tiny $W_0^c$};
		\draw[arrows=<->] (-0.15,1)--(-0.15,1.3)node[left,pos=0.35]{\small $d_0^c$};
		\draw[ thick, fill=green!30!white, pattern color=blue] (0,1.3) rectangle (1,1.5) node [pos=.5] {\tiny $W_{11}$};
		\draw[arrows=<->] (-0.15,1.3)--(-0.15,1.5)node[left,pos=0.7]{\small $d_{11}$};
		
		\draw[ thick, fill=blue!30!white, pattern color=blue] (1,0) rectangle (2,0.3) node [pos=.5] {\tiny $W_0^c$};
		\draw[arrows=<->] (2.15,0)--(2.15,0.3)node[right,pos=0.5]{\small $d_0^c$};
		\draw[ thick, fill=orange!30!white, pattern color=blue] (1,0.3) rectangle (2,1) node [pos=.5] {\tiny $W_{22}$};
		\draw[arrows=<->] (2.15,0.3)--(2.15,1)node[right,pos=0.5]{\small $d_{22}$};
		
		\draw[thick] (1,0) node[below]{$Y_2$};
		\draw (0.5, 1.5) node[above]{\tiny $X_1$};
		\draw (1.8, 1) node[above]{\tiny  $(X_2)^{\alpha_{12}}_{\gamma}$};
		\end{scope}
		\end{tikzpicture}
		\caption{\it \small Signal partition depiction for $\alpha_{12}\geq \alpha_{11}+\alpha_{22},\alpha_{21}\leq \alpha_{11}+\alpha_{22}$, where where $\delta=\alpha_{21}-\alpha_{11},\gamma=\alpha_{12}-\alpha_{22}$. }\label{fig:case_3}
	\end{figure}
	In this regime, the sum-GDoF value is 
	{\small
		\begin{align}
		\mathcal{D}_{\Sigma,\iclc}=\min\Big(\alpha_{21}+\pi,\frac{2\alpha_{12}+\alpha_{21}+\pi}{3},\mathcal{D}_{\Sigma,\bc}\Big)
		\end{align}
	}
	\begin{itemize}
		\item When $\pi\leq \alpha_{12}-\alpha_{21}$, the first bound is tight. The achievable scheme is as follows: $W_{11},W_{22},W_{01}^p$ carry $\alpha_{21}-\alpha_{22},\alpha_{22},\pi$ GDoF respectively and they are encoded into independent Gaussian codebooks producing codewords $X_{11},X_{22},X_{01}^p$ with power $\E|X_{11}|^2=1,\E|X_{22}|^2=1-P^{-\alpha_{22}},\E|X_{01}^p|^2=P^{-\alpha_{22}}$. When decoding, User $1$ decodes $X_{22}$ first with SINR value $\sim P^{\alpha_{22}}$. Then, it subtracts the reconstructed codeword $X_{22}$ and acts as a multiple access receiver to jointly decode $X_{11}$ and $X_{01}^p$. The GDoF region for this multiple access channel is the following.
		\begin{align}
			\{(d_{11},d_{01}^p):d_{11}\leq\alpha_{11},d_{11}+d_{01}^p\leq \alpha_{12}-\alpha_{22} \}
		\end{align}
		Since $d_{11}=\alpha_{21}-\alpha_{22}\leq\alpha_{11},d_{11}+d_{01}^p=\alpha_{21}-\alpha_{22}+\pi\leq\alpha_{12}-\alpha_{22}$ belongs to the GDoF region, $X_{11},X_{01}^p$ can be successfully decoded at User $1$. For User $2$, it successively decodes $X_{11},X_{22}$, whose SINR values are $\sim P^{\alpha_{21}-\alpha_{22}}=P^{d_{11}},\sim P^{\alpha_{22}}=P^{d_{22}}$ respectively. Therefore $W_{11},W_{22}$ are decoded successfully decoded at User $2$.
		\item When $\alpha_{12}-\alpha_{21}\leq\pi\leq\pi^*$, where according to Corollary 1 we have $\pi^*=\alpha_{12}+2\alpha_{21}-3\max(\alpha_{11},\alpha_{22})$, the second bound is tight. The achievable scheme is as follows:  Messages $W_{11},W_{22},W_{01}^p,W_{02}^p$  carry $(2\alpha_{21}+\alpha_{12}-3\alpha_{22}-\pi)/3,(3\alpha_{22}+\alpha_{12}-\alpha_{21}-\pi)/3,(2\alpha_{12}-2\alpha_{21}+\pi)/3,(\alpha_{21}-\alpha_{12}+\pi)/3$ GDoF respectively. They are encoded into independent Gaussian codebooks producing codewords $X_{11},X_{22},X_{01}^p,X_{02}^p$ with powers $\E|X_{11}|^2=1-P^{-d_{11}},\E|X_{22}|^2=1-P^{-d_{22}},\E|X_{01}^p|^2=P^{-\alpha_{22}}$, $\E|X_{02}^p|^2=P^{-d_{11}-d_{22}-d_0^c}=P^{-(2\alpha_{21}+\alpha_{12}-\pi)/3}$. $W_0^c$ carries $(\alpha_{21}-\alpha_{12}+\pi)/3$ GDoF and is encoded into a vector Gaussian codebook $X_0^c=(X_{01}^c,X_{02}^c)$ with power covariance matrix $\mbox{\tt Diag}(P^{-d_{11}}-P^{-(2\alpha_{21}+\alpha_{12}-\pi)/3},P^{-d_{22}}-P^{-\alpha_{22}})$. The transmitted symbols are $X_1=X_{11}+X_{01}^c+X_{02}^p,X_2=X_{22}+X_{02}^c+X_{01}^p$. 
		When decoding, User $1$ decodes $W_{22},W_0^c$ successively while treating everything else as noise. Their SINR values are $\sim P^{d_{22}},\sim P^{\alpha_{22}-d_{22}}=P^{d_0^c}$. After this, User $1$ subtracts the reconstructed codewords $X_{22},X_0^c$. Then it acts as a multiple access receiver to jointly decode $W_{11}$ and $W_{01}^p$ while treating the remaining signal as noise. The GDoF region for this multiple access channel is the following.
		\begin{align}
		\{(d_{11},d_{01}^p):d_{11}\leq\alpha_{11},d_{11}+d_{01}^p\leq\alpha_{12}-\alpha_{22}\}
		\end{align}
		Since $d_{11}=(2\alpha_{21}+\alpha_{12}-3\alpha_{22}-\pi)/3\leq\alpha_{21}-\alpha_{22}\leq\alpha_{11},d_{11}+d_{01}^p\leq \alpha_{12}-\alpha_{22}$ belongs to this GDoF region, it follows that $W_{22},W_0^c$ can be successfully decoded.
		User $2$ decodes $X_{11},X_0^c,X_{22},X_{02}^p$ successively, whose SINR values are $\sim P^{d_{11}},\sim P^{\alpha_{21}-d_{11}-\alpha_{22}}=P^{d_0^c},\sim P^{\alpha_{22}-\alpha{21}+(2\alpha_{21}+\alpha_{12}-\pi)/3}=P^{d_{22}},\sim P^{\alpha_{21}-(2\alpha_{21}+\alpha_{12}-\pi)/3}=P^{d_{02}^p}$ respectively. See Figure \ref{fig:case_3} for an illustration. Note that cooperation capability beyond $\pi^*$ is redundant because with $\pi^*$ cooperation the broadcast GDoF are already achieved.
	\end{itemize} 
	
	\subsubsection*{\bf Case 4: $\alpha_{12}\geq\alpha_{11}+\alpha_{22},\alpha_{21}\geq\alpha_{11}+\alpha_{22}$}
	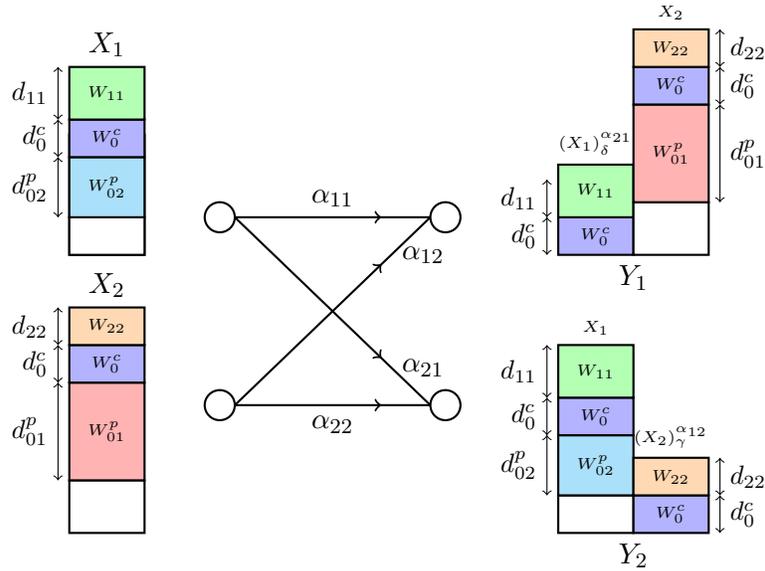
\begin{figure}[htb]
		\centering
		\begin{tikzpicture}[scale=1.0, baseline=(current bounding box.center)]
		\begin{scope}[shift={(0,0.5)}]
		\draw[ thick] (0,0) rectangle (1,1.6);
		\draw (0.5, 2.5) node[above]{${X}_1$};
		\draw[ thick, fill=white, pattern color=blue] (0,0) rectangle (1,0.5);
		\draw[ thick, fill=cyan!30!white, pattern color=blue] (0,0.5) rectangle (1,1.3) node [pos=.5] {\tiny $W_{02}^p$};
		\draw[arrows=<->] (-0.15,0.5)--(-0.15,1.3)node[left,pos=0.5]{\small $d_{02}^p$};
		\draw[ thick, fill=blue!30!white, pattern color=blue] (0,1.3) rectangle (1,1.8) node [pos=.5] {\tiny$W_0^c$};
		\draw[arrows=<->] (-0.15,1.3)--(-0.15,1.8)node[left,pos=0.5]{\small $d_0^c$};
		\draw[ thick, fill=green!30!white, pattern color=blue] (0,1.8) rectangle (1,2.5) node [pos=.5] {\tiny$W_{11}$};
		\draw[arrows=<->] (-0.15,1.8)--(-0.15,2.5)node[left,pos=0.5]{\small $d_{11}$};
		\end{scope}

		\begin{scope}[shift={(-1,-3.2)}]
		\draw (1.5, 3) node[above]{${X}_2$};
		\draw[ thick, fill=white, pattern color=blue] (1,0) rectangle (2,0.7);
		\draw[ thick, fill=red!30!white, pattern color=blue] (1,0.7) rectangle (2,2)node [pos=.5] {\tiny $W_{01}^p$};
		\draw[arrows=<->] (0.85,0.7)--(0.85,2)node[left,pos=0.5]{\small $d_{01}^p$};
		\draw[ thick, fill=blue!30!white, pattern color=blue] (1,2) rectangle (2,2.5) node [pos=.5] {\tiny $W_0^c$};
		\draw[arrows=<->] (0.85,2)--(0.85,2.5)node[left,pos=0.5]{\small $d_0^c$};
		\draw[ thick, fill=orange!30!white, pattern color=blue] (1,2.5) rectangle (2,3) node [pos=.5] {\tiny$W_{22}$};
		\draw[arrows=<->] (0.85,2.5)--(0.85,3)node[left,pos=0.5]{\small $d_{22}$};
		\end{scope}		
		\draw[thick] (2,1) circle(2mm);
		\draw[thick] (5,1) circle(2mm);
		\draw[thick] (2,-1.5) circle(2mm);
		\draw[thick] (5,-1.5) circle(2mm);
		
		\draw[thick,arrowmid] (2.2,1)--(4.8,1)node[above,pos=0.5]{$\alpha_{11}$};
		\draw[thick,arrowmid] (2.2,-1.5)--(4.8,-1.5)node[below,pos=0.5]{$\alpha_{22}$};
		\draw[thick,arrowmid] (2.2,1)--(4.8,-1.5)node[right,pos=0.8]{$\alpha_{21}$};
		\draw[thick,arrowmid] (2.2,-1.5)--(4.8,1)node[right,pos=0.8]{$\alpha_{12}$};
		
		\begin{scope}[shift={(6.5,0.5)}]
		\draw[ thick, fill=blue!30!white, pattern color=blue] (0,0) rectangle (1,0.5) node [pos=.5] {\tiny $W_0^c$};
		\draw[arrows=<->] (-0.15,0)--(-0.15,0.5)node[left,pos=0.5]{\small $d_0^c$};
		\draw[ thick, fill=green!30!white, pattern color=blue] (0,0.5) rectangle (1,1.2)node [pos=.5] {\tiny $W_{11}$};
		\draw[arrows=<->] (-0.15,0.5)--(-0.15,1)node[left,pos=0.5]{\small $d_{11}$};
		
		\draw[ thick, fill=white, pattern color=blue] (1,0) rectangle (2,0.7);
		\draw[ thick, fill=red!30!white, pattern color=blue] (1,0.7) rectangle (2,2)node [pos=.5] {\tiny $W_{01}^p$};
		\draw[arrows=<->] (2.15,0.7)--(2.15,2)node[right,pos=0.5]{\small $d_{01}^p$};
		\draw[ thick, fill=blue!30!white, pattern color=blue] (1,2) rectangle (2,2.5) node [pos=.5] {\tiny $W_0^c$};
		\draw[arrows=<->] (2.15,2)--(2.15,2.5)node[right,pos=0.5]{\small $d_0^c$};
		\draw[ thick, fill=orange!30!white, pattern color=blue] (1,2.5) rectangle (2,3) node [pos=.5] {\tiny$W_{22}$};
		\draw[arrows=<->] (2.15,2.5)--(2.15,3)node[right,pos=0.5]{\small $d_{22}$};
		
		\draw[thick] (1,0) node[below]{$Y_1$};
		\draw (0.5, 1.2) node[above]{\tiny $(X_1)^{\alpha_{21}}_{\delta}$};
		\draw (1.5, 3) node[above]{\tiny $X_2$};
		\end{scope}
		
		\begin{scope}[shift={(6.5,-3.2)}]
		\draw[ thick, fill=white, pattern color=blue] (0,0) rectangle (1,0.5);
		\draw[ thick, fill=cyan!30!white, pattern color=blue] (0,0.5) rectangle (1,1.3) node [pos=.5] {\tiny $W_{02}^p$};
		\draw[arrows=<->] (-0.15,0.5)--(-0.15,1.3)node[left,pos=0.5]{\small $d_{02}^p$};
		\draw[ thick, fill=blue!30!white, pattern color=blue] (0,1.3) rectangle (1,1.8) node [pos=.5] {\tiny$W_0^c$};
		\draw[arrows=<->] (-0.15,1.3)--(-0.15,1.8)node[left,pos=0.5]{\small $d_0^c$};
		\draw[ thick, fill=green!30!white, pattern color=blue] (0,1.8) rectangle (1,2.5) node [pos=.5] {\tiny$W_{11}$};
		\draw[arrows=<->] (-0.15,1.8)--(-0.15,2.5)node[left,pos=0.5]{\small $d_{11}$};
		
		\draw[ thick, fill=blue!30!white, pattern color=blue] (1,0) rectangle (2,0.5) node [pos=.5] {\tiny$W_0^c$};
		\draw[arrows=<->] (2.15,0)--(2.15,0.5)node[right,pos=0.5]{\small $d_0^c$};
		\draw[ thick, fill=orange!30!white, pattern color=blue] (1,0.5) rectangle (2,1) node [pos=.5] {\tiny$W_{22}$};
		\draw[arrows=<->] (2.15,0.5)--(2.15,1)node[right,pos=0.5]{\small $d_{22}$};
		
		\draw[thick] (1,0) node[below]{$Y_2$};
		\draw (0.5, 2.5) node[above]{\tiny $X_1$};
		\draw (1.5, 1) node[above]{\tiny $(X_2)^{\alpha_{12}}_{\gamma}$};
		\end{scope}
		\end{tikzpicture}
		\caption{\it \small Signal partition in the regime $ \alpha_{12},\alpha_{21}\geq \alpha_{11}+\alpha_{22}$, where $\delta=\alpha_{21}-\alpha_{11},\gamma=\alpha_{12}-\alpha_{22}$.}\label{fig:case_4}
	\end{figure}
	In this regime, we have
	\begin{align}
	\mathcal{D}_{\Sigma,\iclc}=\min\Big(\alpha_{11}+\alpha_{22}+\pi,\frac{2\alpha_{12}+2\alpha_{21}-\alpha_{11}-\alpha_{22}+\pi}{3},
	\mathcal{D}_{\Sigma,\bc}\Big) \label{case5}
	\end{align}
	\begin{itemize}
		\item When $\pi\leq \alpha_{21}-\alpha_{11}-\alpha_{22}$, the first bound is active, which is achieved by letting $W_{11},W_{22},W_{02}^p$ carry $\alpha_{11},\alpha_{22},\pi$ GDoF respectively. They are encoded into independent Gaussian codebooks $X_{11},X_{22},X_{02}^p$ with power $\E|X_{11}|^2=1-P^{-\alpha_{11}},\E|X_{22}|^2=1,\E|X_{02}^p|^2=P^{-\alpha_{11}}$. The transmitted symbols are $X_1=X_{11}+X_{02}^p,X_2=X_{22}$. When decoding, User $1$ decodes $X_{22},X_{11}$ successively, whose SINR values are $\sim P^{\alpha_{12}-\alpha_{11}},\sim P^{\alpha_{11}}$ respectively. Since $d_{22}=\alpha_{22}\leq\alpha_{12}-\alpha_{11},d_{11}=\alpha_{11}$, messages $W_{22},W_{11}$ can be decoded successfully. User $2$ decodes $X_{11},X_{02}^p,X_{22}$ successively, whose SINR values are $\sim P^{\alpha_{11}},\sim P^{\alpha_{21}-\alpha_{11}-\alpha_{22}},\sim P^{\alpha_{22}}$ respectively. Since $d_{11}=\alpha_{11},d_{02}^p=\pi\leq\alpha_{21}-\alpha_{11}-\alpha_{22},d_{22}=\alpha_{22}$, messages $W_{11},W_{02}^p,W_{22}$ can be decoded successfully. 
		
		\item When $\alpha_{21}-\alpha_{11}-\alpha_{22}\leq \pi\leq \alpha_{12}+\alpha_{21}-2\alpha_{11}-2\alpha_{22}$, the first bound is still active and is achieved by letting $W_{11},W_{22},W_{01}^p,W_{02}^p$ carry $\alpha_{11},\alpha_{22},\pi+\alpha_{11}+\alpha_{22}-\alpha_{21},\alpha_{21}-\alpha_{11}-\alpha_{22}$ GDoF respectively. They are encoded into independent Gaussian codebooks $X_{11},X_{22},X_{01}^p,X_{02}^p$ with power $\E|X_{11}|^2=1-P^{-\alpha_{11}},\E|X_{22}|^2=1-P^{-\alpha_{22}},\E|X_{01}^p|^2=P^{-\alpha_{22}},\E|X_{02}^p|^2=P^{-\alpha_{11}}$. The transmitted symbols are $X_1=X_{11}+X_{02}^p,X_2=X_{22}+X_{01}^p$. When decoding, User $1$ decodes $X_{22},X_{01}^p,X_{11}$ successively, whose SINR values are $\sim P^{\alpha_{22}},\sim P^{\alpha_{12}-\alpha_{22}-\alpha_{11}},\sim P^{\alpha_{11}}$ respectively. Since $d_{22}=\alpha_{22},d_{01}^p=\pi+\alpha_{11}+\alpha_{22}-\alpha_{21}\leq\alpha_{12}-\alpha_{22}-\alpha_{11},d_{11}=\alpha_{11}$, $X_{22},X_{01}^p,X_{11}$ can be decoded successfully. User $2$ proceeds similarly by decoding $X_{11},X_{02}^p,X_{22}$ successively. 
		\item When $\alpha_{12}+\alpha_{21}-2\alpha_{11}-2\alpha_{22}\leq\pi\leq\pi^*$, where according to Corollary 1 we have $\pi^*=\alpha_{12}+\alpha_{21}+\alpha_{11}+\alpha_{22}-3\max(\alpha_{11},\alpha_{22})$, the second bound is tight. It is achieved as follows: $W_{11},W_{22},W_{01}^p,W_{02}^p$ carry $(\alpha_{12}+\alpha_{21}-2\alpha_{22}+\alpha_{11}-\pi)/3,(\alpha_{12}+\alpha_{21}-2\alpha_{11}+\alpha_{22}-\pi)/3,(2\alpha_{12}+\pi-\alpha_{11}-\alpha_{22}-\alpha_{21})/3,(2\alpha_{21}+\pi-\alpha_{11}-\alpha_{22}-\alpha_{12})/3$ GDoF respectively and are encoded into independent Gaussian codebooks $X_{11},X_{22},X_{01}^p,X_{02}^p$ with powers $\E|X_{11}|^2=1-P^{-d_{11}},\E|X_{22}|^2=1-P^{-d_{22}},\E|X_{01}^p|^2=P^{-\alpha_{22}},\E|X_{02}^p|^2=P^{-\alpha_{11}}$. $W_0^c$ carries $(2\alpha_{11}+2\alpha_{22}+\pi-\alpha_{12}-\alpha_{21})/3$ GDoF and is encoded into a vector Gaussian codebook $X_0^c=(X_{01}^c,X_{02}^c)$ with power covariance matrix $\mbox{\tt Diag}(P^{-d_{11}}-P^{-\alpha_{11}},P^{-d_{22}}-P^{-\alpha_{22}})$. The transmitted symbols are $X_1=X_{11}+X_{01}^c+X_{02}^p,X_2=X_{22}+X_{02}^c+X_{01}^p$. User $1$ decodes $W_{22},W_0^c$ successively while treating everything else as noise, whose SINR values are $\sim P^{d_{22}},\sim P^{\alpha_{22}-d_{22}}=P^{d_0^c}$ respectively. After this, User $1$ subtracts the contribution of codewords $X_{22},X_0^c$ and then acts as a multiple access receiver by jointly decoding $W_{11}$ and $W_{01}^p$ while treating the remaining signals as noise. The GDoF region for this multiple access channel is the following.
		\begin{align}
		\{(d_{11},d_{01}^p):d_{11}\leq\alpha_{11},d_{11}+d_{01}^p\leq\alpha_{12}-\alpha_{22} \}
		\end{align}
		Since $d_{11}=(\alpha_{12}+\alpha_{21}-2\alpha_{22}+\alpha_{11}-\pi)/3\leq (\alpha_{12}+\alpha_{21}-2\alpha_{22}+\alpha_{11}-(\alpha_{12}+\alpha_{21}-2\alpha_{11}-2\alpha_{22}))/3\leq\alpha_{11},d_{11}+d_{01}^p=\alpha_{12}-\alpha_{22}$ belongs to the GDoF region, $W_{11},W_{01}^p$ can be decoded successfully.
		User $2$ proceeds similarly. See Figure \ref{fig:case_4} for an illustration.
	\end{itemize}

	\subsection{Full-duplex Setting} \label{ach:fdstrong}
	
	In this section we consider the achievability for the full-duplex setting. Before presenting the complete proof, let us use our example ($\alpha_{11}=\alpha_{22}=2,\alpha_{12}=5,\alpha_{21}=3$) to convey the main insights. Here we have $\mathcal{D}_{\Sigma,\iclc}'=\min(3+\pi,3+\frac{\pi}{2},\frac{13+\pi}{3},6)$. The bounds $\mathcal{D}_{\Sigma,\ic}+\pi$ and $ \frac{\mathcal{D}_{3e}+\pi}{3}$ are redundant. To achieve the broadcast channel bound ($\mathcal{D}_{\Sigma,\bc}=6$), the GDoF in the conference link is $\pi=6$, which means our proposed scheme is no more efficient than \cite{Arash_Jafar_cooperation}. This is because in our scheme, $d_0^c=1,d_{01}^p=3,d_{02}^p=1$, which requires $\frac{\pi}{2}\geq d_{01}^p=3$. Hence $1$ DoF in the $W_{02}$ conference link is wasted because $d_{02}\leq d_{02}^p+d_0^c=2$. We can see that under full-duplex setting, one cooperation link is fully wasted in the mixed interference regime, but in the strong interference regime, one cooperation link is  partially wasted.
	
	In the full-duplex setting, first of all, the achievable schemes even for one cooperative bit to buy one over-the-air bit or half over-the-air bit become nontrivial as one of the cooperation links is partially wasted for some $\pi$ values. Hence we will discuss it in a bit more detail. On the other hand, the achievable scheme for achieving the $1/3$ bound (when the bound is active) in the full-duplex setting is the same as the corresponding scheme for  half-duplex setting. In general, we also consider the 4 cases. Similarly, $\pi\leq\pi^+$ is assumed.
	\subsubsection*{\bf Case 1: $\alpha_{12}\leq\alpha_{11}+\alpha_{22},\alpha_{21}\leq\alpha_{11}+\alpha_{22},\alpha_{12}+\alpha_{21}\leq\alpha_{11}+\alpha_{22}+\max(\alpha_{11},\alpha_{22})$}	
	In this regime, the sum-GDoF is 
	\begin{align}
	\mathcal{D}_{\Sigma,\iclc}'=\min\Big(\alpha_{21}+\frac{\pi}{2},\mathcal{D}_{\Sigma,\bc}\Big)
	\end{align}
	\begin{itemize}
		\item When $\frac{\pi}{2}\leq\alpha_{12}-\alpha_{21}$, the first bound is active. The achievability is the same as Case $1$ in the half-duplex setting to achieve the first bound, except $d_{01}^p=\pi/2$ here.
		\item When $\alpha_{12}-\alpha_{21}\leq\frac{\pi}{2}\leq\frac{\pi^+}{2}$,  where according to Corollary $2$ we have $\pi^+=2\alpha_{12}-2\max(\alpha_{11},\alpha_{22})$, the first bound is still active and is achieved by letting $W_{11},W_{22},W_{01}^p,W_{02}^p$ carry $\alpha_{11}-\frac{\pi}{2},\alpha_{12}-\alpha_{11},\frac{\pi}{2},\frac{\pi}{2}+\alpha_{21}-\alpha_{12}$ GDoF respectively. They are encoded into independent Gaussian codebooks producing codewords $X_{11},X_{22},X_{01}^p,X_{02}^p$ with powers $\E|X_{11}|^2=1-P^{-d_{11}-d_{22}},\E|X_{22}|^2=1-P^{-d_{11}-d_{22}},\E|X_{01}^p|^2=P^{-d_{11}-d_{22}},\E|X_{02}^p|^2=P^{-d_{11}-d_{22}}$. The transmitted signals are $X_1=X_{11}+X_{02}^p,X_2=X_{22}+X_{01}^p$. When decoding, Receiver $1$ uses successive interference cancellation to decode $X_{22},X_{11},X_{01}^p$ successively, whose SINR values are $\sim P^{\alpha_{12}},\sim P^{\alpha_{11}-\alpha_{12}+d_{11}+d_{22}}=P^{d_{11}},\sim P^{\alpha_{12}-d_{11}-d_{22}}=P^{\frac{\pi}{2}}$. Therefore, $W_{22},W_{11},W_{01}^p$ can be successfully decoded. User $2$ acts as a multiple access receiver, it jointly decodes $X_{11}$ and $X_{22}$, while the noise floor due to $X_{02}^p$ is $P^{\alpha_{21}-\alpha_{12}+\frac{\pi}{2}}$, the GDoF region for this multiple access channel is the following.
		
		\begin{align}
		\{(d_{11},d_{22}):d_{22}\leq\alpha_{22}-(\alpha_{21}-\alpha_{12}+\frac{\pi}{2}),d_{11}+d_{22}\leq\alpha_{12}-\frac{\pi}{2}\}
		\end{align}
		Since $d_{22}=\alpha_{12}-\alpha_{11}\leq\alpha_{22}-\alpha_{21}+\max(\alpha_{11},\alpha_{22})=\alpha_{22}-(\alpha_{21}-\alpha_{12}+\frac{\pi^+}{2})\leq\alpha_{22}-(\alpha_{21}-\alpha_{12}+\frac{\pi}{2}),d_{11}+d_{22}=\alpha_{12}-\frac{\pi}{2}=\alpha_{21}-d_{02}^p$, $W_{22},W_{11}$ can be decoded successfully.
		Then User $2$ subtracts the contribution of $X_{11},X_{22}$ and decodes $X_{02}^p$, whose SINR is $\sim P^{\alpha_{21}-d_{11}-d_{22}}=P^{d_{02}^p}$, so $W_{02}^p$ is decoded successfully. 
	\end{itemize}
	\subsubsection*{\bf Case 2: $\alpha_{12}\leq\alpha_{11}+\alpha_{22},\alpha_{21}\leq\alpha_{11}+\alpha_{22},\alpha_{12}+\alpha_{21}\geq\alpha_{11}+\alpha_{22}+\max(\alpha_{11},\alpha_{22})$}
	\begin{itemize}
		\item $2\alpha_{21}\leq\alpha_{11}+\alpha_{22}+\max(\alpha_{11},\alpha_{22})$\\
		In this regime, the sum-GDoF value is 
		\begin{align}
		\mathcal{D}_{\Sigma,\iclc}'=\min\Big(\alpha_{21}+\frac{\pi}{2},\mathcal{D}_{\Sigma,\bc}\Big)
		\end{align}
		\begin{itemize}
			\item When $\frac{\pi}{2}\leq\alpha_{11}+\alpha_{22}-\alpha_{21}$, the first bound is active. The achievability is the same as in Case $1$ above to achieve the first bound for the corresponding $\pi$ value.
			\item When $\alpha_{11}+\alpha_{22}-\alpha_{21}\leq\frac{\pi}{2}\leq\frac{\pi^+}{2}$, where according to Corollary $2$ we have $\pi^+=2\alpha_{12}-2\max(\alpha_{11},\alpha_{22})$, the first bound is still active and is achieved by letting $W_{11},W_{22},W_{01}^p,W_{02}^p,W_0^c$ carry $\alpha_{11}-\frac{\pi}{2},\alpha_{12}-\alpha_{21}+\alpha_{22}-\frac{\pi}{2},\frac{\pi}{2},\alpha_{21}-\alpha_{12}+\frac{\pi}{2},\alpha_{21}-\alpha_{11}-\alpha_{22}+\frac{\pi}{2}$ GDoF respectively. Messages $W_{11},W_{22},W_{01}^p,W_{02}^p$ are encoded into independent Gaussian codewords $X_{11},X_{22},X_{01}^p,X_{02}^p$ with powers $\E|X_{11}|^2=1-P^{-d_{11}},\E|X_{22}|^2=1-P^{-d_{22}},\E|X_{01}^p|^2=P^{\alpha_{11}-\alpha_{12}-d_{11}},\E|X_{02}^p|^2=P^{\alpha_{22}-\alpha_{21}-d_{22}},$ respectively. Message $W_0^c$ is encoded into a vector Gaussian codeword $X_0^c=(X_{01}^c,X_{02}^c)$ with covariance matrix $\mbox{\tt Diag}(P^{-d_{11}}-P^{\alpha_{22}-\alpha_{21}-d_{22}},P^{-d_{22}}-P^{\alpha_{11}-\alpha_{12}-d_{11}})$. The transmitted symbols are $X_1=X_{11}+X_{01}^c+X_{02}^p,X_1=X_{22}+X_{02}^c+X_{01}^p$. When decoding, User $1$ decodes $X_{22},X_0^c,X_{11},X_{01}^p$ successively, with SINR values $\sim P^{d_{22}},\sim P^{\alpha_{12}-\alpha_{11}-d_{22}}=P^{d_0^c},\sim P^{d_{11}}, \sim P^{\alpha_{11}-d_{11}}=P^{\frac{\pi}{2}}=P^{d_{01}^p}$, respectively. Therefore $W_{22},W_0^c,W_{11},W_{01}^p$ can be successfully decoded at User $1$. User $2$ proceeds similarly by decoding $X_{11},X_0^c,X_{22},X_{02}^p$ successively. The signal partition depiction is similar to Figure \ref{fig:case_2} except that codewords' power levels are changed. It can be checked that $d_{01}=d_{01}^p=\frac{\pi}{2},d_{02}=d_{02}^p+d_0^c=2\alpha_{21}-\alpha_{12}-\alpha_{11}-\alpha_{22}+\pi\leq 2\alpha_{21}-\alpha_{12}-\alpha_{11}-\alpha_{22}+\frac{\pi^+}{2}+\frac{\pi}{2}\leq2\alpha_{21}-\alpha_{11}-\alpha_{22}-\max(\alpha_{11},\alpha_{22})+\frac{\pi}{2}\leq\frac{\pi}{2}$. 
		\end{itemize}
		\item $2\alpha_{21}\geq\alpha_{11}+\alpha_{22}+\max(\alpha_{11},\alpha_{22})$\\
		The sum-GDoF value in this regime is 
		\begin{align}
		\mathcal{D}_{\Sigma,\iclc}'=\min\Big(\alpha_{21}+\frac{\pi}{2},\frac{\alpha_{11}+\alpha_{12}+\alpha_{21}+\alpha_{21}+\pi}{3},\mathcal{D}_{\Sigma,\bc}\Big)
		\end{align}	
		\begin{itemize} 
			\item When $\frac{\pi}{2}\leq \alpha_{11}+\alpha_{22}+\alpha_{12}-2\alpha_{21}$, the first bound is active. The achievable scheme is identical to $2\alpha_{21}\leq\alpha_{11}+\alpha_{22}+\max(\alpha_{11},\alpha_{22})$ for the same $\pi$ value.   
			\item When $\alpha_{11}+\alpha_{22}-2\alpha_{12}+\alpha_{21}\leq\frac{\pi}{2}\leq\frac{\pi^+}{2}$, where according to Corollary $2$ we have $\pi^+=2\alpha_{12}+2\alpha_{21}-\alpha_{11}-\alpha_{22}-3\max(\alpha_{11},\alpha_{22})$, the second bound is active, whose achievability is the same as the  achievable scheme for the corresponding bound in Case $2$ in the half-duplex setting.
		\end{itemize}
	\end{itemize}

	\subsubsection*{\bf Case 3: $\alpha_{12}\geq\alpha_{11}+\alpha_{22},\alpha_{21}\leq\alpha_{11}+\alpha_{22}$}
	\begin{itemize}
		\item $\alpha_{12}\geq2\alpha_{21}-\max(\alpha_{11},\alpha_{22})$\\
		In this regime,  the sum-GDoF value is 
		\begin{align}
		\mathcal{D}_{\Sigma,\iclc}'=\min\Big(\alpha_{21}+\frac{\pi}{2},\mathcal{D}_{\Sigma,\bc}\Big)
		\end{align}
		
		\begin{itemize}
			\item When $\frac{\pi}{2}\leq \alpha_{12}-\alpha_{21}$, the first bound is tight, and its achievability is identical to first bound in Case 3 under the half-duplex setting except $d_{01}^p=\frac{\pi}{2}$ here. 
			\item When $\alpha_{12}-\alpha_{21}\leq\frac{\pi}{2}\leq \frac{\pi^+}{2}$,  where according to Corollary $2$ we have $\pi^+=2\alpha_{12}-2\max(\alpha_{11},\alpha_{22})$, the first bound is active and is achieved as follows: Messages $W_{11},W_{22},W_{01}^p,W_{02}^p$ carry $\alpha_{12}-\alpha_{22}-\frac{\pi}{2},\alpha_{22}+\alpha_{12}-\alpha_{21}-\frac{\pi}{2},\frac{\pi}{2},\alpha_{21}-\alpha_{12}+\frac{\pi}{2}$ GDoF respectively. They are encoded into independent Gaussian codewords $X_{11},X_{22},X_{01}^p,X_{02}^p$ with powers $\E|X_{11}|^2=1-P^{-d_{11}},\E|X_{22}|^2=1-P^{-d_{22}},\E|X_{01}^p|^2=P^{-\alpha_{22}},\E|X_{02}^p|^2=P^{-d_{11}-d_{22}-d_0^c}=P^{-\alpha_{12}+\frac{\pi}{2}}$. $W_0^c$ carries $\alpha_{21}-\alpha_{12}+\frac{\pi}{2}$ GDoF and is encoded into a vector Gaussian codebook $X_0^c=(X_{01}^c,X_{02}^c)$ with  covariance matrix $\E|X_0^c|^2=\mbox{\tt Diag}(P^{-d_{11}}-P^{-\alpha_{12}+\frac{\pi}{2}},P^{-d_{22}}-P^{-\alpha_{22}})$.  The transmitted symbols are $X_1=X_{11}+X_{01}^c+X_{02}^p,X_2=X_{22}+X_{02}^c+X_{01}^p$. When decoding, User $1$ decodes $X_{22},X_0^c$ successively while treating everything else as noise. The SINR values for $X_{22},X_0^c$ are $\sim P^{d_{22}},\sim P^{\alpha_{22}-d_{22}}=P^{d_0^c}$ respectively. After subtracting the contribution of $X_{22},X_0^c$, it jointly decodes $X_{11}$ and $X_{01}^p$. The GDoF region for this multiple access channel is the following.
			\begin{align}
			\{(d_{11},d_{01}^p):d_{11}\leq\alpha_{11},d_{22}+d_{01}^p\leq \alpha_{12}-\alpha_{22}\}
			\end{align}
			Since $d_{11}=\alpha_{12}-\alpha_{22}-\frac{\pi}{2}\leq\alpha_{12}-\alpha_{22}-\frac{\pi^+}{2}=\max(\alpha_{11},\alpha_{22})-\alpha_{22}\leq\alpha_{11},d_{11}+d_{01}^p=\alpha_{12}-\alpha_{22}$ belongs to the GDoF region of the multiple access channel, $W_{11},W_{01}^p$ can be decoded successfully. User 2 successively decodes $X_{11},X_0^c,X_{22},X_{02}^p$, whose SINR values are $\sim P^{d_{11}},\sim P^{\alpha_{21}-d_{11}-\alpha_{22}}=P^{\alpha_{21}-\alpha_{12}+\frac{\pi}{2}}=P^{d_0^c},\sim P^{\alpha_{22}-\alpha_{21}+\alpha_{12}-\frac{\pi}{2}}=P^{d_{22}},\sim P^{\alpha_{21}-\alpha_{12}+\frac{\pi}{2}}=P^{d_{02}^p}$ respectively. The signal partition depiction is similar to Figure \ref{fig:case_3} except that the codewords' power levels are changed. Note that $d_{01}=d_{01}^p=\frac{\pi}{2}, d_{02}=d_{02}^p+d_0^c=2\alpha_{21}-2\alpha_{12}+\pi\leq2\alpha_{21}-2\alpha_{12}+\frac{\pi^+}{2}+\frac{\pi}{2}=2\alpha_{21}-\alpha_{12}-\max(\alpha_{11},\alpha_{22})+\frac{\pi}{2}\leq\frac{\pi}{2}$. 
		\end{itemize}
		\item $\alpha_{12}\leq2\alpha_{21}-\max(\alpha_{11},\alpha_{22})$\\
		In this regime, the sum-GDoF value is 
		\begin{align}
		\mathcal{D}_{\Sigma,\iclc}'=\min\Big(\alpha_{21}+\frac{\pi}{2},\frac{2\alpha_{12}+\alpha_{21}+\pi}{3},\mathcal{D}_{\Sigma,\bc}\Big)
		\end{align}
		\begin{itemize}
			\item When $\frac{\pi}{2}\leq2\alpha_{12}-2\alpha_{21}$, the first bound is active. The achievable schemes are identical to those for $\alpha_{12}\geq2\alpha_{21}-\max(\alpha_{11},\alpha_{22})$ for the same $\pi$ value.
			\item When $2\alpha_{12}-2\alpha_{21}\leq\frac{\pi}{2}\leq\frac{\pi^+}{2}$, where according to Corollary $2$ we have $\pi^+=\alpha_{12}+2\alpha_{21}-3\max(\alpha_{11},\alpha_{22})$, the second bound is active, and its achievability is the same as the corresponding $1/3$ factor bound scheme in Case 3 of the half-duplex setting.
		\end{itemize}
	\end{itemize}
	\subsubsection*{\bf Case 4: $\alpha_{12}\geq\alpha_{11}+\alpha_{22},\alpha_{21}\geq\alpha_{11}+\alpha_{22}$}
	\begin{table}[h]
		\centering
		\begin{adjustbox}{angle=0,width=\columnwidth,center}
			\begin{tabular}{|c|c|c|c|c|}
				\hline
				\multirow{2}{*}{Regimes}                                                                                                                                    & \multirow{2}{*}{}                                                                                                     & \multirow{2}{*}{Codewords' GDoF and Corresponding Power}                                                                                                                                                                                                                                                                                                                                                                                                                                                                                                                          & \multicolumn{2}{c|}{Received Power}                                                                                                                                                                                                                                                                       \\ \cline{4-5} 
				&                                                                                                                       &                                                                                                                                                                                                                                                                                                                                                                                                                                                                                                                                                                                   & User $1$                                                                                                                                            & User $2$                                                                                                                                            \\ \hline
				\multirow{3}{*}{\begin{tabular}[c]{@{}c@{}}$\alpha_{12}\geq M,$\\
						$\alpha_{21}\geq M,$\\$\alpha_{12}\geq\alpha_{21}+$\\ $\min(\alpha_{11},\alpha_{22})$ \\ ($\alpha_{12}\geq\alpha_{21}$ is\\ assumed)\end{tabular}} & {\begin{tabular}[c]{@{}c@{}}$\frac{\pi}{2}\leq\alpha_{21}-M$\\  $\mathcal{D}_{\Sigma,\iclc}=M+\pi$\end{tabular}}                                                                                   & \begin{tabular}[c]{@{}c@{}}$X_{11}: d_{11}=\alpha_{11}$,$E|X_{11}|^2=1-P^{-\alpha_{11}}$\\ $X_{22}: d_{22}=\alpha_{22}$,$E|X_{22}|^2=1-P^{-\alpha_{22}}$\\
					$X_{01}^p:d_{01}^p=\frac{\pi}{2}$,$E|X_{01}^p|^2=P^{-\alpha_{22}}$\\ $X_{02}^p:d_{02}^p=\frac{\pi}{2}$,$E|X_{02}^p|^2=P^{-\alpha_{11}}$\end{tabular}          & \begin{tabular}[c]{@{}c@{}} $X_{22}:\sim P^{\alpha_{12}}$\\ $X_{01}^p:\sim P^{\alpha_{12}-\alpha_{22}}$\\$X_{11}:\sim P^{\alpha_{11}}$\\ $X_{02}^p:\sim P^{0}$\end{tabular}                                              & \begin{tabular}[c]{@{}c@{}}$X_{11}:\sim P^{\alpha_{21}}$\\ $X_{02}^p:\sim P^{\alpha_{21}-\alpha_{11}}$\\$X_{22}:\sim P^{\alpha_{22}}$\\ $X_{01}^p:\sim P^{0}$\\\end{tabular}                                              \\ \cline{2-5} 
				& \begin{tabular}[c]{@{}c@{}}$\alpha_{21}-M\leq\frac{\pi}{2}$\\ $\leq\alpha_{12}-M$ \\ 
					
					$\mathcal{D}_{\Sigma,\iclc}=\alpha_{21}+\frac{\pi}{2}$ \end{tabular} & \begin{tabular}[c]{@{}c@{}}$X_{11}: d_{11}=\alpha_{11}$,$E|X_{11}|^2=1-P^{-\alpha_{11}}$\\ $X_{22}: d_{22}=\alpha_{22}$,$E|X_{22}|^2=1-P^{-\alpha_{22}}$\\ $X_{01}^p:d_{01}^p=\frac{\pi}{2}$,$E|X_{01}^p|^2=P^{-\alpha_{22}}$\\ $X_{02}^p:d_{02}^p=\alpha_{21}-M$,$E|X_{02}^p|^2=P^{-\alpha_{11}}$\end{tabular}                                                                                                                                                                                                                                    & \begin{tabular}[c]{@{}c@{}}$X_{22}:\sim P^{\alpha_{12}}$\\ $X_{01}^p:\sim P^{\alpha_{12}-\alpha_{22}}$\\ $X_{11}:\sim P^{\alpha_{11}}$\\   $X_{02}^p:\sim P^{0}$\end{tabular}                       & \begin{tabular}[c]{@{}c@{}}$X_{11}:\sim P^{\alpha_{21}}$\\$X_{02}^p:\sim P^{\alpha_{21}-\alpha_{11}}$\\ $X_{22}:\sim P^{\alpha_{22}}$\\ $X_{01}^p:\sim P^{0}$ \end{tabular}                       \\ \cline{2-5}

				& \begin{tabular}[c]{@{}c@{}}$\alpha_{12}-M\leq\frac{\pi}{2}$\\ $\leq\alpha_{12}-\max(\alpha_{11},\alpha_{22})$\\  $\mathcal{D}_{\Sigma,\iclc}=\alpha_{21}+\frac{\pi}{2}$ \end{tabular} & \begin{tabular}[c]{@{}c@{}}$X_{11}: d_{11}=\alpha_{12}-\alpha_{22}-\frac{\pi}{2}$,$E|X_{11}|^2=1-P^{-d_{11}}$\\ $X_{22}: d_{22}=\alpha_{12}-\alpha_{11}-\frac{\pi}{2}$,$E|X_{22}|^2=1-P^{-d_{22}}$\\ $X_{01}^p:d_{01}^p=\frac{\pi}{2}$,$E|X_{01}^p|^2=P^{-\alpha_{22}}$\\ $X_{02}^p:d_{02}^p=\alpha_{21}-\alpha_{12}+\frac{\pi}{2}$,$E|X_{02}^p|^2=P^{-\alpha_{11}}$\\ $X_0^c:d_{0}^c=M-\alpha_{12}+\frac{\pi}{2}$,$E|X_0^c|^2=$\\ $\mbox{\tt Diag}(P^{-d_{11}}-P^{-\alpha_{11}},P^{-d_{22}}-P^{-\alpha_{22}})$\end{tabular} & \begin{tabular}[c]{@{}c@{}}$X_{22}:\sim P^{\alpha_{12}}$\\$X_{0}^c:\sim P^{\alpha_{12}-d_{22}}$\\ $X_{01}^p:\sim P^{\alpha_{12}-\alpha_{22}}$\\ $X_{11}:\sim P^{\alpha_{11}}$\\$X_{02}^p:\sim P^{0}$ \end{tabular} & \begin{tabular}[c]{@{}c@{}}$X_{11}:\sim P^{\alpha_{21}}$\\$X_{0}^c:\sim P^{\alpha_{21}-d_{11}}$\\$X_{02}^p:\sim P^{\alpha_{21}-\alpha_{11}}$\\  $X_{22}:\sim P^{\alpha_{22}}$\\ $X_{01}^p:\sim P^{0}$ \end{tabular} \\ \hline
			\end{tabular}
		\end{adjustbox}
		\caption{\small \it The achievable scheme for Case 4 under the condition $\alpha_{12}\geq\alpha_{21}+\min(\alpha_{11},\alpha_{22})$.}
		\label{tab:case5}
	\end{table}
	\begin{itemize}
		\item $\alpha_{12}\geq\alpha_{21}+\min(\alpha_{11},\alpha_{22})$\\ 
		The achievable scheme for this subcase is shown in Table \ref{tab:case5}. In this regime the sum-GDoF value is 
		\begin{align}
		\mathcal{D}_{\Sigma,\iclc}'=\min\Big(\alpha_{11}+\alpha_{22}+\pi,\alpha_{21}+\frac{\pi}{2},\mathcal{D}_{\Sigma,\bc}\Big)
		\end{align}
		\begin{itemize}
			\item  When $\frac{\pi}{2}\leq \alpha_{21}-\alpha_{11}-\alpha_{22}$, the first bound is active. Messages $W_{11},W_{22},W_{01}^p,W_{02}^p$ carry $\alpha_{11},\alpha_{22},\frac{\pi}{2},\frac{\pi}{2}$ GDoF respectively. They are encoded into independent Gaussian codebooks producing codewords $X_{11},X_{22},X_{01}^p,X_{02}^p$ with powers $\E|X_{11}|^2=1-P^{-\alpha_{11}},\E|X_{22}|^2=1-P^{-\alpha_{22}},\E|X_{01}^p|^2=P^{-\alpha_{22}},\E|X_{02}^p|^2=P^{-\alpha_{11}}$. The transmitted signals are $X_1=X_{11}+X_{02}^p,X_2=X_{22}+X_{01}^p$. When decoding, User $1$ decodes $X_{22},X_{01}^p,X_{11}$ successively, whose SINR values are $\sim P^{\alpha_{22}},\sim P^{\alpha_{12}-\alpha_{11}-\alpha_{22}},\sim P^{\alpha_{11}}$. Since $d_{22}=\alpha_{22},d_{01}^p=\frac{\pi}{2}\leq \alpha_{21}-\alpha_{11}-\alpha_{22}\leq\alpha_{12}-\alpha_{11}-\alpha_{22}$, messages $W_{22},W_{01}^p,W_{11}$ can be decoded successfully. User $2$ proceeds similarly by decoding $X_{11},X_{02}^p,X_{22}$ successively.
			\item When $\alpha_{21}-\alpha_{11}-\alpha_{22}\leq\frac{\pi}{2}\leq \alpha_{12}-\alpha_{11}-\alpha_{22}$, the second bound is active. $W_{11},W_{22},W_{01}^p,W_{02}^p$ carry $\alpha_{11},\alpha_{22},\frac{\pi}{2},\alpha_{21}-\alpha_{11}-\alpha_{22}$ GDoF respectively. They are encoded into independent Gaussian codebooks producing codewords $X_{11},X_{22},X_{01}^p,X_{02}^p$ with powers $\E|X_{11}|^2=1-P^{-\alpha_{11}},\E|X_{22}|^2=1-P^{-\alpha_{22}},\E|X_{01}^p|^2=P^{-\alpha_{22}},\E|X_{02}^p|^2=P^{-\alpha_{11}}$. For decoding, User $1$ decodes $X_{22},X_{01}^p,X_{11}$ successively while User $2$ decodes $X_{11},X_{02}^p,X_{22}$ successively. The distinction between $\alpha_{21}-\alpha_{11}-\alpha_{22}\leq\frac{\pi}{2}\leq \alpha_{12}-\alpha_{11}-\alpha_{22}$ and $\frac{\pi}{2}\leq \alpha_{21}-\alpha_{11}-\alpha_{22}$ is that message $W_{02}^p$ carries different GDoF values.
			\item When $\alpha_{12}-\alpha_{11}-\alpha_{22}\leq\frac{\pi}{2}\leq\frac{\pi^+}{2}$,  where according to Corollary $2$ we have $\pi^+=2\alpha_{12}-2\max(\alpha_{11},\alpha_{22})$, the second bound is still active.  Messages $W_{11},W_{22}$, $W_{01}^p,W_{02}^p$ carry $\alpha_{12}-\alpha_{22}-\frac{\pi}{2},\alpha_{12}-\alpha_{11}-\frac{\pi}{2},\frac{\pi}{2},\alpha_{21}-\alpha_{12}+\frac{\pi}{2}$ GDoF respectively. They are encoded into independent Gaussian codebooks producing codewords $X_{11},X_{22},X_{01}^p,X_{02}^p,X_0^c=(X_{01}^c,X_{02}^c)$ with powers $\E|X_{11}|^2=1-P^{-d_{11}},\E|X_{22}|^2=1-P^{-d_{22}},\E|X_{01}^p|^2=P^{-\alpha_{22}},\E|X_{02}^p|^2=P^{-\alpha_{11}}$. $W_0^c$ carries $\alpha_{11}+\alpha_{22}-\alpha_{12}+\frac{\pi}{2}$ GDoF and is encoded into a vector Gaussian codebook $X_0^c=(X_{01}^c,X_{02}^c)$ with power covariance $\E|X_0^c|^2=\mbox{\tt Diag}(P^{-d_{11}}-P^{-\alpha_{11}},P^{-d_{22}}-P^{-\alpha_{22}})$. The transmitted symbols are $X_1=X_{11}+X_{01}^c+X_{02}^p,X_2=X_{22}+X_{02}^c+X_{01}^p$. User $1$ decodes $W_{22},W_0^c$ successively while treating everything else as noise, the SINR values are $\sim P^{d_{22}},\sim P^{\alpha_{22}-d_{22}}=P^{\alpha_{11}+\alpha_{22}+\frac{\pi}{2}-\alpha_{12}}=P^{d_0^c}$ respectively. After this User $1$ subtracts the contribution of $X_{22},X_0^c$, and it acts as a multiple access receiver by jointly decoding $W_{11}$ and $W_{01}^p$ while treating the remaining signals as noise. The GDoF region of this multiple access channel is the following.
			\begin{align}
			\{(d_{11},d_{01}^p):d_{11}\leq\alpha_{11},d_{11}+d_{01}^p\leq\alpha_{12}-\alpha_{22} \}
			\end{align}
			Since $d_{11}=\alpha_{12}-\alpha_{22}-\frac{\pi}{2}\leq\alpha_{11},d_{11}+d_{01}^p=\alpha_{12}-\alpha_{22}$ belongs to the GDoF region, $W_{11},d_{01}^p$ can be decoded successfully. User $2$ proceeds similarly. The signal partition depiction is similar to Figure \ref{fig:case_4} except that codewords' power levels are changed. It can be checked that $d_{01}=d_{01}^p=\frac{\pi}{2},d_{02}=d_{02}^p+d_0^c=\alpha_{21}-2\alpha_{12}+\alpha_{11}+\alpha_{22}+\pi\leq\alpha_{21}-2\alpha_{12}+\alpha_{11}+\alpha_{22}+\frac{\pi^+}{2}+\frac{\pi}{2}=\alpha_{21}-\alpha_{12}+\alpha_{11}+\alpha_{22}-\max(\alpha_{11},\alpha_{22})+\frac{\pi}{2}=\alpha_{21}-\alpha_{12}+\min(\alpha_{11},\alpha_{22})+\frac{\pi}{2}\leq\frac{\pi}{2}$.
		\end{itemize}
		\item $\alpha_{12}\leq\alpha_{21}+\min(\alpha_{11},\alpha_{22})$\\
		The sum-GDoF value is 
		\begin{align}
		\mathcal{D}_{\Sigma,\iclc}'=\min\Big(\alpha_{11}+\alpha_{22}+\pi,\alpha_{21}+\frac{\pi}{2},\frac{2\alpha_{12}+2\alpha_{21}-\alpha_{11}-\alpha_{22}+\pi}{3},\mathcal{D}_{\Sigma,\bc}\Big)
		\end{align}
		\begin{itemize}
			\item When $\frac{\pi}{2}\leq2\alpha_{12}-\alpha_{21}-\alpha_{11}-\alpha_{22}$, the achievability for the first and second bounds are the corresponding scheme as the regime $\alpha_{12}\geq\alpha_{21}+\min(\alpha_{11},\alpha_{22})$ for the same $\frac{\pi}{2}$ value.
			\item When $2\alpha_{12}-\alpha_{21}-\alpha_{11}-\alpha_{22}\leq\frac{\pi}{2}\leq\frac{\pi^+}{2}$, where according to Corollary $2$ we have $\pi^+=\alpha_{12}+\alpha_{21}+\alpha_{11}+\alpha_{22}-3\max(\alpha_{11},\alpha_{22})$ the third bound is active. The achievability is the same as the corresponding scheme in Case 4 of the half-duplex setting.
		\end{itemize}
	\end{itemize}
	
	\section{Conclusion}
	The aligned image sets approach of  \cite{Arash_Jafar}, and the sum-set inequalities of \cite{Arash_Jafar_sumset} are utilized to characterize the sum-GDoF of two user interference channel with limited cooperation, both in half-duplex setting and full-duplex setting, which bridges the gap between the interference channel and broadcast channel. The sum-GDoF value is characterized for arbitrary parameter regimes. Promising directions for future work include extensions to include more users and more messages, e.g., the $X$ channel setting \cite{Jafar_Shamai}. Notably, the $2$ user $X$ channel setting turns out to be quite straightforward. For the $X$ channel with limited cooperation, both with half or full-duplex operation, it is easy to see that $\mathcal{D}_{\Sigma,\xlc}=\min(\mathcal{D}_{\Sigma,\x}+\pi,\mathcal{D}_{\Sigma,\bc})$. The converse is trivial because the sum-GDoF of the non-cooperative messages are bounded by $\mathcal{D}_{\Sigma,\x}$ and the sum-GDoF of the cooperative messages are bounded by $\pi$, so the total sum-GDoF cannot exceed $\mathcal{D}_{\Sigma,\x}+\pi$. Also, the broadcast channel is still an outer bound. For achievability, let us first consider the weak interference channel regime, where  $\max(\alpha_{12},\alpha_{21})\leq\min(\alpha_{11},\alpha_{22})$. From  Theorem $2$ in \cite{Arash_Jafar_cooperation} we know that in the weak interference regime, $\mathcal{D}_{\Sigma,\x}=\mathcal{D}_{\Sigma,\ic}$, which means X channel boils down to the interference channel with message $W_{11},W_{22}$ from the sum-GDoF perspective. Therefore  $\mathcal{D}_{\Sigma,\xlc}=\min(\mathcal{D}_{\Sigma,\x}+\pi,\mathcal{D}_{\Sigma,\bc})=\min(\mathcal{D}_{\Sigma,\ic}+\pi,\mathcal{D}_{\Sigma,\bc})$ whose achievability is implied by Theorem 1 and Theorem 2 in this work. The strong interference regime maps to the weak interference regime by relabeling the parameters so the sum-GDoF are established for that as well. This leaves just the mixed interference regime. But from Theorem $2$ in \cite{Arash_Jafar_cooperation}, we know that $\mathcal{D}_{\Sigma,\x}=\mathcal{D}_{\Sigma,\bc}$ in the mixed interference regime, i.e., cooperation has no gain in the mixed interference regime. Thus, the sum-GDoF of the $2$ user $X$ channel with limited cooperation are easily characterized and turn out to be much simpler than the $2$ user interference channel. However, we expect that going beyond $2$ users will be challenging for the $X$ channel as well. Going further, the benefits of  limited receiver cooperation under finite precision CSIT are also of interest \cite{HsiangWang_DavidTse_RX}, as are other models of cooperation, such as in-band cooperation \cite{Viswanath_source,Viswanath_desination} which have previously been studied primarily under the idealized assumption of perfect CSIT.

	\bibliography{Thesis}
\end{document}